\documentclass[11pt]{article}
\usepackage[utf8]{inputenc}
\usepackage[T1]{fontenc}
\usepackage{lmodern}
\usepackage[DIV=11]{typearea} 
\usepackage{enumitem}
\usepackage{microtype}

\usepackage{amssymb}
\usepackage{amsmath}
\usepackage{amsthm}
\usepackage{thmtools}
\usepackage{thm-restate}

\usepackage{xcolor}
\usepackage{xspace}
\usepackage{xfrac}

\usepackage{comment}
\usepackage{algorithm}
\usepackage{algorithmic}

\newtheorem{theorem}{Theorem}[section]
\newtheorem*{theorem*}{Theorem}

\newtheorem{lemma}[theorem]{Lemma}
\newtheorem{claim}[theorem]{Claim}
\newtheorem{corollary}[theorem]{Corollary}

\newtheorem{observation}[theorem]{Observation}

\newtheorem{definition}{Definition}

\newcommand{\ED}{\operatorname{ED}}

\title{Approximating LCS and Alignment Distance over Multiple Sequences}

\author{%
Debarati Das%
\thanks{University of Copenhagen.
	Work supported by Basic Algorithms Research Copenhagen (BARC), grant 16582 from the VILLUM Foundation.
	Email: \texttt{debaratix710@gmail.com}.
}
\and
Barna Saha%
\thanks{University of California Berkeley, USA.
	Email: \texttt{barnas@berkeley.edu}. Work supported partly by NSF 1652303, 1909046, and HDR TRIPODS 1934846 grants, and an Alfred P. Sloan Fellowship.
}
}

\date{}

\begin{document}
\maketitle
\setcounter{page}{0}
\thispagestyle{empty}
\begin{abstract}
We study the problem of aligning multiple sequences with the goal of finding an alignment that either maximizes the number of aligned symbols (the longest common subsequence (LCS) problem), or minimizes the number of unaligned symbols (the alignment distance aka the complement of LCS). Multiple sequence alignment is a well-studied problem in bioinformatics and is used routinely to identify regions of similarity among DNA, RNA, or protein sequences to detect functional, structural, or evolutionary relationships among them. It is known that  exact computation of LCS or alignment distance of $m$ sequences each of length $n$ requires $\Theta(n^m)$ time unless the Strong Exponential Time Hypothesis is false. However, unlike the case of two strings, fast algorithms to approximate LCS and alignment distance of multiple sequences is lacking in the literature. In this paper, we make significant progress towards that direction.
\begin{itemize}
\item If the LCS of $m$ sequences each of length $n$ is $\lambda n$ for some $\lambda \in [0,1]$, then in $\tilde{O}_m(n^{\lfloor\frac{m}{2}\rfloor+1})$\footnote{In the context of multiple sequence alignment, we use $\tilde{O}_m$ to hide factors like $c^m\log^m{n}$ where $c$ is a constant.} time, we can return a common subsequence of length at least $\frac{\lambda^2 n}{2+\epsilon}$ for any arbitrary constant $\epsilon >0$.  In contrast, for two strings, the best known subquadratic algorithm may return a common subsequence of length $\Theta(\lambda^4 n)$. 
\item It is possible to approximate the alignment distance within a factor of two in time $\tilde{O}_m(n^{\lceil\frac{m}{2}\rceil})$ by splitting the $m$ sequences into two (roughly) equal sized groups, computing the alignment distance in each group and then combining them by using triangle inequality. However, going ``below'' two approximation requires breaking the triangle inequality barrier which is a major  challenge in this area. No  such algorithm with a running time of $O(n^{\alpha m})$ for any $\alpha < 1$ is known. 

If the alignment distance is $\theta n$, then we design an algorithm that approximates the alignment distance within an approximation factor of $\left(2-\frac{3\theta}{16}+\epsilon\right)$ in $\tilde{O}_m(n^{\lfloor\frac{m}{2}\rfloor+2})$ time. Therefore, if $\theta$ is a constant (i.e., for large alignment distance), we get a below-two approximation in $\tilde{O}_m(n^{\lfloor\frac{m}{2}\rfloor+2})$ time. Moreover, we show if just one out of $m$ sequences is $(p,B)$-pseudorandom then, we can get a below-two approximation in $\tilde{O}_m(nB^{m-1}+n^{\lfloor \frac{m}{2}\rfloor+3})$ time irrespective of $\theta$. In contrast, for two strings, if one of them is $(p,B)$-pseudorandom then only an $O(\frac{1}{p})$ approximation is known in $\tilde{O}(nB)$ time. 
\end{itemize}
\end{abstract}

\newpage
\section{Introduction}
Given $m$ sequences each of length $n$, we are interested to find an alignment that either maximizes the number of aligned characters (the longest common subsequence problem (LCS)), or minimizes the number of unaligned characters (the minimum alignment distance problem, aka the complement of LCS). These problems are extremely well-studied, are known to be notoriously hard, and form the cornerstone of {\em multiple sequence alignment} \cite{THG:94,p:92,G97}, which according to a recent survey in Nature is one of the most widely used modeling methods in biology \cite{VMN:14}. Long back in 1978, the multi-sequence LCS problem (and therefore, the minimum alignment distance problem) was shown to be NP Hard \cite{m:78}. Moreover, for any constant $\delta > 0$, the multi-sequence LCS cannot be approximated within $n^{1-\delta}$ unless $P=NP$ \cite{JL:95}. These hardness results hold even under restricted conditions such as for sequences over relatively small alphabet \cite{bck:20}, or with certain structural properties \cite{BBJTV:12}. Various other multi-sequence based problems such as finding the median or center string are shown ``hard'' by reduction from the minimum alignment distance problem \cite{NicolasRivals05}. 

From a fine-grained complexity viewpoint, an $O(n^{m-\epsilon})$ algorithm to compute LCS or alignment distance of $m$ sequences for any constant $\epsilon >0$ will refute the Strong Exponential Time Hypothesis (SETH) \cite{abw:15}. On the other hand, a basic dynamic programming solves these problems in time $O(mn^m)$. This raises the question whether we can solve these problems faster in $O(n^{m/\alpha})$ time for $\alpha >1$ by allowing approximation. The approximation vs running time trade-off for $m=2$ has received extensive attention over the last two decades with many recent breakthroughs \cite{LMS98,BEKMRRS03,BJKK04,BES06,AKO10,AO12,BEGHS18,CDGKS18,Rub18-blog,BR19,KS20,AN:20}. Andoni and Nosaztki show that in $\tilde{O}(n^{1+\epsilon})$ time, edit distance (also alignment distance) can be approximated within $O(f(\epsilon))$ factor (where $f(\epsilon)$ goes to infty as $\epsilon$ decreases)  \cite{AN:20}, whereas a $3+\epsilon$ approximation is possible in time $\tilde{O}(n^{1.6+\epsilon})$ \cite{GRS:20}. These results build upon the work of Chakraborty, Das, Goldenberg, Kouck{\'{y}} and Saks that gives the first constant factor approximation for edit distance in subquadratic time \cite{CDGKS18}. 

There are numerous works to approximate LCS as well \cite{HSSS:19,RSSS:19,hss:19,RS:20}. It is trivial to get an $\frac{1}{|\Sigma|}$ approximation in linear time irrespective of the number of sequences where $\Sigma$ denotes the alphabet. For binary alphabets and $m=2$, Rubinstein and Song show a slight improvement over this bound \cite{RS:20}. However, as argued in \cite{HSSS:19}, even for DNAs that consist of only four symbols, one may be interested in approximating the number of ``blocks of
nucleotides'' that the DNAs have in common where
each block carries some meaningful information about
the genes. In this case, every block can be seen as a
symbol of the alphabet and thus the size of the alphabet
is large. In general, it is possible to get a $\sqrt{n}$ approximation for LCS between two strings by sampling. Recent works have shown how to break the $\sqrt{n}$ barrier \cite{HSSS:19, BD21}. Specifically the celebrated result of Rubinstein, Seddighin, Song and Sun provides an $O(\lambda^3)$ approximation of LCS when the optimum LCS has size $\lambda n$ in $\tilde{O}(n^{1.95})$ time \cite{RSSS:19}. The later work ingeniously extends the concept of triangle inequality to {\em birthday triangle inequality}, and also employs the framework of \cite{CDGKS18}. It uses the $O(\lambda n^2)$ time algorithm for LCS of two strings \cite{hs:77} in the small $\lambda$ regime to attain their overall subquadratic running time.

In contrast, for $m \geq 3$, the landscape of approximation vs running time trade-off is much less understood. For LCS of three or more strings, no nontrivial approximation algorithm exists in the literature, whereas for the minimum alignment distance problem, an $c$ approximation in $\tilde{O}_m(n^{\lceil \frac{m}{c}\rceil})$ time is possible by creating $c$ groups of roughly equal size, computing the alignment distance in each group and then combining them by a simple application of the triangle inequality. In the later case, going below the $c$ approximation requires {\em breaking the triangle inequality barrier} which is a major bottleneck in this area.

Most results for $m=2$ do not generalize to multiple string settings. The sampling based algorithm for approximating LCS deteriorates fast with $m$, and the exact $\tilde{O}(\lambda n^2)$ time algorithm to compute LCS for $m=2$ translates only to an $\tilde{O}_m(\lambda n^m)$ algorithm giving an insignificant gain. The framework of \cite{CDGKS18} that has been a key to the development of approximation algorithms for $m=2$ would give significantly weaker bounds compared to the trivial algorithm for approximating alignment distance. 
However surprisingly for a closely related problem where given a set of strings, the objective is to find the shortest string containing each input string (famously know as {\em shortest superstring problem}), there exists a simple greedy algorithm that in time $O(n\log n)$ can compute a superstring that is $\alpha$ (where $2\le \alpha\le 3$) times longer than the shortest common superstring~\cite{BJLTY91,L90}.

LCS and equivalently, alignment distance remain one of the fundamental measures of sequence similarity for multiple sequences. Their applications are vivid and broad - ranging from identifying genetic similarities among species \cite{zb:07}, discovering common markers among cancer-causing genes \cite{aravanis2017next}, to efficient information retrieval from aligning domain-ontologies and detection of clone code segments \cite{lwb:12}. Interested readers may refer to the chapter entitled ``Multi String Comparison-the Holy Grail'' of the book \cite{G97} for a comprehensive study on this topic. Given their importance, the state of the art algorithmic results for these problems are unsatisfactory. 
In this paper, we provide several new results for LCS and alignment distance over multiple sequences. 

\paragraph*{Contributions}
\begin{description}[leftmargin=0pt]
\item[LCS of Multiple Sequences.]
Let $\mathcal{L}(s_1,\dots,s_m)$ denote the length of LCS of $m$ strings $s_1,s_2,..,s_m$. We show if $\mathcal{L}(s_1,\dots,s_m)$
\noindent$=\lambda n$ for some $\lambda \in [0,1]$, then we can return a common subsequence of length $\frac{\lambda^2 n}{2+\epsilon}$ in time $\tilde{O}_m(n^{\lfloor m/2 \rfloor+1})$. To contrast, we can get a quadratic algorithm for $m=3$ with $\frac{\lambda}{2+\epsilon}$ approximation (for any arbitrary small constant $\epsilon>0$), whereas the best known bound for $m=2$ case may return a subsequence of length $\Theta(\lambda^4 n)$ in $\tilde{O}(n^{1.95})$ time \cite{RSSS:19}.

\begin{theorem}
\label{thm:lcs}
Given $m$ strings $s_1,\dots,s_m$ of length $n$ over some alphabet set $\Sigma$ such that $\mathcal{L}(s_1,\dots,s_m)$
\noindent$=\lambda n$, where $\lambda\in[0,1]$, there exists an algorithm that for any arbitrary small constant $\epsilon>0$ computes an $\frac{\lambda}{2+\epsilon}$ approximation of $\mathcal{L}(s_1,\dots,s_m)$ in time $\tilde{O}_m(n^{\lfloor m/2 \rfloor +1})$.
\end{theorem}

\item[Minimizing Alignment Distance of Multiple Sequences.]
Let $\mathcal{A}(s_1,\dots,s_m)=n-\mathcal{L}(s_1,\dots,s_m)$ denote the optimal alignment distance of $m$ strings $s_1, s_2,..., s_m$ of length $n$. We show if $\mathcal{A}(s_1,\dots,s_m)=\theta n$ then for any arbitrary small constant $\epsilon >0$, it is possible to obtain a $c(1-\frac{3\theta}{32}+\epsilon)$ approximation\footnote{We will assume $c$ is even for simplicity. But all the algorithms work equally well if $c$ is odd.} in time $\tilde{O}_m(n^{\lceil m/c\rceil +2})$ time. 
\begin{theorem}
\label{thm:align1}
Given $m$ strings $s_1,\dots,s_m$ of length $n$ over some alphabet set $\Sigma$ such that $\mathcal{A}(s_1,\dots,s_m)=\theta n$, where $\theta\in(0,1)$, there exists an algorithm that for any arbitrary small constant $\epsilon>0$ computes a $(2-\frac{3\theta}{16}+\epsilon)$ approximation of $\mathcal{A}(s_1,\dots,s_m)$ in time $\tilde{O}_m(n^{\lfloor m/2 \rfloor+2})$. Moreover, for any integer $c>0$, there exists an algorithm that computes a $c(1-\frac{3\theta}{32}+\epsilon)$ approximation of $\Delta(s_1,\dots,s_m)$ in time $\tilde{O}_m(n^{\lceil m/c \rceil+2})$.
\end{theorem}

\noindent
For constant $\theta$,
the above theorem asserts that there exists an algorithm that breaks the triangle inequality barrier and computes a truly below $2$-approximation of $\mathcal{A}(s_1,\dots,s_m)$ in time $\tilde{O}_m(n^{\lfloor m/2 \rfloor+2})$. To get a below $c$-approximation (assume $c$ is even), we divide the input strings into $\frac{c}{2}$ groups each containing at most $\lceil\frac{2m}{c}\rceil$ strings. Then for each group we compute a below $2$-approximation of the alignment distance in time $\tilde{O}_m(n^{\lceil m/c \rceil+2})$.
 Finally, we apply triangle inequality $\frac{c}{2}$ times to combine these groups to get a $\frac{c}{2}(2-\frac{3\theta}{16}+\epsilon)=c(1-\frac{3\theta}{32}+\frac{\epsilon}{2})$ approximation.

\noindent{\bf If there is one pseudorandom string:} However, if we have one pseudorandom string out of $m$ strings, and the rest of $m-1$ strings are chosen adversarially, then irrespective of $\theta$, we can break the triangle inequality barrier!

\begin{definition}[$(p,B)$-pseudorandom]
Given a string $s$ of length $n$ and parameters $p,B\ge 0$ where $p$ is a constant, we call $s$ a $(p,B)$-pseudorandom string if for any two disjoint $B$ length substrings/subsequence $x,y$ of $s$, $\mathcal{A}(x,y)\ge pB$.
\end{definition}

We have the following theorem.
\begin{theorem}
\label{thm:align2}
Given a $(p,B)$-pseudorandom string $s_1$, and $m-1$ adversarial strings $s_2,\dots,s_m$ of length $n$, there exists an algorithm that for any arbitrary small constant $\epsilon>0$ computes $(2-\frac{3p}{512}+\epsilon)$ approximation of $\mathcal{A}(s_1,\dots,s_m)$ in time $\tilde{O}_m(nB^{m-1}+ n^{ \lfloor m/2\rfloor+3})$.
\end{theorem}

The theorem can be extended to get a $c(1-\frac{3p}{1024}+\epsilon)$ approximation in $\tilde{O}_m(nB^{\lceil 2m/c \rceil-1}+n^{\lceil m/c \rceil+3})$ time as discussed earlier.

\noindent{\bf What do we know in the two strings case?:} Let us contrast this result to what is known for $m=2$ case \cite{AK:12,K:19,BSS20}. When one of the two strings is $(p,B)$-pseudorandom, Kuszmaul gave an algorithm that runs in time $\tilde{O}(nB)$ time but only computes an $O(\frac{1}{p})$ approximation to edit distance \cite{K:19}. Boroujeni, Seddighin and Seddighin consider a different random model for string generation under which they give a $(1+\epsilon)$ approximation in subquadratic time \cite{BSS20}.
While their model captures the case when one string in generated uniformly at random, it does not extend to pseudorandom strings. Moreover, in order to apply their technique to multi-string setting, {\em we would need  all but one string to be generated according to their model.} In fact, there is no result in the two strings case that breaks the triangle inequality barrier and provides below-2 approximation when one of the strings is pseudorandom. Interestingly, our below-2 approximation algorithm for the large distance regime can further be extended to obtain the desired result with just {\em one pseudorandom string} for any distance regime. 
We stress that this is one of the important contributions of our work and is technically involved. 
\end{description}
\subsection{Technical Overview}
\label{sec:overview}
\paragraph*{Notation.} We use the following notations throughout the paper. Given $m$ strings $s_1,\dots,s_m$, each of length $n$ over some alphabet set $\Sigma$, the longest common subsequence (LCS) of $s_1,\dots,s_m$, denoted  by $LCS(s_1,\dots,s_m)$ is one of the longest sequences that is present in each $s_i$. Define $\mathcal{L}(s_1,\dots,s_m)=|LCS(s_1,\dots,s_m)|$. 
The optimal alignment distance (AD) of $s_1,\dots,s_m$, denoted by $\mathcal{A}(s_1,\dots,s_m)$ is $n-\mathcal{L}(s_1,\dots,s_m)$. 

For a given string $s$, $s[i]$ represents the $i$th character of $s$ and $s[i,j]$ represents the substring of $s$ starting at index $i$ and ending at index $j$. Given a LCS $\sigma$ of $s_1,\dots,s_m$ define $\sigma(s_j)\subseteq[n]$ be the set of indices such that for each $k\in \sigma(s_j)$, $s_j[k]$ is aligned in $\sigma$
and $\bar{\sigma}(s_j)\subseteq [n]$ be the set of indices of the characters in $s_j$ that are not aligned in $\sigma$. 
Define the \emph{alignment cost} of $\sigma$ to be $|\bar{\sigma}(s_1)|$ and the \emph{cumulative alignment cost} of $\sigma$ to be $\sum_{j=1}^{m}|\bar{\sigma}(s_j)|= m|\bar{\sigma}(s_1)|$. Given a set $T\subseteq [n]$ and a string $s$, let $s^T$ denote the subsequence of $s$ containing characters with indices in $T$.

Given a string $s$, we define a window $w$ of size $d$ of $s$ to be a substring of $s$ having length $d$. 
Given $m$ strings $s_1,\dots,s_m$, we define a $m$-window tuple to be a set of $m$ windows denoted by $(w_1,\dots,w_m)$, where $w_j$ is a window of string $s_j$.

Given two characters $a,b\in \Sigma$, $a\circ b$ represents the concatenation of $b$ after $a$.
Given two string $x,y$, $x\circ y$
represents the concatenation of string $y$ after $x$. For notational simplicity we use $\tilde{O}_m$ to hide factors like $c^m\log^m n$, where $c$ is a constant. Moreover we use $\tilde{O}$ to hide polylog factors. 

\subsubsection{Breaking the Triangle Inequality Barrier for Large Alignment Distance and Approximating LCS}
We first give an overview of our algorithms leading to Theorem~\ref{thm:lcs} (Section~\ref{sec:multilcs}) and Theorem~\ref{thm:align1} (Section~\ref{sec:large}). Let us consider the problem of minimizing the alignment distance. Given $m$ (say $m$ is even) sequences $s_1, s_2,..., s_m$ each of length $n$, partition them into two groups $G_1=\{s_1, s_2,...,s_{m/2}\}$ and $G_2=\{s_{m/2+1},..,s_{m}\}$. Suppose the optimum alignment distance of the $m$ sequences is $d=\theta n$. With each alignment, we can associate a set of indices of $s_1$ that are not aligned in that alignment. Let $\sigma^*$ be an optimum alignment and $\bar{\sigma^*}(s_1)$ be that set.  We have $|\bar{\sigma}(s_1)|=d$. Let $\mathcal{X}_1=\{(\sigma_i, \bar{\sigma_i}(s_1))\}$ denote all possible alignments $\sigma_i$ of $G_1$ of cost at most $d$, $|\bar{\sigma_i}(s_1)|\leq d$. Then $(\sigma,\bar{\sigma}(s_1)) \in \mathcal{X}_1$. Therefore, if we can (i) find all possible alignments $\mathcal{X}_1$, and (ii) for each $(\sigma_i, \bar{\sigma_i}(s_1)) \in \mathcal{X}_1$ can verify if that is a valid alignment of $G_2$, we can find an optimal alignment.

Unfortunately, it is possible that $|\mathcal{X}_1|=\sum_{l \leq d}{{n}\choose{l}}$ which is prohibitively large. Therefore, instead of trying to find all possible alignments, we try to find a {\em cover} for $\mathcal{X}_1$ using a few alignments $(\tau_j, \bar{\tau_j}(s_1))$, $j=1,2,..,k$ such that for any $(\sigma_i, \bar{\sigma_i}(s_1)) \in \mathcal{X}_1$, there exists a $(\tau_j, \bar{\tau_j}(s_1))$ with large $|\bar{\sigma_i}(s_1) \cap \bar{\tau_j}(s_1)|$. In fact, one of the key ingredients of our algorithm is to show such a covering exists and can be obtained in time (roughly) $n^{|G_1|}$. With just $k=\frac{4}{\theta}$ alignments, we show it is possible to cover $\mathcal{X}_1$ such that for any $(\sigma_i, \bar{\sigma_i}(s_1)) \in \mathcal{X}_1$, there exists a $(\tau_j, \bar{\tau_j}(s_1))$ having $|\bar{\sigma_i}(s_1) \cap \bar{\tau_j}(s_1)|\geq \frac{3 \theta^2 n}{16}$. 

The algorithm to compute the covering starts by finding any optimal alignment $(\sigma_1, \bar{\sigma_1}(s_1))$ of $G_1$. Next it finds another alignment $(\sigma_2, \bar{\sigma_2}(s_1))$ of cost at most $d$ which is {\em farthest} from $(\sigma_1, \bar{\sigma_1}(s_1))$, that is $|\bar{\sigma_1}(s_1) \cap \bar{\sigma_2}(s_1)|$ is minimized. If $|\bar{\sigma_1}(s_1) \cap \bar{\sigma_2}(s_1)|\sim |\bar{\sigma_2}(s_1)|$, then it stops. Otherwise, it finds another alignment $(\sigma_3, \bar{\sigma_3}(s_1))$ such that $|(\bar{\sigma_1}(s_1) \cup \bar{\sigma_2}(s_1)) \cap \bar{\sigma_3}(s_1)|$ is minimized. We show the process terminates after at most $\frac{4}{\theta}$ rounds.

Suppose without loss of generality, $|\bar{\sigma^*}(s_1) \cap \bar{\tau_1}(s_1)| \geq \frac{3 \theta^2 n}{16}$. Given $\bar{\tau_1}(s_1), \bar{\tau_2}(s_1),..., \bar{\tau_k}(s_1)$, for each $(\tau_i,\bar{\tau_i}(s_1))$, we find an alignment $(\rho_i,\bar{\rho_i}(s_1))$ of $G_2 \cup s_1$ of cost at most $d$ such that $\bar{\rho_i}(s_1)$ is {\em nearest} to $\bar{\tau_i}(s_1)$, that is $|\bar{\rho_i}(s_1) \cap \bar{\tau_i}(s_1)|$ is maximized. Then, we must have $|\bar{\rho_i}(s_1) \cap \bar{\tau_1}(s_1)| \geq |\bar{\sigma^*}(s_1) \cap \bar{\tau_1}(s_1)| \geq \frac{3 \theta^2 n}{16}$. 
Our alignment cost is 
$\min_j{(|\bar{\tau_j}(s_1) \cup \bar{\rho_j}(s_1)|)} \leq |\bar{\tau_1}(s_1) \cup \bar{\rho_1}(s_1)| \leq 2d-\frac{3 \theta^2 n}{16}=d(2-\frac{3\theta}{16})$
giving the desired below-2 approximation when $\theta$ is a constant.

Of course, there are two main parts in this algorithm that we have not elaborated; given a set of indices $T$ of $s_1$, and a group of strings $G$, we need to find an alignment of cost at most $d$ of $G \cup s_1$ that is farthest from (nearest to) $T$. In general, any application that needs to compute multiple diverse (or similar) alignments can be benefited by such subroutines. We can use dynamic programming to solve these problems; however, it is to keep in mind - an alignment that has minimum cost may not necessarily be the farthest (or nearest). Thus, we need to check all possible costs up to the threshold $d$ to find such an alignment.

Our algorithm for obtaining a $\frac{\lambda}{(2+\epsilon)}$ approximation for multi-sequence LCS is nearly identical to the above, and in fact simpler. This helps us to improve the running time slightly (contrast Theorem~\ref{thm:lcs} with Theorem~\ref{thm:align1}). Moreover, the result holds irrespective of the size of LCS.

\subsubsection{Approximating Alignment Distance with just One Pseudorandom String.}
Next we consider the case where the input consists of a single $(p,B)$ pseudorandom string and $m-1$ adversarial strings each of length $n$. We give an overview of our algorithm that returns a below $2$ approximation of the optimal alignment distance (even for small regime) proving Theorem~\ref{thm:align2}.

In most of the previous literature for computing edit distance of two strings, the widely used framework first partitions both the input strings into windows (a substring) and finds distance between all pairs of windows. Then using dynamic program all these subsolutions are combined to find the edit distance between the input strings. In this scenario instead of taking two arbitrary strings if one input is $(p,B)$ pseudorandom, then as any pair of disjoint windows of the  pseudorandom string have large edit distance, if we consider a window from the adversarial string then by triangular inequality,  there exists at most one window in the pseudorandom string with which it can have small edit distance ($\le \frac{pB}{4}$). We call this low cost match between an adversarial string window and a pseudorandom string window a \emph{unique match}. Notice if we can identify one such unique match that is part of an optimal alignment, we can put restriction on the indices where the rest of the substrings can be matched. This observation still holds for multiple strings but only when we compare a pair of windows, one from the pseudorandom string and the other from an  adversarial string. Thus it is not obvious how we can extend this restriction on pairwise matching to a matching of $m$-window tuples as, $(m-1)$-window tuples come from $(m-1)$ different adversarial strings and they can be very different. Another drawback of this approach is that, here the algorithm aims to identify only those matchings  which have low cost i.e. $< \frac{pB}{4}$. Hence the best approximation ratio we can hope for is $O(1/p)$ which can be a large constant. 

Therefore to shed the approximation factor below $2$, we also need to find a good approximation of the cost of pair of windows having distance $\ge \frac{pB}{4}$. We call a matching with cost $\ge \frac{pB}{4}$ a \emph{large cost match}. As the unique match property fails here if we compute the cost trivially for all pairs of windows having large cost, we can not hope for a better running time. Fortunately, as $p$ is a constant $pB/4=\Omega(B)$ and thus we can use our large alignment distance approximation algorithm to get a nontrivial running time while ensuring below $2$ approximation. Though this simple idea seems promising, if we try to compute an approximation over all large distance $m$-window tuples the running time will become roughly $\tilde{O}_m(n^{\frac{11m}{16}})$. 
We show this with an example. Start by partitioning each string into windows of length $n^{\frac{5}{8}}$ (for simplicity assume the windows are disjoint and $B\le n^{\frac{5}{8}}$). Hence there are $n^{\frac{3}{8}}$ windows in each string. Now there can be as many as $n^{\frac{3m}{8}}$ many $m$-window tuples of large cost. If we evaluate each of them using our large alignment distance algorithm then time taken for each tuple is roughly $\tilde{O}_m(n^{\frac{5m}{16}})$. Hence total time required is $\tilde{O}_m(n^{\frac{11m}{16}})$. Note we can not keep the window size arbitrarily large as to find a single unique match the required time is $\tilde{O}_m(n^{\frac{5m}{8}})$ for exact computation. We later show the total unique matches that we evaluate is at most $\tilde{O}_m(1)$. Hence we get a running time bound $\tilde{O}_m(n^{\frac{11m}{16}})$.

To further reduce the running time, instead of evaluating all window tuples having large cost match, we restrict the computation by estimating the cost of only those tuples that maybe necessary to compute an optimal alignment. This is challenging as we do not have any prior information of the optimal alignment. For this purpose we use an adaptive strategy where depending on the unique matches computed so far, we perform a restricted search to estimate the cost of window tuples having large alignment distance. Moreover the length of these windows are also decided adaptively. Next we give a brief overview of the three main steps of our algorithm. In Step 1, we provide the construction of windows of the input strings that will be used as input in Step 2 and 3. In Step 2, we estimate the alignment cost of the $m$-window tuples such that the matching is unique i.e. the optimal alignment distance is at most $pB/4$. In step 3, we further find an approximation of the cost of relevant $m$-window tuples having large distance i.e. $\ge pB/4$.

\paragraph{Step 1.} 

We describe the construction of windows of the input strings in two stages. In stage one, we follow a rather straightforward strategy similar to the one used in \cite{GRS:20} and partition the $(p,B)$ pseudorandom string into $\frac{n}{\beta}$ disjoint windows each of size $\beta$ (except the right most one). Here $\beta=max(B,\sqrt{n})$. For the rest of the strings we generate a set of overlapping variable sized windows. If the distance threshold parameter is $\theta$ and the error tolerance parameter is $\epsilon$, then for each adversarial strings we generate windows of size $\{(\beta-\theta \beta), (1+\epsilon)(\beta-\theta \beta),(1+\epsilon)^2(\beta-\theta \beta),\dots, (\beta+\theta \beta) \}$ and from starting indices in $\{1,\epsilon\theta\beta+1,2\epsilon\theta\beta+1,\dots \}$. These windows are fed as an input to Step 2 of our algorithm.

\begin{figure}[tp]
    \centering
    \includegraphics[scale=0.5]{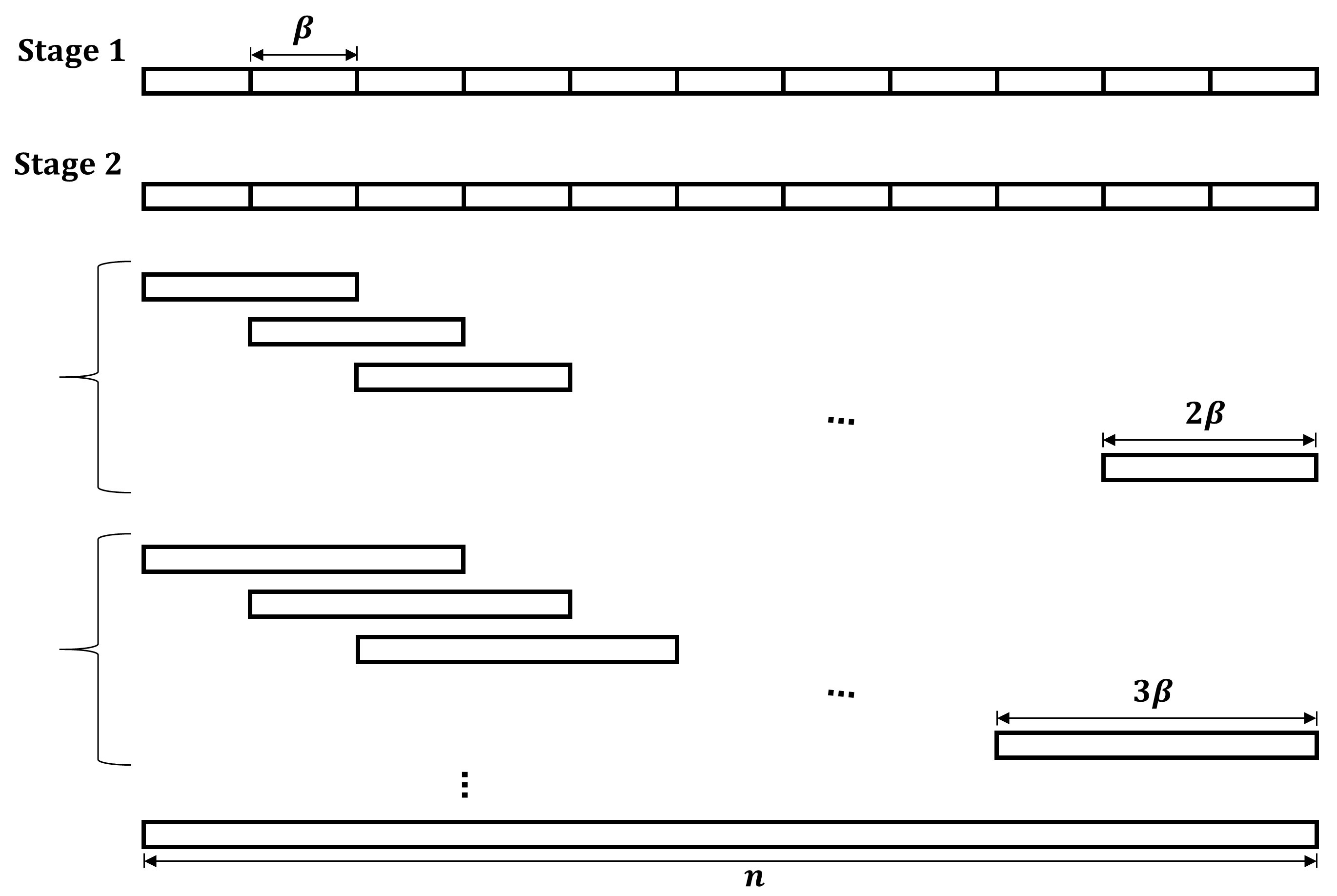}
    \caption{Construction of windows for the $(p,B)$-pseudorandom string $s_1$}
    \label{fig:fig1}
\end{figure}

The second stage is much more involved where the strategy significantly differs from the previous literature. The primary difference here is that instead of just considering a fixed length partition of the pseudorandom string, we take variable sizes that are multiple of $\beta$ i.e. $\{\beta,2\beta,\dots,n\}$ and the windows start from indices in $\{1,\beta+1,2\beta+1,\dots\}$. Next for each adversarial string, we adaptively try to guess a set of useful substrings where the large cost windows of the pseudorandom string are matched under the optimal alignment (we fix one for the analysis purpose). These guesses are guided by the unique low cost matches found in Step 2. Next for each such useful substring and each length in $\{\beta,2\beta,\dots,n\}$, we create a set of overlapping windows as described in stage one. Note the windows created in this stage are used as an input to Step 3 of the algorithm. 

\begin{figure}[tp]
    \centering
    \includegraphics[scale=0.7]{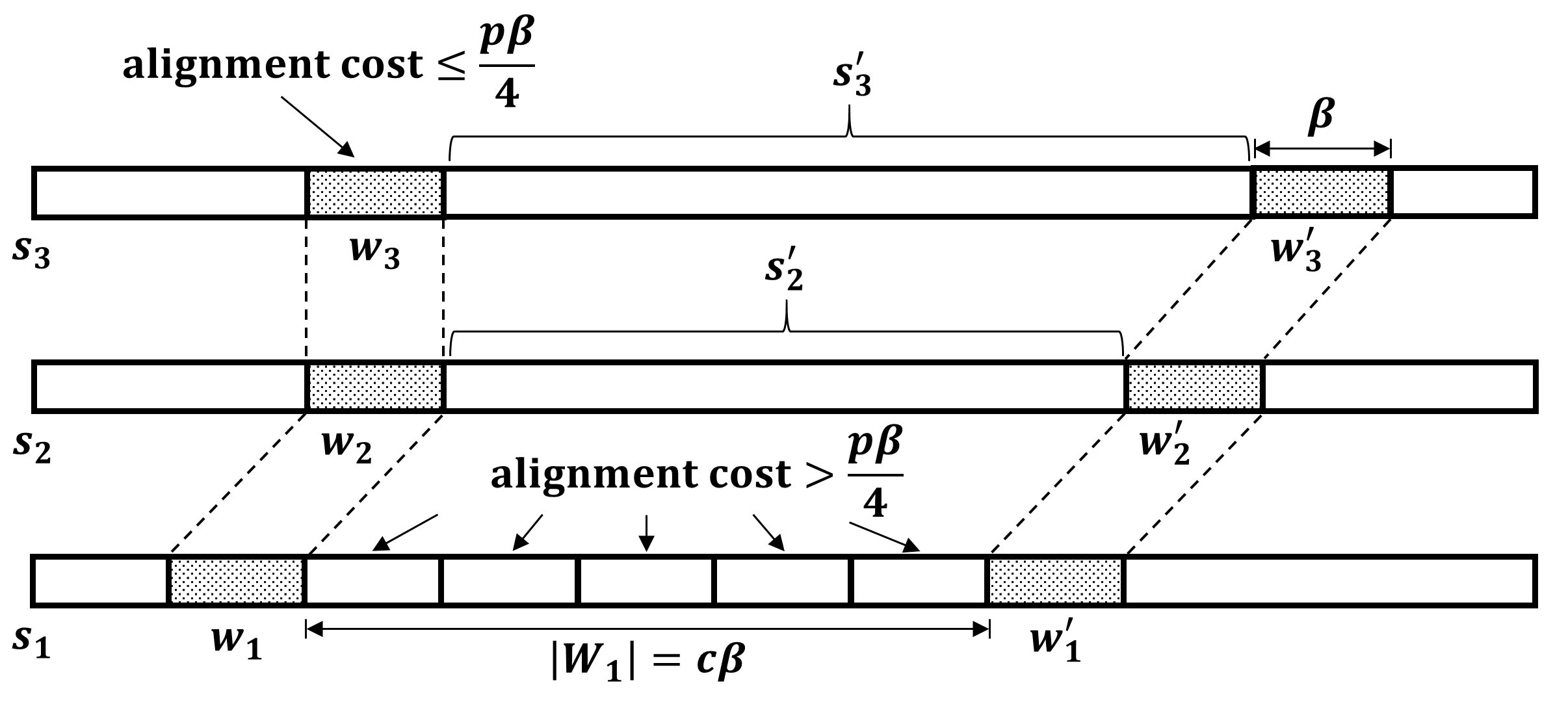}
    \caption{An example of large cost match}
    \label{fig:fig2}
\end{figure}

We explain the motivation behind the variable sized window partitioning for the pseudorandom string and the restricted   window construction for the adversarial strings with an example(see Figure \ref{fig:fig2}).

We start with three input strings where $s_1$ is $(p,B)$-pseudorandom. We divide all three strings in windows (for simplicity assume these windows are disjoint) of size $\beta$. For the analysis purpose fix an optimal alignment where window $w_1$ is aligned with window $w_2,w_3$ and window $w'_1$ is aligned with window $w'_2,w'_3$. Moreover their costs are $\le\frac{p\beta}{4}$. Hence we get an estimation of the alignment cost of these window tuples at Step 2. Also assume that all the windows appearing between $w_1$ and $w'_1$ in $s_1$ has alignment cost $>\frac{p\beta}{4}$ but $\le \frac{\beta}{2}$. Hence, for these windows we can not use a trivial maximum cost of $\beta$ in order to get a below two approximation. Here we estimate the cost of these windows using Algorithm LargeAlign(). Now if we compute this estimation separately for each window between $w_1$ and $w'_1$ in $s_1$, as there can be as many as $\sqrt{n}$ (assume $\beta=\sqrt{n})$ such windows and each call to algorithm LargeAlign() takes time $O(n^{m/4})$, the total running time will be $O(n^{3m/4})$. To further reduce the running time, instead of considering all small $\beta$ size windows separately we consider the whole substring between $w_1$ and $w'_1$ as one single window $W_1$ and compute its cost estimation. Observe as we do not have the prior information about the optimal alignment (and therefore $w_1$ and $w'_1$) we try all possible lengths in $\beta,2\beta,\dots,n$. The overall idea here is to use a window length so that we can represent a whole substring with large optimal alignment distance that lies between two windows having low cost unique match in the optimal alignment with a single window. Given the matchings among $w_1,w_2,w_3$ and 
$w_1',w_2',w_3'$ present in the optimal alignment it is enough to find a match of $W_1$ in the substring $s'_2$ and $s'_3$ and we construct the windows in $s_2$ and $s_3$ accordingly. 

\paragraph{Step 2.}

In step 2, we start with a set of windows, each of size $\beta$, generated from strings $s_1,\dots,s_m$. Our objective is to identify all $m$-window tuples that have optimal alignment distance $\le \frac{p\beta}{4}$. Here we use the fact that for every adversarial window there exists at most one window in the pseudorandom string that has distance $\le \frac{p\beta}{4}$. We start the algorithm by computing for each window of the pseudorandom string, the set of windows from each $s_j$ that are at distance $\le \frac{p\beta}{4}$. Notice for two string case, if for a pseudorandom window this set size is $\ge \frac{16}{p\epsilon}$, then as at most one adversarial window can be matched with cost $\le \frac{p\beta}{4}$ in the optimal alignment, we can estimate a cost of $\beta$ (maximum cost) for the pseudorandom window and this still gives a $(1+\epsilon)$ approximation of the alignment cost. For multiple strings though we can not use a similar idea; for a pseudorandom window $w$ we can not take a trivial cost estimation if for only one adversarial string, there are many windows which are close to $w$. 

We explain with an example.

\begin{figure}[tp]
    \centering
    \includegraphics[scale=0.47]{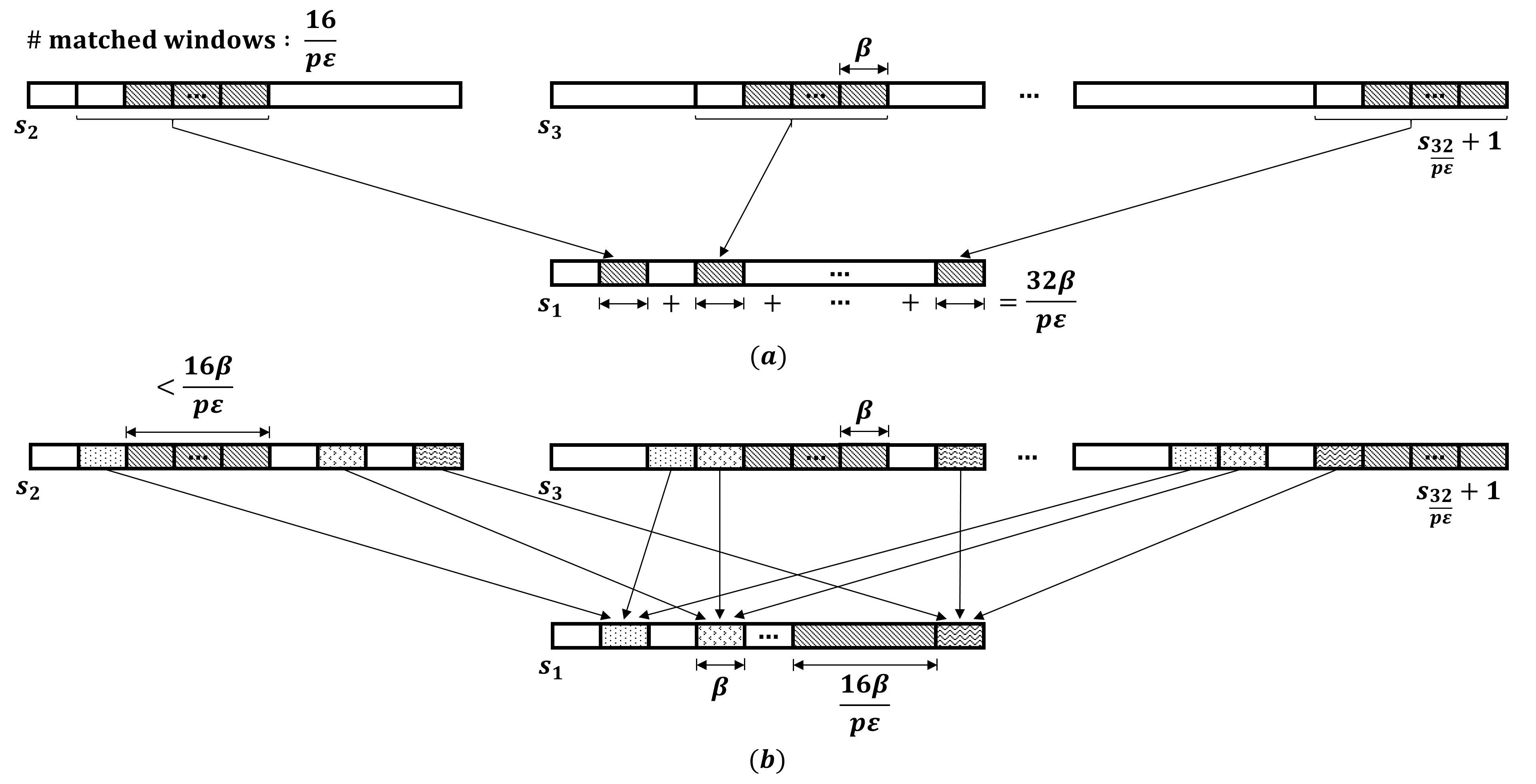}
    \caption{(a) Each hatch window of $s_1$ is matched with $\frac{16}{p\epsilon}$ windows of one adversarial string. Deleting all these windows deletes $\frac{32\beta}{p\epsilon}$ characters from each string. (b) The optimal alignment deletes $<\frac{16\beta}{p\epsilon}$ characters from each string.}
    \label{fig:fig3}
\end{figure}

Consider $k$ windows $w_1,\dots,w_k$ (where $k\ge \frac{32}{p\epsilon}$) from the pseudorandom string such that for each $w_j$, there are $\frac{16}{p\epsilon}$ windows from the adversarial string $s_{j+1}$ that are at distance $\le \frac{p\beta}{4}$ with $w_j$ and for every other string there is exactly one window with distance $\le \frac{p\beta}{4}$. Here following the above argument if for each of the $k$ pseudorandom windows, we take a trivial cost estimation of $\beta$ then the total cost will be $k\beta \ge \frac{32\beta}{p\epsilon}$. Whereas the optimal cost can be $\frac{16\beta}{p\epsilon}$(check Figure \ref{fig:fig3} (a)).

Therefore for every window $w_i$ of the pseudorandom string, we count the total number of windows in all adversarial strings that are at distance $\le \frac{p\beta}{4}$ and if the count is at least $\frac{16m}{p\epsilon}$, then we take a trivial cost estimation of $|w_i|$ for $w_i$. (Note here for two different windows of the pseudorandom string, these sets of close windows from the adversarial strings are disjoint.) Otherwise if the count is small let $\mathcal{W}_j$ be the set of windows from string $s_j$ that are close to $w_i$, then we can bound $|\mathcal{W}_2|\times \dots \times |\mathcal{W}_m|\le \left[(|\mathcal{W}_2|\times \dots\times |\mathcal{W}_m|)/(m-1)\right]^{m-1} \leq  (\frac{32}{p\epsilon})^m=\tilde{O}_m(1)$ and for each choice of $m$-window tuples in $w_i\times \mathcal{W}_2\times \dots\times \mathcal{W}_m$, we find the alignment cost in time $\tilde{O}_m(\beta^{m})$. As there are $n/\beta$ disjoint windows in the pseudorandom string we can bound the total running time by $\tilde{O}_m(n\beta^{m-1})$.

\paragraph{Step 3.}
In this step, our objective is to find a cost estimation for all the windows of the pseudorandom string that have large alignment cost in the optimal alignment. Consider Figure \ref{fig:fig2}, and let $W_1$ be such a window. Also assume that it can not be extended to left or right, i.e. both $w_1$ and $w'_1$ have small cost match which we already have identified in Step 2. Let the length of $W_1$ is $c\beta$ where $c \geq 1$ and assume no $\beta$ length window of $W_1$ has small cost match. Note, we have the assurance that window $W_1$ is generated in Step 1. Next to find the match of $W_1$, instead of checking whole $s_2$ and $s_3$, we consider the substring between $w_2$ and $w'_2$ in $s_2$ and $w_3$ and $w'_3$ in $s_3$. Now if the sum of the length of all the substrings is large (i.e. $|s'_2|+|s'_3|\ge 5m|W_1|=10|W_1|$), we can claim that the cost of $W_1$ in the optimal alignment is very large and we can take a trivial cost estimation of $|W_1|$. Otherwise, we can ensure that the total choices for $m$-window tuple that need to be evaluated for $W_1$ is at most $\tilde{O}_m(1)$ and for each of them, we calculate a below 2 approximation of the cost using Algorithm LargeAlign(). In the algorithm, as we don't know the optimal alignment, for every window of the pseudorandom string generated in Step 1 stage two, we assume it to be large cost and check whether the $\beta$ length window appearing just before and after this window has a small cost unique match. If not we discard it and consider a larger window.  Otherwise we perform the above step to find its cost estimation. Notice, we ensure that for every window of the pseudorandom string, the algorithm provides $<2$ cost approximation. Moreover for each window, we evaluate at most $\tilde{O}_m(1)$ $m$-window tuples where if they have small cost then we compute the cost exactly in time $\tilde{O}_m(\beta^m)$ (as the window size is $\beta$) and otherwise if the cost is large we use algorithm LargeAlgin() that computes an approximation in time $\tilde{O}_m(n^{m/2})$ (here the window length can be as large as $n$). As the number of windows generated for string $s_1$ is polynomial in $n$, we get the required running time bound.
\paragraph*{Organization} In Section~\ref{sec:large}, we give the algorithm for minimizing alignment distance when the distance is large. Section~\ref{sec:pseudo} provides the details of the below-2 approximation algorithm for alignment distance when one string is $(p, B)$ pseudorandom. In Section~\ref{sec:multilcs}, we give our $\frac{\lambda}{2+\epsilon}$ approximation algorithm for multi-sequence LCS.

\sloppy
\section{Below-$2$ Approximation for Multi-sequence Alignment Distance}
\label{sec:large}
In this section, we provide an algorithm LargeAlign() that given $m$ strings $s_1,\dots,s_m$ each of length $n$, such that $\mathcal{A}(s_1,\dots,s_m)=\theta n$, where $\theta\in(0,1)$ computes a $(2-\frac{3\theta}{16}+\epsilon)$ approximation of $\mathcal{A}(s_1,\dots,s_m)$ in time $\tilde{O}_m(n^{\lfloor m/2\rfloor+2})$. Notice when $\theta=\Omega(1)$, this implies a below $2$ approximation of $\mathcal{A}(s_1,\dots,s_m)$. 


\begin{theorem*} [\ref{thm:align1}]
Given $m$ strings $s_1,\dots,s_m$ of length $n$ over some alphabet set $\Sigma$ such that $\mathcal{A}(s_1,\dots,s_m)=\theta n$, where $\theta\in(0,1)$, there exists an algorithm that for any arbitrary small constant $\epsilon>0$ computes a $(2-\frac{3\theta}{16}+\epsilon)$ approximation of $\mathcal{A}(s_1,\dots,s_m)$ in time $\tilde{O}_m(n^{\lfloor m/2 \rfloor+2})$. Moreover, there exists an algorithm that computes a $c(1-\frac{3\theta}{32}+\epsilon)$ approximation of $\Delta(s_1,\dots,s_m)$ in time $\tilde{O}_m(n^{\lceil m/c \rceil+2})$.
\end{theorem*}

As we do not have any prior knowledge of $\theta$, instead of proving the theorem directly, we solve the following gap version for a given fixed threshold $\theta$. We define the gap version as follows.

\noindent{\bf GapMultiAlignDist$(s_1,\dots,s_m,\theta,c)$}: Given $m$ strings $s_1,\dots,s_m$ of length $n$ over some alphabet set $\Sigma$, $\theta\in (0,1)$ and a constant $c>1$, decide whether $\mathcal{A}(s_1,\dots,s_m)\le \theta n$ or $\mathcal{A}(s_1,\dots,s_m)> c\theta n$. More specifically if $\mathcal{A}(s_1,\dots,s_m)\le \theta n$ we output $1$, else if $\mathcal{A}(s_1,\dots,s_m)> c\theta n$ we output $0$, otherwise output any arbitrary answer. 

\begin{theorem}
\label{thm:align1gap}
Given $m$ strings $s_1,\dots,s_m$ of length $n$ over some alphabet set $\Sigma$ and a parameter $\theta\in(0,1)$, there exists an algorithm that computes GapMultiAlignDist$(s_1,\dots,s_m,\theta,(2-\frac{3\theta}{16}))$ in time $\tilde{O}_m(n^{\lfloor m/2\rfloor+2})$.
\end{theorem}

\begin{proof}[Proof of Theorem \ref{thm:align1} from Theorem \ref{thm:align1gap}]
Let us consider an arbitrary small constant $\epsilon' >0$, and fix a sequence of parameters $\theta_0, \theta_1, \dots$ as follows: for $i=0,1,\dots,\log_{1+\epsilon'} n$, $\theta_i=1/(1+\epsilon')^i$. Find the largest $i$ such that GapMultiAlignDist$(s_1,\dots,s_m,\theta_i,(2-\frac{3\theta}{16}))=1$. Let $\mathcal{A}(s_1,\dots,s_m)=\theta n$. Then there exists a $\theta_i$, such that $\theta_i n\ge \theta n> \theta_{i+1} n$ as $\theta_i=(1+\epsilon')\theta_{i+1}$. In this case the algorithm outputs a value at most $(2-\frac{3\theta}{16})\theta_i n\le (2-\frac{3\theta}{16}+\epsilon'(2-\frac{3\theta}{16}))\theta n$. As $\theta\ge 1/n$, by appropriately scaling $\epsilon'$, we get the desired running time of Theorem \ref{thm:align1}. 
\end{proof}

The rest of the section is dedicated towards proving Theorem \ref{thm:align1gap}. Before providing the the algorithm for computing $GapMultiAlignDist(s_1,\dots,s_m,\theta_i,(2-\frac{3\theta}{16}))$, we first outline another two algorithms that will be used as subroutines in our main algorithm.

\subsection{Finding Alignment with Maximum Deletion Similarity}
\label{sec:maxdel}

Given $m$ strings $s_1,\dots,s_m$ of length $n$, a set $S\subseteq[n]$ and a parameter $0\le d\le n$, our objective is to compute an alignment $\sigma_n$ of $s_1,\dots,s_m$ with alignment cost at most $d$ such that $|\bar{\sigma}_n(s_1)\cap S|$ is maximized. 

\begin{theorem}
Given $m$ strings $s_1,\dots,s_m$, each of length $n$, a set $S\subseteq[n]$ and a parameter $0\le d\le n$, there exists an algorithm that computes a common alignment $\sigma_n$ such that  $\sum_{k\in [m]}|\bar{\sigma}_n(s_k)|\le dm$ and $ |\bar{\sigma}_n(s_1)\cap S|$ is maximized in time $\tilde{O}(2^mmn^{m+1})$.
\end{theorem}

We refer to the total number of unaligned characters $\sum_{k\in [m]}|\bar{\sigma}_n(s_k)|$ as the cumulative alignment cost. Our algorithm MaxDelSimilarAlignment() uses dynamic program to compute $\sigma_n$. The dynamic program matrix $D$ is defined as follows: $D_{i_1,\dots,i_m,x}$ where, $i_1,\dots,i_m\in[n]$ and $x\in[mn]$. It represents in a common subsequence of $s_1[1,i_1],\dots,s_m[1,i_m]$ with cumulative cost at most $x$ what is the maximum overlap possible between the unaligned characters of $s_1[1,i_1]$ with $S$.

We can compute $D_{i_1,\dots,i_m,x}$ recursively as follows. We consider the following four different cases and output the one that maximizes the overlap.

\noindent{\textbf{Case 1.}} First consider the scenario where all the $i_j$th characters of string $s_j$s are not the same and therefore we can not align them. Moreover let $i_1\in S$. Hence not aligning the $i_1$th character of $s_1$ will increase the overlap by one.
Hence to compute the maximum overlap consider each possible subsets (except the null set) of the $m$, $C \subset [m]/\phi$, input strings and for each string $s_j \in C$, delete the $i_j$th character and find the maximum overlap of these modified substrings. Moreover add one to this overlap if $s_1$ is selected in the subset. Also ensure that the cumulative alignment cost is at most $x$. Output the one that maximizes the overlap. Formally if $\exists j\in [2,m]$ such that $s_j[i_j]\neq s_1[i_1]$ and $i_1\in S$ we can define $D_{i_1,\dots,i_m,x}$ as follows.

 \begin{equation}
 D_{i_1,\dots,i_m,x}= 
max
  \begin{cases}
 \begin{cases}
 {max}_{\begin{subarray} CC\subseteq[m]\setminus{\phi}\\ x'\le x-|C|\\ 1\in C \end{subarray}} D_{i'_1,\dots,i'_m,x'}+1  & i'_j=i_j-1 \text{ if } j\in C \\
& \text{ else } i'_j=i_j 
 \end{cases}\\
\begin{cases}
 {max}_{\begin{subarray} CC\subseteq[m]\setminus{\phi}\\ x'\le x-|C|\\ 1\notin C \end{subarray}} D_{i'_1,\dots,i'_m,x'}  & i'_j=i_j-1 \text{ if } j\in C \\
& \text{ else } i'_j=i_j 
 \end{cases}
\end{cases}
\end{equation}

\noindent{\textbf{Case 2.}} Next consider the scenario where all the $i_j$th characters of string $s_j$s are not the same and $i_1\notin S$. Here also we do the same as above except the difference that as $i_1\notin S$, deleting the $i_1$th character of $s_1$ does not increase the count of the unaligned characters given by $S$. Formally if $\exists j\in [2,m]$ such that $s_j[i_j]\neq s_1[i_1]$ and $i_1\notin S$ we can define $D_{i_1,\dots,i_m,x}$ as follows.

 \begin{equation}
 D_{i_1,\dots,i_m,x}= 
\begin{cases}
 {max}_{\begin{subarray} CC\subseteq[m]\setminus{\phi}\\ x'\le x-|C| \end{subarray}} D_{i'_1,\dots,i'_m,x'}  & i'_j=i_j-1 \text{ if } j\in C \\
& \text{ else } i'_j=i_j 
 \end{cases}
\end{equation}

\noindent{\textbf{Case 3.}} Now consider the case where $s_1[i_1]=\dots=s_m[i_m]$ and $i_1\in S$. Here we have the option of both aligning and not aligning $s_1[i_1]$ whereas the second option will increase the count of the overlap by one. Hence to compute the maximum overlap, consider the following options and take the one that maximizes the overlap: 1) Align $s_1[i_1],\dots, s_m[i_m]$, and count the maximum overlap of the intersection of the indices of the unaligned characters of $s_1[1,i_1-1]$ and $S$ in an alignment of $s_1[1,i_1-1],\dots,s_m[1,i_m-1]$ of cumulative cost $\le x$. 2) consider all possible subsets (except the null set) of the $m$ input strings and for each string $s_j$ present in the subset, delete the $i_j$th character and find the maximum overlap of these modified substrings while ensuring that the total cumulative cost is at most $x$. Moreover add one to this overlap if $s_1$ is selected in the subset. Therefore if $s_1[i_1]=\dots=s_m[i_m]$ and $i_1\in S$ we can define $D_{i_1,\dots,i_m,x}$ as follows.

 \begin{equation}
 D_{i_1,\dots,i_m,x}= 
  max
  \begin{cases}
  {max}_{x'\le x}D_{i_1-1,\dots,i_m-1,x'} \\
  
 \begin{cases}
 {max}_{\begin{subarray} CC\subseteq[m]\setminus{\phi}\\ x'\le x-|C|\\ 1\in C \end{subarray}} D_{i'_1,\dots,i'_m,x'}+1  & i'_j=i_j-1 \text{ if } j\in C \\
& \text{ else } i'_j=i_j 
 \end{cases}\\
 \begin{cases}
 {max}_{\begin{subarray} CC\subseteq[m]\setminus{\phi}\\ x'\le x-|C|\\ 1\notin C  \end{subarray}} D_{i'_1,\dots,i'_m,x'}  & i'_j=i_j-1 \text{ if } j\in C \\
& \text{ else } i'_j=i_j 
 \end{cases}
 \end{cases}
\end{equation}

\noindent{\textbf{Case 4.}} Consider the case where $s_1[i_1]=\dots=s_m[i_m]$ and $i_1\notin S$. Here we do the same as Case 3, except the difference that as $i_1\notin S$, deleting the $i_1$th character of $s_1$ does not increase the count of the overlap. Formally $D_{i_1,\dots,i_m,x}$ can be defined as follows.

 \begin{equation}
 D_{i_1,\dots,i_m,x}= 
 max
 \begin{cases}
 {max}_{x'\le x}D_{i_1-1,\dots,i_m-1,x'} \\
     \begin{cases}
 {max}_{\begin{subarray} CC\subseteq[m]\setminus{\phi}\\ x'\le x-|C| \end{subarray}} D_{i'_1,\dots,i'_m,x'}  & i'_j=i_j-1 \text{ if } j\in C \\
& \text{ else } i'_j=i_j 
 \end{cases}
 \end{cases}
\end{equation}

 \noindent{\bf Backtracking to Construct $\sigma_n$.} We can assume there is a new character $\#$ appended at the end of all $m$ strings which are aligned in $\sigma_n$. The subsequence $\sigma_n$ corresponds to $max_{x\le dm}D_{n+1,\dots,n+1,x}$. After computing $D$ we use backtracking to compute $\sigma_n$. We start with the entry $max_{x\le d}D_{n+1,\dots,n+1,x}$ and backtrack over $D$ to find the entry on which we build the entry $D_{n,\dots,n,x}$. We continue this process until we reach $D_{0,\dots,0,0}$. If 
 $D(i_1,\dots,i_m,x)$ is computed from  $D(i'_1,\dots,i'_m,x')$ then either (i) $x'=x$ and $\forall \ell\in[m]$, $i'_\ell<i_\ell$, or (ii) $x' < x$ then $\exists \ell\in[m]$, such that $i'_\ell<i_\ell$. In this process, we retrieve an ordered set $A$ of at most $(mn+1)$ entries $(i_1^1,\dots,i_m^1,x^1),\dots,(i_1^{mn+1},\dots,i_m^{mn+1},x^{mn+1})$ where each entry corresponds to a $(m+1)$-tuple.
 
  
  Construct a set $A'_e\subseteq A$ that contains the tuple having the first occurrence of $i_1^j$ in set $A$ (last during the backtracking) for all $j$. Similarly define set $A'_s\subseteq A$ that contains the tuple having the last occurrence of $i_1^j$ in $A$ (first during the backtracking) for all $j$.
  If we have $(\ell,i_2,\dots,i_m,x)\in A'_e$ and $(\ell+1,i'_2,\dots,i'_m,x')\in A'_s$ for some $\ell$ such that $x'=x$, then $s_1[\ell]$ is aligned in $\sigma_n$. Let $L=\{\ell_1,\dots,\ell_p=n+1\}$ denote the set of indices of all the aligned characters of $s_1$. Define $\bar{L}=[n]\setminus L$, the set of indices of the unaligned characters of $s_1$. Define $\sigma_n=s_1[\ell_1]\circ s_1[\ell_2]\circ\dots\circ s_1[\ell_p]$.
  For the purpose of our main algorithm the subroutine MaxDelSimilarAlignment() returns the set $\bar{L}$.

\paragraph{Running Time Analysis.} As the cumulative alignment cost of $s_1,\dots,s_m$ is bounded by $mn$, there are $mn^{m+1}$ entries in the dynamic program matrix $D$. To compute each entry of $D$, the algorithm checks at most $2^mmn$ previously computed values and take their maximum. This can be done in time $O(2^mmn)$. Hence the total running time is $O(2^mm^2n^{m+2})$. We can further reduce the running time in the following way. Notice if $D(i_1,\dots,i_m,x)$th entry is computed from  $D(i'_1,\dots,i'_m,x')$th entry then, if $x'=x$ then $\forall \ell \in [m]$, $i'_\ell<i_\ell$ and if $x'<x$ then $\exists \ell\in[m]$, such that $i'_\ell<i_\ell$. Hence for any pair $(i_1,\dots,i_m,x)$ and $(i_1,\dots,i_m,x')$ if $x\neq x'$ then the corresponding two entries can be computed independent of each other. Hence instead of computing each entry of $\{(i_1,\dots,i_m,x); x\in[mn]\}$ separately we perform a combined operation on all previously computed entries on which any of the entry in $\{(i_1,\dots,i_m,x); x\in[mn]\}$ depends. Note there are $2^mmn$ of them and computation takes time $O(2^mmn)$. As there are $n^m$ different choices for $(i_1,\dots,i_m)$, the total running time is $O(2^mmn^{m+1})$.
Moreover for any $i_1,\dots,i_m\in[n]$ and $x\in[mn]$, if $D_{i_1,\dots,i_m,x}=D_{i'_1,\dots,i'_m,x'}$ (or $=D_{i'_1,\dots,i'_m,x'}+1$), the algorithm stores the tuple $(i'_1,\dots,i'_m,x')$ for entry $D_{i_1,\dots,i_m,x}$. Hence after computing $D$, the backtrack process to compute $L,\bar{L},\sigma_n$ can be performed in time $O(m^2n)$. Hence the running time of Algorithm MaxDelSimilarAlignment() is $\tilde{O}(2^mmn^{m+1})$.

If $d=\theta n$ where $\theta\in(0,1)$, then as we know from every string no more than $\theta n$ characters will be deleted, using \cite{Ukkonen85} we can claim the following.

\begin{corollary}
\label{thm:maxdeltheta}
Given $m$ strings $s_1,\dots,s_m$, each of length $n$, a set $S\subseteq[n]$ and a parameter $d=\theta n$, where $\theta\in (0,1)$ there exists an algorithm that computes a common alignment $\sigma_n$ such that  $\sum_{k\in [m]}|\bar{\sigma}_n(s_k)|\le dm$ and $ |\bar{\sigma}_n(s_1)\cap S|$ is maximized in time $\tilde{O}(2^m\theta^mmn^{m+1})$.
\end{corollary}


\subsection{Finding Alignment with Minimum Deletion Similarity}
\label{sec:mindel}

Given $m$ strings $s_1,\dots,s_m$ of length $n$, $q$ sets $S_1,\dots, S_q\subseteq[n]$ and a parameter $0\le d\le n$, our objective is to compute an alignment $\sigma_n$ of $s_1,\dots,s_m$ with alignment cost at most $d$ such that $|\cup_{j\in [q]}(\bar{\sigma}_n(s_1)\cap S_j)|$ is minimized.

\begin{theorem}
\label{thm:thm1}
Given $m$ strings $s_1,\dots,s_m$, each of length $n$, $q$ sets $S_1,\dots, S_q\subseteq[n]$ and a parameter $0\le d\le n$, there exists an algorithm that computes a common alignment $\sigma_n$ such that $\sum_{k\in [m]}|\bar{\sigma}_n(s_k)|\le dm$ and $|\cup_{j\in [q]}(\bar{\sigma}_n(s_1)\cap S_j)|$ is minimized in time $\tilde{O}(2^mmn^{m+1})$.
\end{theorem}

\noindent Our algorithm MinDelSimilarAlignment() uses dynamic programming just like the one used in finding alignment with maximum deletion similarity to compute $\sigma_n$. As the technique and the analysis are similar to the one used in Section \ref{sec:maxdel}, we move the details to Appendix \ref{sec:mindelappendix}.

\noindent If $d=\theta n$ where $\theta\in(0,1)$, again using \cite{Ukkonen85} we can claim the following.

\begin{corollary}
\label{thm:mindeltheta}
Given $m$ strings $s_1,\dots,s_m$, each of length $n$, $q$ sets $S_1,\dots, S_q\subseteq[n]$ and a parameter $0\le d\le \theta n$, there exists an algorithm that computes a common alignment $\sigma_n$ such that $\sum_{k\in [m]}|\bar{\sigma}_n(s_k)|\le dm$ and $|\cup_{j\in [q]}(\bar{\sigma}_n(s_1)\cap S_j)|$ is minimized in time $\tilde{O}(2^m\theta^mmn^{m+1})$.
\end{corollary}

\subsection{Algorithm for $(2-\frac{3\theta}{16}+\epsilon)$-approximation of $\mathcal{A}(s_1,\dots,s_m)$}

To compute GapMultiAlignDist$(s_1,\dots,s_m,\theta,(2-\frac{3\theta}{16}))$ (Algorithm \ref{alg:dist1}), the procedure calls MultiAlign$(s_1,\dots,s_m, \theta)$ (Algorithm \ref{alg:dist2}) that returns a string $\sigma$. If $\sigma$ is a null string it outputs $0$ and otherwise it outputs $1$. 

\begin{algorithm}
	\begin{algorithmic}[1]
		\REQUIRE Strings $s_1,\dots,s_m$ of length $n$; parameter $\theta\in (0,1)$ 
		
		\ENSURE Computes $GapMultiAlignDist(s_1,\dots,s_m,\theta,(2-\frac{3\theta}{16}))$ 
	
		\STATE $L\gets MultiAlign(s_1,\dots,s_m,\theta)$

	    \IF{$L\neq \phi$}
	    \RETURN $1$
	    \ELSE \RETURN $0$
        \ENDIF
	
		\caption{{GapMultiAlignDist} $(s_1,\dots,s_m,\theta)$}
		\label{alg:dist1}
	\end{algorithmic}
\end{algorithm}

We show if $\mathcal{A}(s_1,\dots,s_m)\le \theta n$, MultiAlign$(s_1,\dots,s_m, \theta)$ computes a common subsequence $\sigma$ such that $|\bar{\sigma}(s_1)|\le (2-\frac{3\theta}{16})\theta n$ and otherwise if $\mathcal{A}(s_1,\dots,s_m)> (2-\frac{3\theta}{16})\theta n$ it computes a null string. Next we describe Algorithm \ref{alg:dist2} MultiAlign$(s_1,\dots,s_m, \theta)$. It starts by partitioning the input strings into two groups $G_1$ and $G_2$ where $G_1$ contains the strings $s_1,\dots,s_{\lceil m/2\rceil}$ and $G_2$ contains the strings $s_1,s_{\lceil m/2\rceil+1},\dots,s_m$. 

Assume $\mathcal{A}(S_1,\dots,s_m)\le \theta n$. Next we state an observation that is used as one of the key elements to conceptualize our algorithm. Any common subsequence of $s_1,\dots,s_m$ is indeed a common subsequence of $G_1$. Therefore, as $\mathcal{A}(s_1,\dots,s_m)\le \theta n$, if we can enumerate all common subsequences of $G_1$ of length at least $n-\theta n$, we generate the optimal alignment as well. Notice after generating each common subsequence of $G_1$, it can be checked whether it is a common subsequence of $G_2$ or not. The main hurdle here is that enumerating all common subsequences of $G_1$ of length at least $n-\theta n$ is time consuming. 

We overcome this barrier by designing Algorithm \ref{alg:dist3} (EnumerateAlignments$(s_1,\dots,s_m,\theta)$) that generates $k$ (where $k=O(1/\theta)$) different sets $L_1,\dots,L_k$ where $L_j\subseteq [n]$, $|L_j|\le \theta n$ and each $L_j$ corresponds to a common subsequence $\sigma_j$ of $G_1$ such that $L_j=\bar{\sigma}_j(s_1)$.
Moreover, we can ensure that either $\exists j\in[k]$ where $|L_j|\le \frac{3\theta n}{4}$ or for any common subsequence $\sigma$ of $G_1$ with $\bar{\sigma}(s_1)\le \theta n$ there exists a $L_i$ where $|\bar{\sigma}(s_1)\cap L_i|\ge \frac{3\theta^2 n}{16}$. 

\begin{algorithm}
	\begin{algorithmic}[1]
		\REQUIRE Strings $s_1,\dots,s_m$ of length $n$; parameter $\theta\in (0,1)$
		
		\ENSURE A common subsequence $\sigma$ of $s_1,\dots,s_m$ such that if $\mathcal{A}(s_1,\dots,s_m)\le \theta n$ then $\bar{\sigma}(s_1)\le (2-\frac{3\theta}{16})\theta n$, else if  $\mathcal{A}(s_1,\dots,s_m)> (2-\frac{3\theta}{16})\theta n$ then $\sigma=\phi$.
	
		\STATE Define set $G_1$ containing strings $s_1,\dots,s_{\lceil m/2\rceil}$ and $G_2$ containing strings $s_1,s_{\lceil m/2\rceil+1}\dots,s_{m}$; $\mathcal{E}\gets \phi$; $A\gets \phi$; $\sigma\gets \phi$

	    \IF{$\mathcal{A}(G_1)\le \theta n$}
	    \STATE $\mathcal{E}\gets EnumerateAlignments(G_1,\theta)$ 
        \ENDIF
	
		\IF{$|\mathcal{A}(G_2)|\le \theta n$}
		\FOR {each $L_i\in \mathcal{E}$}
	
		\STATE $L'_i \gets MaxDelSimilarAlignment(G_2,L_i,\theta n)$
		
		
		\IF{$|L_i\cup L'_i|\le (2-\frac{3 \theta}{16})\theta n$}

		\STATE $\sigma \gets s_1[i_1]\circ\dots \circ s_1[i_p]$   (where, $[n]\setminus (L_i\cup L'_i)=\{i_1,\dots, i_p\}$)  
		
		\ENDIF
	
		\ENDFOR
		\ENDIF
		\RETURN $\sigma$
		\caption{{MultiAlign} $(s_1,\dots,s_m,\theta)$}
		\label{alg:dist2}
	\end{algorithmic}
\end{algorithm}

Algorithm \ref{alg:dist3} (EnumerateAlignments()) starts by computing a LCS $\sigma_1$ of $G_1$ such that $|\bar{\sigma_1}(s_1)|\le \theta n$. Let $L_1=\bar{\sigma_1}(s_1)$. If $|L_1|\le \frac{3\theta n}{4}$ return $L_1$. Otherwise it calls the algorithm MinDelSimilarAlignment() to compute a common subsequence $\sigma_2$ of $G_1$ of cost at most $\theta n$ such that $L_2\cap L_1$, where $L_2=\bar{\sigma_2}(s_1)$ is minimized. At the $i$th step given $i-1$ sets $L_1,\dots,L_{i-1}$, the algorithm computes an alignment $\sigma_i$ of $G_1$ with $L_i$ being the set of indices of unaligned characters of $s_1$ such that the intersection of $L_i$ with $\cup_{j\in[i-1]}L_j$ is minimized. 
The algorithm continues with this process until it reaches a round $k$ such that 
$|\cup_{i\in [k-1]}(L_k(s_1)\cap L_i(s_1))|\ge \frac{\theta^2n(k-1)}{4}$. 
Let $L_1,\dots,L_k$ be the sets generated. Output all theses sets.

Next for each $L_i$ returned by Algorithm \ref{alg:dist3}, call the algorithm MaxDelSimilarAlignment() to find an alignment $\sigma'$ of $G_2$ of cost at most $\theta n$ such that the intersection of $L_i$ and $L'_i=\bar{\sigma'}(s_1)$ is maximized. If $|L_i\cup L'_i|\le (2-\frac{3\theta}{16})\theta n$, define $\sigma=s_1[i_1]\circ \dots \circ s_1[i_p]$ where, $[n]\setminus (L_i\cup L'_i)=\{i_1,\dots, i_p\}$.
Output $\sigma$. 

\begin{algorithm}
	\begin{algorithmic}[1]
		\REQUIRE Strings $s_1,\dots,s_m$ of length $n$; parameter $\theta \in (0,1)$ 
		
	\ENSURE A set $\mathcal{E}=\{L_1,\dots L_k\}$ where $L_j\subseteq[n]$, $|L_j|\le \theta n$ and $L_j=\bar{\sigma}_j(s_1)$ where $\sigma_j$ is a common subsequence of $s_1,\dots,s_m$ and for any common subsequence $L$ such that $|\bar{L}(s_1)|\le \theta n$, $\exists i\in [k]$ where $(\bar{L}(s_1)\cap L_i)\ge \frac{3\theta^2n}{16}$
	
		\STATE  $\mathcal{E}\gets \phi$; $f\gets 0$; $i\gets 1$
	
	    \STATE $\sigma\gets LCS(G_1)$
	    \STATE $\mathcal{E}\gets \mathcal{E}\cup \bar{\sigma}(s_1)$

	    \IF{$|LCS(G_1)|< n- \frac{3\theta n}{4}$}

	     \WHILE{$f\neq 1$}
		
		\STATE $i\gets i+1$
	
		\STATE $L_i \gets MinDelSimilarAlignment(G_1,L_1(s_1)\cup\dots\ \cup L_{i-1}(s_1),\theta n)$

		\IF{$|\cup_{j\in [i-1]}(L_i(s_1)\cap L_j(s_1))|\ge \frac{\theta^2n(i-1)}{4}$}
		\STATE $f\gets 1$

	\ELSE
	\STATE $\mathcal{E}\gets \mathcal{E}\cup L_i$  
\ENDIF
		\ENDWHILE
		
		\ENDIF
		
		\RETURN $\mathcal{E}$
		\caption{{EnumerateAlignments} $(s_1,\dots,s_m,\theta )$}
		\label{alg:dist3}
	\end{algorithmic}
\end{algorithm}

We now prove two crucial lemmas to establish the correctness.
\begin{lemma}
\label{lem:lem1}
Given strings $s_1,\dots,s_{|G_1|}$ of length $n$ such that $\mathcal{A}(s_1,\dots,s_{|G_1|})\le \theta n$, where $\theta\in (0,1)$, there exists an algorithm that computes $k\le 4/\theta$, different sets $L_1,\dots,L_k \subseteq [n]$ each of size at most  $\theta n$ such that $\forall j\in[k]$, there exists a common subsequence $\sigma_j$ of $G_1$ where, $L_j=\bar{\sigma}_j(s_1)$ and one of the following is true.
\begin{enumerate}
    \item  $\exists j\in[k]$ such that $|L_j|\le \frac{3\theta n}{4}$. 
    \item For any common subsequence $\sigma$ of $G_1$ with $\bar{\sigma}(s_1)\le \theta n$ there exists a $L_i$, where $|\bar{\sigma}(s_1)\cap L_i|\ge \frac{3\theta^2 n}{16}$. The running time of the algorithm is $\tilde{O}(2^mmn^{|G_1|+1})$.
\end{enumerate}
\end{lemma}

\begin{proof}
Let $\sigma'=LCS(s_1,\dots,s_{|G_1|})$. If $|\sigma'|\ge n-\frac{3\theta n}{4}$, then $|\bar{\sigma'}(s_1)|\le\frac{3\theta n}{4}$ and we satisfy condition 1.  Otherwise assume $|\bar{\sigma'}(s_1)|>\frac{3\theta n}{4}$.
Let $L_1,\dots,L_k$ be the sets returned by Algorithm \ref{alg:dist3}. By construction, for each $L_i$ there exists a common subsequence $\sigma_i$ of $G_1$ where $L_i=\bar{\sigma}_i(s_1)$.
Note every $L_i$ has size at least $\frac{3\theta n}{4}$. 
Moreover if the algorithm does not terminate at round $i$, then $|L_i(s_1)\setminus \{L_1(s_1)\cup\dots\cup L_{i-1}(s_1)\}|\ge \frac{3\theta n}{4}-\frac{\theta^2n(i-1)}{4}$. Hence after $k$ steps we have 
\begin{align*}
  |\cup_{i\in[k]}L_i(s_1)|
  &\ge \frac{3\theta n}{4}+(\frac{3\theta n}{4}-\frac{\theta^2n}{4})+\dots+(\frac{3\theta n}{4}-\frac{(k-1)\theta^2n}{4})
  \\
  &=\frac{3k\theta n}{4}-\frac{\theta^2n}{4}(1+2+\dots+(k-1))
 \\ 
  &= \frac{3k\theta n}{4}-\frac{k(k-1)\theta^2 n}{8}
\\
  &> \frac{3k\theta n}{4}-\frac{k^2\theta^2 n}{8}
\end{align*}

Substituting $k=4/\theta$, we get $|\cup_{i\in[k]}L_i(s_1)|>n$. 
Now if the algorithm stops at round $i<4/\theta$, then we know for each common subsequence $\sigma$ of $G_1$ of length at least $n-\theta n$ if $\sigma\notin \{L_1,\dots,L_{i-1}\}$, $|\cup_{j\in [i-1]}(\sigma(s_1)\cap L_j(s_1))|\ge \frac{(i-1)\theta^2n}{4}$. Hence there exists at least one $j\in [i-1]$ such that $|L_j(s_1)\cap \sigma(s_1)|\ge \frac{\theta^2n}{4}$. Otherwise if the algorithm runs for $4/\theta$ rounds then $\cup_{i\in[4/\theta]}L_i(s_1)=[n]$. Hence for each common subsequence $\sigma$ of cost in $[\frac{3\theta n}{4},\theta n]$, $\bar{\sigma}(s_1)$ will have intersection at least $\frac{3\theta^2n}{16}$ with at least one $L_i$. 

As $|k|\le 4/\theta$ the algorithm runs for at most $4/\theta$ rounds where at the $i$th round it calls MinDelSumilarAlignment() with strings in $G_1$, and sets $L_1,\dots,L_{i-1}$ (where $i\le 4/\theta$) and parameter $\theta n$. By Corollary \ref{thm:mindeltheta} each call to MinDelSumilarAlignment() takes time $\tilde{O}(2^{|G_1|}\theta^{|G_1|}|G_1|n^{|G_1|+1})$. Hence the total running time taken is $\tilde{O}(2^{|G_1|}|G_1|n^{|G_1|+1})$.
\end{proof}

We set $|G_1|=\lceil \frac{m}{2}\rceil\le \lfloor m/2 \rfloor +1$ to obtain a running time of $\tilde{O}(2^{\lfloor m/2\rfloor+1}mn^{\lfloor m/2\rfloor+2})$.

\begin{lemma}
\label{lem:lem2}
Given $\mathcal{A}(s_1,\dots,s_m)\le \theta n$, Algorithm \ref{alg:dist2} MultiAlign($s_1,\dots, s_m, \theta$) generates a string $\sigma$, such that 
$\sigma$ is a common sequence of $s_1,\dots,s_m$ and the alignment cost of $\sigma$ is at most $(2-\frac{3\theta}{16})\theta n$. Moreover Algorithm \ref{alg:dist2} runs in time $\tilde{O}_m(n^{\lfloor \frac{m}{2} \rfloor+2})$.
\end{lemma}
\begin{proof}
Let $\eta$ be some LCS of $s_1,\dots,s_m$ such that $|\bar{\eta}(s_1)|\le \theta n$. First assume $\mathcal{A}(G_1)\le \frac{3\theta n}{4}$. Then Algorithm \ref{alg:dist3} computes a LCS $\sigma_1$ of $G_1$ and returns the set $L_1=|\bar{\sigma_1}(s_1)|$ where $|L_1|\le \frac{3\theta n}{4}$ to Algorithm \ref{alg:dist2}. Next Algorithm \ref{alg:dist2} calls MaxDelSimilarAlignmemnt($G_2,L_1,\theta n$) which returns a set $L'_1=\bar{\sigma'}(s_1)$, where $\sigma'$ is a common subsequence of $G_2$ and $|L'_1|\le \theta n$. Notice $\sigma \gets s_1[i_1]\circ\dots \circ s_1[i_p]$   (where, $[n]\setminus (L_i\cup L'_i)=\{i_1,\dots, i_p\}$) is a common subsequence of $s_1,\dots,s_m$ and $|\bar{\sigma}(s_1)|\le (|L_1|+|L'_1|)\le (\frac{3\theta n}{4}+\theta n)=(2-\frac{1}{4}) \theta n\le (2-\frac{\theta}{4}) \theta n$ as $\theta\in(0,1)$. Hence Algorithm \ref{alg:dist2} computes a common subsequence $\sigma$ of $s_1,\dots,s_m$ such that the alignment cost of $\sigma$ is at most $(2-\frac{\theta}{4})\theta n$. 

Next assume $\mathcal{A}(G_1)> \frac{3\theta n}{4}$. Then by Lemma \ref{lem:lem1}, Algorithm \ref{alg:dist3} computes a set $L_i=\bar{\sigma_i}(s_1)$ where $\sigma_i$ is a common subsequence of $G_1$, $|L_i|\le \theta n$ and $L_i\cap \bar{\eta}(s_1)\ge \frac{3\theta^2 n}{16}$. Notice as $\eta$ is a common subsequence of $G_2$, when  Algorithm \ref{alg:dist2} calls MaxDelSimilarAlignmemnt($G_2,L_i,\theta n$), it returns a set $L'_1=\bar{\sigma'}(s_1)$, where $\sigma'$ is a common subsequence of $G_2$, $|L'_1|\le \theta n$ and $|L_i\cap L'_1|\ge \frac{3\theta^2 n}{16}$. Therefore $|L_i \cup L'_1|\le \theta n+\theta n-\frac{3\theta^2 n}{16}=(2-\frac{3\theta n}{16})\theta n$, and Algorithm \ref{alg:dist2} computes a common subsequence $\sigma$ of $s_1,\dots,s_m$ such that the alignment cost of $\sigma$ is at most $(2-\frac{3\theta}{16})\theta n$. 


Next we analyze the running time of Algorithm \ref{alg:dist2}. First we compute $LCS(G_1)$ which takes time $\tilde{O}(2^{\lfloor m/2\rfloor+1}n^{\lfloor m/2\rfloor+1})$ using the classic dynamic program 
algorithm (note $|G_1|=\lceil \frac{m}{2}\rceil \le \lfloor \frac{m}{2}\rfloor+1$). Next we call $EnumerateAlignments(s_1,\dots,s_{m/2},\theta)$. By Lemma \ref{lem:lem1} this takes time $\tilde{O}(2^{\lfloor m/2\rfloor+1 }mn^{\lfloor m/2\rfloor+2})$. Moreover it returns at most $O(1/\theta)$ sets and for each of them Algorithm \ref{alg:dist2} calls $MaxDelSimilarAlignment()$ on $G_2$. As $|G_2|\leq \lfloor \frac{m}{2}\rfloor+1$ and each set has size at most $\theta n$, each call takes time $\tilde{O}(2^{\lfloor m/2\rfloor+1}\theta^{\lfloor m/2\rfloor +1}mn^{\lfloor m/2\rfloor+2})$. Hence total time taken is $\tilde{O}(2^{\lfloor m/2\rfloor+1}mn^{\lfloor m/2\rfloor+2})$. Next each union of $L_i$ and $L'_i$ and corresponding $\sigma$ can be computed in time $O(n)$. Hence the running time of Algorithm \ref{alg:dist2} is $\tilde{O}(2^{\lfloor m/2\rfloor+1}mn^{\lfloor m/2\rfloor+2})=\tilde{O}_m(n^{\lfloor \frac{m}{2} \rfloor+2})$.
\end{proof}

\begin{lemma}
\label{lem:gaplcsmain}
Algorithm \ref{alg:dist1} computes $GapMultiAlignDist(s_1,\dots,s_m,\theta,(2-\frac{3\theta}{16}))$ in time $\tilde{O}_m(n^{\lfloor \frac{m}{2} \rfloor+2})$.
\end{lemma}
\begin{proof}
First assume the case where $\mathcal{A}(s_1,\dots, s_m)\le \theta n$. from Lemma \ref{lem:lem2} we have Algorithm \ref{alg:dist2} returns a common subsequence $\sigma$ of $s_1,\dots,s_m$ such that the alignment cost of $\sigma$ is at most $(2-\frac{3\theta}{16})\theta n$. Hence,  Algorithm \ref{alg:dist1} outputs 1. Next assume $\mathcal{A}(s_1,\dots,s_m)> (2-\frac{3\theta}{16})\theta n$. In this case in Algorithm \ref{alg:dist2}, for each set $L_i\in \mathcal{E}$, $|L_i\cup L'_i|>(2-\frac{3\theta}{16})\theta n$. Hence Algorithm \ref{alg:dist2} returns a null string and Algorithm \ref{alg:dist1} outputs 0. The bound on the running time is directly implied by the running time bound of Algorithm \ref{alg:dist2}.
\end{proof}

\section{Below-$2$ Approximation for Multi-sequence Alignment Distance with One Pseudorandom String}
\label{sec:pseudo}
In the last section we present an algorithm that given $m$ strings, computes a truly below 2 approximation of the optimal alignment distance of the input strings provided the distance is large i.e. $\Omega(n)$. In this section we use this algorithm as a black box and show given a $(p,B)$-pseudorandom string $s_1$, and $m-1$ adversarial strings $s_2,\dots,s_m$, there exists an algorithm that for any arbitrary small constant $\epsilon>0$ computes $(2-\frac{3p}{512}+\epsilon)$ approximation of $\mathcal{A}(s_1,\dots,s_m)$ in time $\tilde{O}_m(n\beta^{m-1}+ n^{ \lfloor m/2\rfloor+3})$. Here $\beta=max(B,\sqrt{n}).$ Notice as the approximation factor is independent of $\theta=\frac{\mathcal{A}(s_1,\dots,s_m)}{n}$, we can assure truly below-2 approximation of the alignment cost for any distance regime. 
Formally we show the following.

\begin{theorem*}[\ref{thm:align2}]
Given a $(p,B)$-pseudorandom string $s_1$, and $m-1$ adversarial strings $s_2,\dots,s_m$ each of length $n$, there exists an algorithm that for any arbitrary small constant $\epsilon>0$ computes $(2-\frac{3p}{512}+89\epsilon)$ approximation of $\mathcal{A}(s_1,\dots,s_m)$ in time $\tilde{O}_m(n\beta^{m-1}+ n^{ \lfloor m/2\rfloor+3})$. Here $\beta=max(B,\sqrt{n}).$
\end{theorem*}

\subsection{Adaptive window decomposition}
\label{sec:windowdecomposition}

Given a string $s$, we define a window $w$ of size $d$ of $s$ to be a substring of $s$ having length $d$. Let $s(w)$ denote the starting index of $w$ in $s$ and $e(w)$ denote the last index of $w$ in $s$. 

As described in Section \ref{sec:overview}, we start by partitioning the $(p,B)$ pseudorandom string $s_1$ into windows. For Algorithm \ref{alg:uniquematch}, we use a fixed window size $\beta$ and for Algorithm \ref{alg:largecostmatch}, we construct windows of variable sizes that are multiples of $\beta$ i.e. $\{\beta,2\beta,\dots,n\}$. For both the algorithms, the windows start from indices in $\{1,\beta+1,2\beta+1,\dots\}$. Here $\beta=max(B,\sqrt{n})$. 


Next we show that given a fixed partition of $s_1$ into disjoint variable sized windows, we can partition $s_2,\dots,s_m$ accordingly so that there exists a series of tuples containing $m$ windows such that these tuples nicely cover the optimal alignment of $s_1,\dots,s_m$ and the sum of the alignment cost of these $m$-tuples is at most $(1+12\epsilon)\mathcal{A}(s_1,\dots,s_m)$ for arbitrary small $\epsilon$. For every adversarial string, though we generate windows covering the whole string, in the algorithm we only use/estimate those windows that are required to compute the cost of the optimal alignment. Even if we do not have any prior knowledge of the optimal alignment, we show the windows that we evaluate contain the required ones.

\paragraph{Window decomposition.}
We start with a given disjoint variable sized window decomposition $\mathcal{J}_1$ of $s_1$ defined as $$\mathcal{J}_1=\{w_1,\dots,w_k\}=\{s_1[1,\dots,d_1],s_1[d_1+1,\dots,d_1+d_2],\dots,s_1[d_1+\dots+d_{k-1},\dots,n]\}.$$
 Here, $|w_i|=d_i\in \{\beta,2\beta,\dots\}$.

\noindent Next for each adversarial string $s_j$ we compute a set of windows $\mathcal{J}_j$. We do this by computing for each $d_i$, a set of windows $\mathcal{J}_j^{d_i}$ and define $\mathcal{J}_j=\cup_{i\in[k]}\mathcal{J}_j^{d_i}$. To compute $\mathcal{J}_j^{d_i}$, take $\tau_{i}=\{0,\frac{1}{d_i},\frac{(1+\epsilon)}{d_i},\dots,1\}$. For each $\tau_i$, compute a set of windows $\mathcal{J}_{j,\tau_i}^{d_i}$ and set $\mathcal{J}_j^{d_i}=\cup_{\tau_i}\mathcal{J}_{j,\tau_i}^{d_i}$. 
For $\tau_i=0$ set $h_{\tau_i}=d_i$ and $\ell_{\tau_i}=d_i$. For $\tau_i=\frac{1}{d_i}$ set $h_{\tau_i}=d_i+1$, $\ell_{\tau_i}=d_i-1$ and for general $\tau_i=\frac{(1+\epsilon)^k}{d_i}$, set $h_{\tau_i}=\lfloor d_i+(1+\epsilon)^{k-1}\rfloor$ and $\ell_{\tau_i}=\lfloor d_i-(1+\epsilon)^{k-1}\rfloor$. Set $\gamma_{\tau_i}=max(1,\lfloor \epsilon \tau_i d_i\rfloor)$. Define $$\mathcal{H}_{j,\tau_i}^{d_i}=\{s_j[1,h_{\tau_i}],s_j[\gamma_{\tau_i}+1,\gamma_{\tau_i}+h_{\tau_i}],s_j[2\gamma_{\tau_i}+1,2\gamma_{\tau_i}+h_{\tau_i}],\dots\}$$

$$\mathcal{L}_{j,\tau_i}^{d_i}=\{s_j[1,\ell_{\tau_i}],s_j[\gamma_{\tau_i}+1,\gamma_{\tau_i}+\ell_{\tau_i}],s_j[2\gamma_{\tau_i}+1,2\gamma_{\tau_i}+\ell_{\tau_i}],\dots\}$$ 
Set $\mathcal{J}_{j,\tau_i}^{d_i}=\mathcal{H}_{j,\tau_i}^{d_i}\cup\mathcal{L}_{j,\tau_i}^{d_i}$.

\paragraph{Multi-window mapping} We call a mapping $\mu: \mathcal{J}_1\rightarrow \mathcal{J}_2\cup \{\perp\}\times \mathcal{J}_3\cup \{\perp\}\times \dots \times \mathcal{J}_m\cup \{\perp\}$ from a single window of $s_1$ to $m-1$ windows in $s_2,\dots,s_m$ monotone if for all $w,w'\in \mathcal{J}_1$ such that $\mu(w)=(w_2,\dots,w_m)\neq \phi$, $\mu(w')=(w'_2,\dots,w'_m)\neq \phi$ and $s(w)< s(w')$ then for all $k\in[2,m]$, $e(w_k)< s(w'_k)$. We abuse the notation for the sake of simplicity and assume $\mathcal{J}_1$ represents the set of windows of $s_1$ such that for each $w\in \mathcal{J}_1$, $\mu(w)\neq \phi$. Also let $next(w)$ represents the window appearing after $w$ and $prev(w)$ represents the window appearing before $w$ in $s_1$. Let $\mu(w)=(w_2,\dots,w_m)$ and $\mu(w')=(w'_2,\dots,w'_m)$. For some $i\in[n]$, we say $i\in extended(w_j)$ if either $i\in w_j$ or there exists a $w'_j$ such that $i\in w'_j$ and if $\ell$ be such that $w'_j[\ell]=i$ then there exists a $k\in[m]$ where $w'_k[\ell]\in w_k$.

For any given monotone mapping $\mu$, we define its alignment cost as:
$$\mathcal{A}(s_1,\mu(s_1))\le \sum_{w\in \mathcal{J}_1}\mathcal{A}(w,\mu(w))+(\sum_{j=2}^m\sum_{i=1}^n(|\#\{w \textit{ such that }  i\in extended(w_j)\}-1|)/m.$$

\begin{lemma}
\label{lem:window1}
Let $s_1,\dots,s_m\in \Sigma^n$ then the following holds:

\begin{enumerate}
    \item For every disjoint decomposition $\mathcal{J}_1$ of $s_1$ and every monotone mapping $\mu: \mathcal{J}_1\rightarrow \mathcal{J}_2\cup\{\perp\}\times \dots \times \mathcal{J}_m\cup\{\perp\}$ we have $\mathcal{A}(s_1,\mu(s_1))\ge \mathcal{A}(s_1,s_2,\dots,s_m)$.
    \item For every disjoint decomposition $\mathcal{J}_1$ of $s_1$, there exists a monotone mapping $\mu: \mathcal{J}_1\rightarrow \mathcal{J}_2\cup\{\perp\}\times \dots \times \mathcal{J}_m\cup\{\perp\}$ satisfying $\mathcal{A}(s_1,\mu(s_1))\le (1+12\epsilon)\mathcal{A}(s_1,\dots,s_m)$. 
\end{enumerate}
\end{lemma}
\begin{proof}
Let $\mu(w)=(w_2,\dots,w_m)$. To prove the first part we apply $\mu$ on $s_1$, to generate a subsequence $L_\mu$ of $s_1$. To do this for each window $w\in \mathcal{J}_1$ (note $w$ can be of variable sizes) if $\mu$ matches a character of $w$ with a character of $w_j$ for each $j\in[2,m]$ then match it. Delete all unmatched characters in $s_1,\dots,s_m$. Next if $w$ is not the last window and $\mu(w)\neq \phi$, if there exists a $j$ such that $e(w_j)-s(next(w_j))>0$ delete all the overlapping characters from $w_j$. Also delete all the characters in rest of the $m-1$ strings to which theses deleted characters are aligned. This gives a common subsequence $L_\mu$ of $s_1,\dots,s_m$. Notice the cost of this alignment is 
$$\mathcal{A}(s_1,\mu(s_1))\le \sum_{w\in \mathcal{J}_1}\mathcal{A}(w,\mu(w))+(\sum_{j=2}^m\sum_{i=1}^n(|\#\{w \textit{ such that }  i\in extended(w_j)\}-1|)/m$$.
Also if $L$ is a LCS of $s_1,s_2,\dots,s_m$ then $|L_{\mu}|\le |L|$. Hence, $\mathcal{A}(s_1,\mu(s_1))\le \mathcal{A}(s_1,\dots,s_m)$.

To prove the second part let $\sigma$ be some LCS of $s_1,\dots,s_m$. Let $s'_i$ be the subsequence of $s_i$ containing  the characters that are aligned under $\sigma$ in $s_i$. Using $\sigma$ we first define a mapping $\mu':\mathcal{J}_1\rightarrow \mathcal{J}_2\cup\{\perp\}\times \mathcal{J}_m\cup\{\perp\}$ of small cost. For a window $w\in \mathcal{J}_1$ let $\mu'(w)=(w'_2,\dots,w'_k)$. As $\mu'$ may not be a monotone map, we further modify it to generate a monotone map $\mu$. 
For a window $w\in \mathcal{J}_1$, if $w\cap \sigma(s_1)=\phi$ set $\mu'(w)=\phi$. Otherwise for each $w$ let $i_s^w$ be the first index of $w$ such that $w[i_s^w]\in \sigma(s_1)$ and $i_\ell^w$ be the last index of $w$ such that $w[i_\ell^w]\in \sigma(s_1)$. For a given index $i\in w$, if $i\in\sigma(s_1)$ let $i(j)$ denotes the index of the character in $s_j$ to which $w[i]$ is aligned. 
For each $k\in[2,m]$ we define $w'_k$ as follows. 
Let $c_w=|w|-|\sigma(w)|$. Let $\tau=\frac{{(1+\epsilon)}^{j}}{|w|}$ be such that ${{(1+\epsilon)}^{j-1}}< c_w\le{{(1+\epsilon)}^{j}}$.
When $(i^w_e(k)-i_s^w(k)+1)\ge |w|$, set $w'_k$ to be the rightmost interval of $\mathcal{H}_{k,\tau}^{|w|}$ that contains $i_s^w(k)$.
Note the interval length of $w'_k$ is at most $|w|+(1+\epsilon)^{j}$. Otherwise when $(i^w_e(k)-i_s^w(k)+1)< |w|$, set $w'_k$ to be the rightmost interval of $\mathcal{L}_{k,\tau}^{|w|}$ that contains $i_s^w(k)$. Note the interval length of $w'_k$ is at least $|w|-(1+\epsilon)^{j}$.

Hence the total cost of $\mu'$ is:
\begin{align}
 \mathcal{A}(w,\mu'(w)) &   \le c_w+\frac{m\epsilon(1+\epsilon)^{j}}{m}+2max_{k}(i_s^w(k)-s(w'_k))\\& \le c_w+3\epsilon\tau|w|\\
 & \le (1+3\epsilon)(1+\epsilon)c_w\\
 & \le (1+7\epsilon)c_w
\end{align}

The second term in Equation $9$ represents the error due to the size estimation of the windows and the third term represents the error due to the shift for the choice of starting indices of the windows. The important point to note here is that as the shift of all the windows in $m$ different strings are always in one direction, the maximum number of aligned characters of $\sigma$, that get unaligned because of the shift is at most $2max_{k}(i_s^w(k)-s(w'_k))\le 2\epsilon \tau |w|$.
Moreover $\tau|w|\le (1+\epsilon)c_w$.

For all $w\in \mathcal{J}_1$, $i_e^w<i_s^{next(w)}$. Let $shift_w=max_k (i_s^w(k)-s(w'_k))$. Hence $max_k|e(w'_k)-s(next(w'_k))|\le shift_w+max_k M(i_e^w(k),i_s^{next(w)}(k))$ where $M(i_e^w(k),i_s^{next(w)}(k))$ are the number of characters in $s_k$ between $i_e^w(k)$ and $i_s^{next(w)}(k)$. Note these symbols are deleted in $\sigma$ from $s_k$. If $shift_w\ge shift_{next(w)}$ we charge $|shift_w-shift_{next(w)}|$ to $w$ else we charge $|shift_w-shift_{next(w)}|$ to $next(w)$. Without loss of generality assume it is charged to $w$. Hence $|shift_w-shift_{next(w)}|\le shift_w\le \epsilon c_w$. Hence
\begin{align*}
\mathcal{A}(s_1,\mu'(s_1))&\le \sum_{w\in \mathcal{J}_1}\mathcal{A}(w,\mu'(w))+(\sum_{j=2}^m \sum_{i=1}^n|\#\{w \textit{ such that } i\in extended(w_j)\}-1|)/m\\
&\le (1+7\epsilon)\mathcal{A}(s_1,\dots,s_m)+4\sum_{w,\mu'(w)\neq \phi} shift_w\\
&\le (1+11\epsilon) \mathcal{A}(s_1,\dots,s_m)
\end{align*}

Next to make $\mu'$ monotone we need to ensure that for each $w,w'\in \mathcal{J}_1$, if $s(w)<s(w')$ then for each $k\in [2,m]$, $e(w_k)<s(w'_k)$. Notice $max_k (e(w_k)-s(w'_k))\le shift_{w'}-shift_w\le shift_{w'}\le \epsilon c_{w'}$. Hence we define $\mu$ by shrinking each $w'_k$ from below by deleting $\epsilon c_{w'}$ characters and charge this to $w'$. Notice such a window already exists and each window is charged at most once. Hence, 
$$\mathcal{A}(s_1,\mu(s_1))=\sum_{w\in \mathcal{J}_1}\mathcal{A}(w,\mu(w)\le \mathcal{A}(s_1,\mu'(s_1))+\sum_{w,\mu'(w)\neq \phi} \epsilon c_w\le (1+12\epsilon) \mathcal{A}(s_1,\dots,s_m).$$

\noindent Notice here for the first equality we use the fact that if a map $\mu$ is monotone then $\mathcal{A}(s_1,\mu(s_1))=\sum_{w\in \mathcal{J}_1}\mathcal{A}(w,\mu(w)$.
\end{proof}

From the definition of mapping $\mu$ notice for each $w\in \mathcal{J}_1$, if we know the cost $c_w$ in advance it is enough to consider $\tau\in\{0,\frac{1}{|w|},\frac{(1+\epsilon)}{|w|},\frac{(1+\epsilon)^2}{|w|},\dots,\theta\}$ and $\gamma_{\tau}=max(1,\lfloor \epsilon \theta |w|\rfloor)$ where $c_w=\theta w$.
Further for each $j\in[2,m]$ if we also know $i_s^w(j)$ and $i_e^w(j)$ we can further restrict the map by restricting each $\mathcal{J}_{j,\tau}^{|w|}$ containing only those windows whose starting index lies in $[i_s^w(j)-|w|,i_e^w(j)+|w|]$. Let this restricted set of windows be $\mathcal{\tilde{J}}_{j,\tau}^{|w|}$. Define $\mathcal{\tilde{J}}_j^{|w|}=\cup_{\tau}\mathcal{\tilde{J}}_{j,\tau}^{|w|}$ and $\mathcal{\tilde{J}}_j=\cup_{w\in \mathcal{J}_1}\mathcal{\tilde{J}}_{j}^{|w|}$.
Then from $\mu$ we can create a restricted map $\mu_r: \mathcal{J}_1\rightarrow \mathcal{\tilde{J}}_2\cup \{\perp\}\times \mathcal{\tilde{J}}_3\cup \{\perp\}\times \dots \times \mathcal{\tilde{J}}_m\cup \{\perp\}$ which can be represented by $|\mathcal{J}_1|=k$ different mappings, one for each $w\in \mathcal{J}_1$ defined as $\mu_r^w: w\rightarrow \mathcal{\tilde{J}}_2^{|w|}\cup\{\perp\}\times\dots\times \mathcal{\tilde{J}}_m^{|w|}\cup\{\perp\}$. As a direct consequence of Lemma \ref{lem:window1} we can make the following claim.

\begin{lemma}
\label{lem:window2}
Let $s_1,\dots,s_m \in \Sigma^n$ then the following holds:

\begin{enumerate}
    \item For every disjoint decomposition $\mathcal{J}_1$ of $s_1$ and every restricted monotone mapping $\mu_r: \mathcal{J}_1\rightarrow \mathcal{\tilde{J}}_2\cup\{\perp\}\times \dots \times \mathcal{\tilde{J}}_m\cup\{\perp\}$ we have $\mathcal{A}(s_1,\mu_r(s_1))\ge \mathcal{A}(s_1,s_2,\dots,s_m)$.
    \item For every disjoint decomposition $\mathcal{J}_1$ of $s_1$, there exists a restricted monotone mapping $\mu_r: \mathcal{J}_1\rightarrow \mathcal{\tilde{J}}_2\cup\{\perp\}\times \dots \times \mathcal{\tilde{J}}_m\cup\{\perp\}$ satisfying $\mathcal{A}(s_1,\mu_r(s_1))\le (1+12\epsilon)\mathcal{A}(s_1,\dots,s_m)$. 
\end{enumerate}
\end{lemma}

\subsection{Computing alignment cost of window tuples}

From the previous section, we observe that if for each $w\in \mathcal{J}_1$ we have a good estimation of the alignment cost of each window tuple in $w\times \mathcal{\tilde{J}}^{|w|}_2\times \dots \times \mathcal{\tilde{J}}^{|w|}_m$, we get a good estimation of $\mathcal{A}(s_1,\dots,s_m)$. 

\paragraph{Estimation of alignment cost of window tuples.} For a window $w\in \mathcal{J}_1$, let $\mathcal{E}_w:w\times \mathcal{\tilde{J}}^{|w|}_2\times \dots \times \mathcal{\tilde{J}}^{|w|}_m\rightarrow \{0,\dots,|w|\}$ be an estimate of the alignment cost $(w,w_2,\dots,w_m)$ such that $\mathcal{E}_w(w,w_2,\dots,w_m)\ge \mathcal{A}(w,w_2,\dots,w_m)$ for all $w\in \mathcal{J}_1$ and $(w_2,\dots,w_m)\in \mathcal{\tilde{J}}^{|w|}_2\times \dots \times \mathcal{\tilde{J}}^{|w|}_m$. Combining all $\mathcal{E}_w$ we define cost estimation $\mathcal{E}:\mathcal{J}_1\times \mathcal{\tilde{J}}_2\times\dots\times\mathcal{\tilde{J}}_m\rightarrow \{0,1,\dots,max_{w\in \mathcal{J}_1}|w|\}$. Given such a cost estimation $\mathcal{E}$ and a restricted monotone map $\mu_r: \mathcal{J}_1\rightarrow \mathcal{\tilde{J}}_2\cup\{\perp\}\times \dots \times \mathcal{\tilde{J}}_m\cup\{\perp\}$, we represent the cost of $\mathcal{E}$ with respect to the map $\mu_r$ as $\mathcal{A}_{\mathcal{E}}(s_1,\mu_r(s_1))$. We define $\mathcal{A}_{\mathcal{E}}(s_1,\dots,s_m)$ as the minimal cost $\mathcal{A}_{\mathcal{E}}(s_1,\mu_r(s_1))$ over all restricted monotone mappings $\mu_r$. Let $\mu_r$ be the monotone map that minimizes $\mathcal{A}_{\mathcal{E}}(s_1,\dots,s_m)$. Then

$$\mathcal{A}_{\mathcal{E}}(s_1,\dots,s_m)= \sum_{w\in \mathcal{J}_1}\mathcal{E}(w,\mu_r(w))\le k\sum_{w\in \mathcal{J}_1}\mathcal{A}(w,\mu_r(w))\le k(1+12\epsilon)\mathcal{A}(s_1,\dots,s_m)$$
Here the cost estimation $\mathcal{E}$ ensures $\sum_{w\in \mathcal{J}_1}\mathcal{E}(w,\mu_r(w))\le k\sum_{w\in \mathcal{J}_1}\mathcal{A}(w,\mu_r(w))$. 


\subsection{Algorithm for computing cost estimation of $m$-window tuples.} 
We dedicate the rest of the section to design a suitable cost estimation $\mathcal{E}$. One challenge here is that at the beginning of the algorithm we do not have any prior knowledge of the restriction we put on $\mathcal{J}_k$ to create $\mathcal{\tilde{J}}_k$. We decide them adaptively while running the algorithm depending on the cost and alignment of the windows in $\mathcal{J}_1$ computed so far.

\paragraph{Window generating function.} In the algorithm we use a window generating function $\mathcal{P}(s,d,\theta)$, that given a string $s$, a window size parameter $d>0$ and a cost threshold $\theta\in[0,1]$ outputs the following set of overlapping windows. 
Take $\tau=\{0,\frac{1}{d},\frac{(1+\epsilon)}{d},\dots,\theta\}$. For each $\tau$, compute a set of windows $\mathcal{J}_{\tau}$ and set $\mathcal{P}(s,d,\theta)=\cup_{\tau}\mathcal{J}_{\tau}$. 
For $\tau=0$ set $h_{\tau}=d$ and $\ell_{\tau}=d$. For $\tau=\frac{1}{d}$ set $h_{\tau}=d+1$, $\ell_{\tau}=d-1$ and for general $\tau=\frac{(1+\epsilon)^k}{d}$, set $h_{\tau}=\lfloor d+\tau d\rfloor$ and $\ell_{\tau}=\lfloor d-\tau d\rfloor$. Set $\gamma_{\tau}=max(1,\lfloor \epsilon \theta d\rfloor)$. 
Define 
$$\mathcal{H}_{\tau}=\{s[1,h_{\tau}],s[\gamma_{\tau}+1,\gamma_{\tau}+h_{\tau}],s[2\gamma_{\tau}+1,2\gamma_{\tau}+h_{\tau}],\dots\}$$

$$\mathcal{L}_{\tau}=\{s[1,\ell_{\tau}],s[\gamma_{\tau}+1,\gamma_{\tau}+\ell_{\tau}],s[2\gamma_{\tau}+1,2\gamma_{\tau}+\ell_{\tau}],\dots\}$$
 
Set $\mathcal{J}_{\tau}=\mathcal{H}_{\tau}\cup\mathcal{L}_{\tau}$.

We now sketch the main algorithm, Algorithm \ref{alg:windowestimate} that takes a $(p,B)$-pseudorandom string $s_1$ and $m$ adversarial string $s_2,\dots,s_m$ as input and outputs a set of $m$-certified tuples each of the form $(w_1,\dots,w_m,c)$. Here $w_j$ is a window of string $s_j$. We call a tuple $(w_1,\dots,w_m,c)$ $m$-certified if $\mathcal{A}(w_1,\dots,w_m)$ is bounded by $c$. 
Our algorithm has two phases that we describe next.

\paragraph{Finding unique match.} In the first phase Algorithm \ref{alg:windowestimate} calls Algorithm \ref{alg:uniquematch} that 
 starts by partitioning the $(p,B)$-pseudorandom string $s_1$ into disjoint windows of size $\beta=max(|B|,\sqrt{n})$. Let $\mathcal{W}_1^s$ be the set of all the windows generated. For each string $s_j$, the algorithm then generates a set of windows $\mathcal{\bar{W}}_j^s=\cup_{\theta\in\{\frac{p}{4},\frac{p}{4(1+\epsilon)},\dots,0\}}\mathcal{P}(s_j,\beta,\theta)$.
For analysis fix an optimal alignment $\sigma$ of $s_1,\dots,s_m$. In the first phase Algorithm \ref{alg:uniquematch} identifies all the windows $w\in \mathcal{W}_1^s$ such that $|\bar{\sigma}(w)|\le \frac{p|w|}{4}$.
Instead of computing the cost directly, we try all $\theta\in \{\frac{p}{4},\frac{p}{4(1+\epsilon)},\frac{p}{4(1+\epsilon)^2},\dots,\frac{1}{|w|}\}$ and try to identify for each $\theta$ if $|\bar{\sigma}(w)|\le \theta |w|$, and output the minimum $\theta$ for which this is true.

If we perform a trivial search by finding for each fixed $\theta$, all the
tuples $(w_1,w_2,\dots,w_m)$ such that $\mathcal{A}(w_1,w_2,\dots,w_m)\le \theta |w|$ where $(w_1,w_2,\dots,w_m)\in \mathcal{W}_1^s\times\mathcal{\bar{W}}_{2,\theta}^s \times \dots \times \mathcal{\bar{W}}_{m,\theta}^s$, as $|\mathcal{W}_1^s|= \frac{n}{\beta}$ and $|\mathcal{\bar{W}}_{j,\theta}^s|=\frac{n}{\epsilon \theta \beta}$, then the total time required will be $\tilde{O}(\epsilon^{1-m}2^{m}mn^m)$. Here using the algorithm from Section \ref{sec:linear} for each tuple we can check if $\mathcal{A}(w_1,w_2,\dots,w_m)\le \theta |w|$ in time $\tilde{O}(2^{m}\theta^m m^2\beta^{m})$. As this naive search technique gives a large time bound, instead of estimating the cost of all tuples in $\mathcal{W}_1^s\times\mathcal{\bar{W}}_{2,\theta}^s \times \dots \times \mathcal{\bar{W}}_{m,\theta}^s$, we perform a restricted search using the following observation.

\begin{observation}
Given $0<\theta\le \frac{p}{4}$ for each $j\in[2,m]$, for every window $\bar{w}\in \mathcal{P}(s_j,\beta,\theta)$, there exists at most one window $w\in\mathcal{W}_1^s$ such that $\mathcal{A}(\bar{w},w)\le \theta |w|$.
\end{observation}

\begin{proof}
For contradiction assume there exists two windows $w,w' \in\mathcal{W}_1^s$ such that $\mathcal{A}(\bar{w},w)\le \theta |w|$ and $\mathcal{A}(\bar{w},w')\le \theta |w'|$. Note $|w|=|w'|=\beta$. Hence by triangular inequality $\mathcal{A}(w,w')\le 2\theta |w|\le \frac{p|w|}{2}$. As by definition $w,w'$ are disjoint, this violates the $(p,B)$-pseudorandom property of $s_1$.
\end{proof}

After creating $\mathcal{W}^s_1,\mathcal{\bar{W}}^s_2,\dots, \mathcal{\bar{W}}^s_m$, for each window $w_i\in \mathcal{W}^s_1$ and each $j\in[2,m]$, Algorithm \ref{alg:uniquematch} computes a set $\mathcal{W}^s_{i,j}\subseteq \mathcal{\bar{W}}^s_j$ containing all the windows $w\in \mathcal{\bar{W}}^s_j$ such that $\mathcal{A}(w_i,w)\le \frac{p|w_i|}{4}$. 
Next the algorithm calls the function Disjoint($\mathcal{{W}}^s_{i,j}$) to compute a maximal subset $\mathcal{\tilde{W}}^s_{i,j}\subseteq \mathcal{{W}}^s_{i,j}$ such that any pair of windows in $\mathcal{\tilde{W}}^s_{i,j}$ are index disjoint. 
This can be calculated using a simple greedy algorithm in time $O(|\mathcal{{W}}^s_{i,j}|^2)$. Then if $\sum_{j\in[2,\dots,m]}|\mathcal{\tilde{W}}^s_{i,j}|\ge \frac{16m}{p\epsilon}$, $\forall \theta$ and $\forall j\in [2,m]$ set $\mathcal{W}^s_{i,j,\theta}= \phi$ and $\mathcal{W}^s_{i,j}= \phi$. 
Otherwise set $\mathcal{W}^s_{i,j,\theta}=\mathcal{W}^s_{i,j}\cap \mathcal{\bar{W}}^s_{j,\theta}$.
For each $\theta$ and every $(w_i,w_2,\dots,w_m)\in w_i\times \mathcal{W}_{i,2,\theta}^s\times \mathcal{W}_{i,3,\theta}^s\times \dots \times \mathcal{W}_{i,m,\theta}^s$ check if $\mathcal{A}(w_i,w_2,\dots,w_m)\le \theta |w_i|$, output $(w_i,w_2,\dots,w_m,\theta|w_i|)$.
The intuition behind this is that if $|\mathcal{\tilde{W}}^s_{i,j}|$ is large many windows of $s_j$ have large cost. Therefore on average we can say that $w_i$ has large cost.

\paragraph{Approximation of large cost windows.} In the second phase, Algorithm \ref{alg:windowestimate} calls Algorithm \ref{alg:largecostmatch} that finds an approximation of the alignment cost of a set of variable sized windows of $s_1$ assuming the cost is large. Let $\beta=max(B,\sqrt{N})$. 
The algorithm starts by creating for each $\ell\in\{1,2,\dots,\frac{n}{\beta}\}$ a set of windows $\mathcal{W}_1^\ell$ of $s_1$ of length $\ell\beta$ defined as follows.

$$\mathcal{W}_1^\ell=\{s_1[1,\ell\beta],s_1[\beta+1,(\ell+1)\beta],s_1[2\beta+1,(\ell+2)\beta],\dots\}$$ 
Next for each window $w_i\in \mathcal{W}_1^\ell$, let $w_p$ be the window preceding $w_i$ and $w_q$ be the window following $w_i$ in $\mathcal{W}_1^s$. Note $w_p, w_q$ are well defined. Next for any $j\in[2,m]$ if either $\mathcal{W}_{p,j}^s=\phi$ or $\mathcal{W}_{q,j}^s=\phi$ or $\sum_{j\in[2,m]}|\mathcal{\tilde{W}}_{p,j}^s|\ge \frac{16m}{p\epsilon}$ or $\sum_{j\in[2,m]}|\mathcal{\tilde{W}}_{q,j}^s|\ge \frac{16m}{p\epsilon}$
discard $w_i$.
Otherwise assuming $w_p$ and $w_q$ has cost $\le \frac{p\beta}{4}$ in an optimal alignment $\sigma$, for each choice of $(\bar{w}_j,\bar{\bar{w}}_j)\in \mathcal{\tilde{W}}_{p,j}^s\times \mathcal{\tilde{W}}_{q,j}^s$, if 
$w_p$ is matched with $\bar{w}_j$ or some window of $\mathcal{{W}}_{p,j}^s$ having an overlap with $\bar{w}_j$ and $w_q$ is matched with $\bar{\bar{w}}_j$ or some window of $\mathcal{{W}}_{q,j}^s$ having an overlap with $\bar{\bar{w}}_j$ in string $s_j$, then window $w_i$ can have a match only in the substring $s_j^{\bar{w},\bar{\bar{w}}}=s_j[s(\bar{w}_j)-\beta,e(\bar{\bar{w}}_j)+\beta]$ of string $s_j$. 
Next if $\sum_{j\in[2,m]}s(\bar{\bar{w}}_j)-e(\bar{w}_j)\ge 5\ell \beta m$, then the alignment cost of $w_i$ is already at least $(3\ell\beta m-\ell\beta (m-1))/m\ge \ell\beta =|w_i|$. As here $\theta=1$, define the set of all possible windows in the substring $s_j^{\bar{w},\bar{\bar{w}}}$ as $\mathcal{W}_{i,j}^{\bar{w},\bar{\bar{w}}}\gets
	    \mathcal{P}(s_j^{\bar{w},\bar{\bar{w}}},\ell\beta,1)$.
Then for every $w_i\in \mathcal{W}_1^\ell$ and every $(w_2,\dots,w_m)\in \mathcal{W}_{i,2}^{\bar{w},\bar{\bar{w}}}\times \dots \times \mathcal{W}_{i,m}^{\bar{w},\bar{\bar{w}}}$ such that $|w_j|\le |w_i|$ for each $j\in[2,m]$
Algorithm \ref{alg:largecostmatch} outputs a certified tuple with trivial maximum cost $|w_i|$. Otherwise define the set of all possible windows in the substring $s_j^{\bar{w},\bar{\bar{w}}}$ as $\mathcal{W}_{i,j}^{\bar{w},\bar{\bar{w}}}\gets
	    \cup_{\theta \in \{1,\dots,\frac{p}{16}\}}\mathcal{P}(s_j^{\bar{w},\bar{\bar{w}}},\ell\beta,\theta)$. Then for every $w_i\in \mathcal{W}_1^\ell$ and every $(w_2,\dots,w_m)\in \mathcal{W}_{i,2}^{\bar{w},\bar{\bar{w}}}\times \dots \times \mathcal{W}_{i,m}^{\bar{w},\bar{\bar{w}}}$,
Algorithm \ref{alg:largecostmatch}
calls algorithm LargeAlign() to compute $(2-\frac{3p}{512})$ approximation of $\mathcal{A}(w_i,w_2,\dots,w_m)$.


\begin{algorithm}
	\begin{algorithmic}[1]
		\REQUIRE Strings $s_1,\dots,s_m$ of length $n$ where $s_1$ is $(p,B)-pseudorandom$; parameter $p\in (0,1)$, $B>0$.
		
		\ENSURE A set $S$ of certified $m$-window tuples $(w_1,\dots,w_m,c)$.
	
	     \STATE $\beta \gets max(B,\sqrt{n})$
	     
	     \STATE $S_1\gets FindUniqueMatch(s_1,\dots,s_m,p,B,\beta)$
	     \STATE $S_1\gets ApproxLargeWindow(s_1,\dots,s_m,p,B,\beta)$
	
	    \STATE $S\gets S_1\cup S_2$
	    
		\RETURN $S$
		\caption{{MultiWindowEstimation} $(s_1,\dots,s_m,p,B)$}
		\label{alg:windowestimate}
	\end{algorithmic}
\end{algorithm}

\begin{algorithm}
	\begin{algorithmic}[1]
		\REQUIRE Strings $s_1,\dots,s_m$ of length $n$ where $s_1$ is $(p,B)-pseudorandom$; parameter $p\in (0,1)$, $B>0$.
		
		\ENSURE A set $S_1$ of certified $m$-window tuples $(w_1,\dots,w_m,c)$. 
		
		\STATE $S_1\gets \phi$
		
		\FOR{$j=2,\dots,m$}
		\STATE $\mathcal{\bar{W}}_j^s\gets \phi$
		\ENDFOR
		\STATE $\mathcal{W}_1^s\gets\{s_1[1,\beta],s_1[\beta+1,2\beta],\dots\}$
		
		\FOR{$w_i\in \mathcal{W}_1^s$}
		\FOR{$j=2,\dots,m$}
		\STATE $\mathcal{W}_{i,j}^s\gets \phi$
		\ENDFOR
		
		\ENDFOR
		
		\FOR{$\theta\in\{\frac{p}{4},\frac{p}{(1+\epsilon)4},\dots,\frac{1}{\beta}\}$}
		
		\FOR{$j=2,\dots,m$}
		
	   \STATE	$\mathcal{\bar{W}}_{j,\theta}^s\gets \mathcal{P}(s_j,\beta,\theta)$
		
	   \STATE	$\mathcal{\bar{W}}_{j}^s\gets \mathcal{\bar{W}}_j^s\cup \mathcal{\bar{W}}_{j,\theta}^s$
	   \ENDFOR
	   \ENDFOR
	   
	   \FOR{$w_i\in \mathcal{W}_1^s$}
	   
	   \FOR{$j=2,\dots,m$}
	   
	   \FOR{$w_j\in \mathcal{\bar{W}}^s_j$}
	   \IF{$\mathcal{A}(w_i,w_j)\le \frac{p\beta}{4}$}
	   \STATE $\mathcal{W}_{i,j}^s\gets\mathcal{W}_{i,j}^s\cup w_j$
	   \ENDIF
	   \ENDFOR
	   \STATE $\mathcal{\tilde{W}}_{i,j}^s\gets Disjoint(\mathcal{W}_{i,j}^s)$
	   \ENDFOR
	   \ENDFOR

	   \FOR{$w_i\in \mathcal{W}_1^s$}
	   
	   \FOR{$j=2,\dots,m$}
	   
	   \FOR{$\theta\in\{\frac{p}{4},\frac{p}{(1+\epsilon)4},\dots,\frac{1}{\beta}\}$}
	   \IF{$\sum_{j\in[2,m]}|\mathcal{\tilde{W}}_{i,j}^s|\ge \frac{16m}{p\epsilon}$}
	   \STATE $\mathcal{W}_{i,j,\theta}^s\gets \phi$
	   
	   \ELSE
	   
	   \STATE $\mathcal{W}_{i,j,\theta}^s\gets \mathcal{W}_{i,j}^s\cap \mathcal{\bar{W}}_{j,\theta}^s$
	   \ENDIF
	   
	   \ENDFOR
	   \STATE $\mathcal{W}_{i,j}^s\gets \cup_\theta \mathcal{W}_{i,j,\theta}^s $
	   \ENDFOR
	   
	   \ENDFOR
	
       \FOR{$w_i\in \mathcal{W}_1^s$}
       
       \FOR{$\theta\in\{\frac{p}{4},\frac{p}{(1+\epsilon)4},\dots,\frac{1}{\beta}\}$}
       
       \FOR{$(w_2,\dots,w_m)\in \mathcal{W}_{i,2,\theta}^s\times \dots,\times \mathcal{W}_{i,m,\theta}^s$}
       
       \IF{$\mathcal{A}(w_i,w_2,\dots,w_m)\le \theta|w_i|$}
       
       \STATE $S_1\gets S_1\cup (w_i,w_2,\dots,w_m, \theta|w_i|)$
       
       \ENDIF
       
       \ENDFOR

       \ENDFOR

       \ENDFOR
       
		\RETURN ${S_1}$
		\caption{{FindUniqueMatch} $(s_1,\dots,s_m,p,B,\beta)$}
		\label{alg:uniquematch}
	\end{algorithmic}
\end{algorithm}

\begin{algorithm}
	\begin{algorithmic}[1]
		\REQUIRE Strings $s_1,\dots,s_m$ of length $n$ where $s_1$ is $(p,B)-pseudorandom$; parameter $p\in (0,1)$, $B>0$.
		
		\ENSURE A set $S_2$ of certified $m$-window tuples $(w_1,\dots,w_m,c)$.
	
	\STATE $S_2\gets \phi$
		
		\FOR{$\ell=1,2,\dots,\frac{n}{\beta}$}
		\STATE $\mathcal{W}_1^\ell\gets\{s_1[1,\ell\beta],s_1[\beta+1,(\ell+1)\beta],s_1[2\beta+1,(\ell+2)\beta],\dots\}$
		\ENDFOR
		
		\STATE $\mathcal{W}_1^{large}\gets \cup_\ell \mathcal{W}_1^\ell$
		
		\FOR{$\ell=1,2,\dots,\frac{n}{\beta}$}
	
	    \FOR{$w_i\in \mathcal{W}_1^\ell$}
	    
	    \STATE $\alpha\gets s(w_i)$
	    
	    \STATE $\gamma \gets e(w_i)$
	    
	    \STATE $w_p\gets w_j\in \mathcal{W}_1^s$ such that $e(w_j)=\alpha -1$
	    
	    \STATE $w_q\gets w_k\in \mathcal{W}_1^s$ such that $s(w_k)=\gamma+1$
	    
	    \IF{$\exists j\in[2,m]$ such that $\mathcal{W}_{p,j}^s= \phi$ or $\mathcal{W}_{q,j}^s= \phi$ or $\sum_{j\in[2,m]}|\mathcal{\tilde{W}}_{p,j}^s|\ge \frac{16m}{p\epsilon}$ or $\sum_{j\in[2,m]}|\mathcal{\tilde{W}}_{q,j}^s|\ge \frac{16m}{p\epsilon}$}
	    
	    \STATE Discard $w_i$

	    \ELSE

	    \FOR{$\bar{w}_2,\bar{\bar{w}}_2,\dots,\bar{w}_m,\bar{\bar{w}}_m\in \mathcal{\tilde{W}}^s_{p,2}\times  \mathcal{\tilde{W}}^s_{q,2}\times \dots\times  \mathcal{\tilde{W}}^s_{p,m}\times  \mathcal{\tilde{W}}^s_{q,m}$}
	    
	    \FOR{$j=2,\dots,m$}
	    
	    \STATE $s_j^{\bar{w},\bar{\bar{w}}}\gets s_j[s(\bar{w}_j)-\beta,e(\bar{\bar{w}}_j)+\beta]$
	    
	    
	    \ENDFOR
	    \IF{$\sum_{j\in[2,m]}s(\bar{\bar{w}}_j)-e(\bar{w}_j)\ge 5\ell \beta m$}
	    
	    \FOR{$j=2,\dots,m$}

	    \STATE $\mathcal{W}_{i,j}^{\bar{w},\bar{\bar{w}}}\gets
	    \mathcal{P}(s_j^{\bar{w},\bar{\bar{w}}},\ell\beta,1)$
	    
	    \ENDFOR
	    
	    \FOR{$(w_2,\dots,w_m)\in \mathcal{W}_{i,2}^{\bar{w},\bar{\bar{w}}}\times \dots \times \mathcal{W}_{i,m}^{\bar{w},\bar{\bar{w}}}$}
	    \IF{$\forall j\in[2,m] |w_j|\le |w_i|$}
	    
	    \STATE $S_2\gets S_2\cup(w_i,w_2,\dots,w_m,|w_i|)$
	    \ENDIF
	    \ENDFOR
	    
	    \ELSE
	    
	    \FOR{$j=2,\dots,m$}

	    \STATE $\mathcal{W}_{i,j}^{\bar{w},\bar{\bar{w}}}\gets
	    \cup_{\theta \in \{1,\dots,\frac{p}{16}\}}\mathcal{P}(s_j^{\bar{w},\bar{\bar{w}}},\ell\beta,\theta)$
	    
	    \ENDFOR
	    
	    \FOR{$(w_2,\dots,w_m)\in \mathcal{W}_{i,2}^{\bar{w},\bar{\bar{w}}}\times \dots \times \mathcal{W}_{i,m}^{\bar{w},\bar{\bar{w}}}$}
	    
	    \STATE $S_2\gets S_2\cup(w_i,w_2,\dots,w_m,LargeAlign(w_i,w_2,\dots,w_m))$
	    
	    \ENDFOR
	    
	    \ENDIF
	    
	    \ENDFOR
	    \ENDIF
	    \ENDFOR
	    \ENDFOR

		\RETURN ${S_2}$
		\caption{{ApproxLargeWindows} $(s_1,\dots,s_m,p,B,\beta)$}
		\label{alg:largecostmatch}
	\end{algorithmic}
\end{algorithm}

\paragraph{Correctness and Running time analysis.} We start by fixing an optimal alignment $\sigma$ of $s_1,\dots s_m$.

\begin{definition}
\label{def:uniquewindow}
Given $p,\epsilon\in(0,1)$ we call a window $w_i\in \mathcal{W}_1^s$ uniquely certifiable if $|w_i|=\beta$, $|\sigma(w_i)|\ge |w_i|-\frac{p|w_i|}{8}$ and $\forall j\in[2,\dots,m]$, $\sum_{j\in[2,m]}|\mathcal{\tilde{W}}_{i,j}^s|< \frac{16m}{p\epsilon}$.
\end{definition}
Notice we can represent every window of $\mathcal{W}_1^{large}$ as a concatenation of a set of consecutive windows of $\mathcal{W}_1^s$. Therefore for each $w_i\in \mathcal{W}_1^{large}$ we can write $w_i=w_u\circ w_{u+1}\circ \dots\circ w_v$ where, $w_u,\dots,w_v\in \mathcal{W}_1^s$. Let $prev(w_i)$ denotes the window in $\mathcal{W}_1^s$ appearing before $w_u$ and $next(w_i)$ denotes the window in $\mathcal{W}_1^s$ appearing after $w_v$.

\begin{definition}
\label{def:largewindow}
Given $p,\epsilon\in(0,1)$ we call a window $w_i\in \mathcal{W}_1^{large}$ where $w_i=w_u\circ w_{u+1}\circ \dots\circ w_v$ large-cost if the following holds.
\begin{enumerate}
    \item $\forall j\in[u,v]$, $|w_j|=\beta$, $w_j$ is not uniquely certifiable.
    \item $prev(w_u)$ and $next(w_v)$ are uniquely certifiable.
\end{enumerate}
\end{definition}

For a given \emph{large-cost} window $w_i\in \mathcal{W}_1^{large}$ where $w_i=w_u\circ w_{u+1}\circ \dots\circ w_v$, Let $\mathcal{D}_i^1\subseteq \{w_u,\dots,w_v\}$ be the set of windows such that for every $w\in \mathcal{D}_i^1$, $\sigma(w)\ge |w|-\frac{p|w|}{8}$ and $\mathcal{D}_i^2\subseteq \{w_u,\dots,w_v\}$ be the set of windows defined as $\mathcal{D}_i^2=\{w_u,\dots,w_v\} \setminus \mathcal{D}_i^1$. Moreover let $prev(w_i)^j$ be the window in string $s_j$ to which  $prev(w_i)$ is matched under $\sigma$. Similarly define $next(w_i)^j$. Note both $prev(w_i)$ and $next(w_i)$ are uniquely certifiable. 

\begin{definition}
\label{def:triapprox}
Given $p,\epsilon\in(0,1)$ we call a window $w_i\in \mathcal{W}_1^{large}$ where $w_i=w_u\circ w_{u+1}\circ \dots\circ w_v$ trivially approximable if the following holds.
\begin{enumerate}
    \item $w_i$ is large-cost.
    \item $\sum_{j\in[2,m]}(s(next(w_i)^j)-e(prev(w_i)^j)-1)\ge 2m|w_i|$.
\end{enumerate}
\end{definition}

\begin{definition}
\label{def:largewindowcertify}
Given $p,\epsilon\in(0,1)$ we call a window $w_i\in \mathcal{W}_1^{large}$ where $w_i=w_u\circ w_{u+1}\circ \dots\circ w_v$ large-cost certifiable if the following holds.
\begin{enumerate}
    \item $w_i$ is large-cost.
    \item $w_i$ is not trivially approximable.
    \item $|\mathcal{D}_i^2|\ge |\mathcal{D}_i^1|$.
\end{enumerate}
\end{definition}
 
We call a \emph{large-cost} \emph{not trivially approximable} window $w_i\in \mathcal{W}_1^{large}$ \emph{large-cost uncertifiable} if $w_i$ is not large-cost certifiable.

\begin{definition}
\label{def:verpartition}
Given a $(p,B)$-pseudorandom string $s_1$ and $m-1$ adversarial strings $s_2,\dots,s_m$ and an optimal alignment $\sigma$ of $s_1,\dots,s_m$ we call a partition $\mathcal{J}_1=\{w_1,\dots,w_k\}$ of $s_1$ into $k$ windows a verified partition if the following holds.
\begin{enumerate}
    \item $\forall j\in[k]$, $|w_j|=c\beta$ where $\beta=max(B,\sqrt{n})$ and $c\in[1,\frac{n}{\beta}]$ is a constant.
    \item Every window $w_j$ is either uniquely certifiable or large-cost.
    \item For any window $w_j$ if $w_j$ is large-cost then $prev(w_j)$ and $next(w_j)$ are uniquely certifiable.
\end{enumerate}

\end{definition}

\begin{claim}
\label{claim:verpartition}
Given a $(p,B)$-pseudorandom string $s_1$ and $m-1$ adversarial strings $s_2,\dots,s_m$ and an optimal alignment $\sigma$ of $s_1,\dots,s_m$, there exists a unique verified partition $\mathcal{J}_1$ of $s_1$. Moreover for every window $w\in \mathcal{J}_1$ if $w$ is uniquely certifiable then $w\in \mathcal{W}_1^s$ and if $w$ is large-cost then $w\in \mathcal{W}_1^{large}$.
\end{claim}

\begin{proof}
From Definition \ref{def:verpartition} we can directly claim about the existence of a verified partition for a given set of strings  $s_1,\dots,s_m$ and an optimal alignment $\sigma$. Next for contradiction assume there exists more than one different verified partition namely $\mathcal{J}_1$ and $\mathcal{J}_2$. Then there exists at least one pair of windows $(w_1,w_2)\in \mathcal{J}_1\times \mathcal{J}_2$ such that $w_1\cap w_2\neq \phi$ and $w_1\neq w_2$. Without loss of generality assume $s(w_1)<s(w_2)$. Trivially by construction $|w_1|,|w_2|>\beta$, hence both are large-cost. Let $w_1=w_1^1\circ\dots\circ w_1^{k_1}$ and $w_2=w_2^1\circ\dots\circ w_2^{k_2}$. As $s(w_1)<s(w_2)$, let $w_1^1\circ\dots\circ w_1^{p}$ be the prefix of $w_1$ that is not a part of $w_2$. By definition as $w_1$ is large cost and $w_1^p\in w_1$, $w_1^p$ is not uniquely certifiable. On the contrary as  $w_1^p=prev(w_2^1)$ by definition it needs to be uniquely certifiable and we get a contradiction. The last point follows trivially from Definition \ref{def:verpartition} and the construction of $\mathcal{W}_1^s$ and $\mathcal{W}_1^{large}$.
\end{proof}

For a given unique verified partition $\mathcal{J}_1$ of $s_1$ and for every window $w\in \mathcal{J}_1$ consider the set of windows $\mathcal{\tilde{J}}_j^{|w|}$ of $s_j$ as defined in Section \ref{sec:windowdecomposition}. For each $w\in \mathcal{J}_1$, if $w$ is \emph{uniquely certifiable} then define $\mathcal{\tilde{W}}_j^{|w|}=\cup_{\theta \in \{\frac{p}{4},\frac{p}{4(1+\epsilon)},\dots,\frac{1}{\beta}\}}\mathcal{W}_{i,j,\theta}^s$ and if $w$ is \emph{large-cost} then $\mathcal{\tilde{W}}_j^{|w|}=\cup_{(\bar{w},\bar{\bar{w}})\in \mathcal{\tilde{W}}^s_{p,j}\times \mathcal{\tilde{W}}^s_{q,j}}\mathcal{W}_{i,j}^{\bar{w},\bar{\bar{w}}}$.

\begin{lemma}
\label{lem:searchspace}
For a given unique verified partition $\mathcal{J}_1$ of $s_1$ for every window $w_i\in \mathcal{J}_1$, $\mathcal{\tilde{J}}_j^{|w_i|}\subseteq \mathcal{\tilde{W}}_j^{|w_i|}$.
\end{lemma}

\begin{proof}
First consider the case where $w_i$ is uniquely certifiable. Then 
$\mathcal{\tilde{W}}_j^{|w_i|}=\cup_{\theta \in \{\frac{p}{4},\frac{p}{4(1+\epsilon)},\dots,\frac{1}{\beta}\}}\mathcal{W}_{i,j,\theta}^s$. 
By definition $|\sigma(w_i)|\ge |w_i|-\frac{p|w_i|}{8}$. Hence if the cost of alignment $\sigma$ restricted to $w_i$ is $\theta |w_i|$ then $\theta\le \frac{p}{8}$. Hence by construction we have $\mathcal{\tilde{J}}_j^{|w_i|}\subseteq \cup_{\theta \in \{\frac{p}{4},\frac{p}{4(1+\epsilon)},\dots,\frac{1}{\beta}\}}\mathcal{W}_{i,j,\theta}^s=\mathcal{\tilde{W}}_j^{|w_i|}$. 

Next given $w_i$ is large cost $\mathcal{\tilde{W}}_j^{|w_i|}=\cup_{(\bar{w},\bar{\bar{w}})\in \mathcal{\tilde{W}}^s_{p,j}\times \mathcal{\tilde{W}}^s_{q,j}}\mathcal{W}_{i,j}^{\bar{w},\bar{\bar{w}}}$.
By definition we know $w_p=prev(w_i)$ and $w_q=next(w_i)$ are uniquely certifiable. Hence  $w_p$ is matched with some window in $\mathcal{W}_{p,j}^s$ in string $s_j$ and $w_q$ is matched with some window in $\mathcal{W}_{q,j}^s$ in string $s_j$ under $\sigma$. Notice as for each $w\in \mathcal{W}_{p,j}^s$ there exists a window $w'\in \mathcal{\tilde{W}}_{p,j}^s$  such that $w\cap w'\neq \phi$ then
$s(w)\ge s(w')-\beta$. Similarly for each $w\in \mathcal{W}_{q,j}^s$ there exists a window $w'\in \mathcal{\tilde{W}}_{q,j}^s$  such that $w\cap w'\neq \phi$. Hence
$s(w)\le s(w')+\beta$. Therefore it is sufficient if we try to match $w_i$ in the substring $s_j^{\bar{w}_j,\bar{\bar{w}}_j}=s_j[s(\bar{w}_j)-\beta,e(\bar{\bar{w}}_j)+\beta]$ of $s_j$ for all choices of $(\bar{w}_j,\bar{\bar{w}}_j)\in \mathcal{\tilde{W}}_{p,j}^s\times \mathcal{\tilde{W}}_{q,j}^s$.
Hence by construction $\mathcal{\tilde{J}}_j^{|w_i|}\subseteq \cup_{(\bar{w},\bar{\bar{w}})\in \mathcal{\tilde{W}}^s_{p,j}\times \mathcal{\tilde{W}}^s_{q,j}}\mathcal{W}_{i,j}^{\bar{w},\bar{\bar{w}}}=\mathcal{\tilde{W}}_j^{|w_i|}$. Notice as a special case for choices of  $(\bar{w}_2,\bar{\bar{w}}_2,\dots , \bar{w}_m,\bar{\bar{w}}_m)$ if $\sum_{j\in[2,m]}s(\bar{\bar{w}}_j)-e(\bar{{w}}_j)\ge 5\ell \beta m$, cost of $w_i$ under $\sigma$ is already $w_i$. Hence if $w_i$ is matched with a window of length $>|w_i|$ in some $s_j$, we can convert the size of the window to $|w_i|$ by removing first few characters. Notice once we perform it for all $s_j$s the cost is still at most $|w_i|$. Hence for this case in our algorithm it is enough to consider windows of length at most $|w_i|$.
\end{proof}

\begin{claim}
\label{claim:cover}
Given a $(p,B)$-pseudorandom string $s_1$ and $m-1$ adversarial strings $s_2,\dots,s_m$, an optimal alignment $\sigma$ of $s_1,\dots,s_m$ and a verified partition $\mathcal{J}_1=\{w_1,\dots,w_k\}$ of $s_1$ 
there exists a restricted monotone mapping $\mu_r: \mathcal{J}_1\rightarrow \cup_{w_i\in \mathcal{J}_1}\mathcal{\tilde{W}}_2^{|w_i|}\cup\{\perp\}\times \dots \times \cup_{w_i\in \mathcal{J}_1}\mathcal{\tilde{W}}_m^{|w_i|}\cup\{\perp\}$ satisfying $\mathcal{A}(s_1,\dots,s_m)\le \mathcal{A}(s_1,\mu_r(s_1))\le (1+12\epsilon)\mathcal{A}(s_1,\dots,s_m)$. 
\end{claim}

\begin{proof}
Claim \ref{claim:cover} directly follows from Lemma \ref{lem:window2} and Lemma \ref{lem:searchspace}
\end{proof}

\begin{lemma}
\label{lem:randomcostapprox}
Given $0<\epsilon\le 1/6$, and an optimal alignment $\sigma$ of $s_1,\dots,s_m$ let the corresponding verified partition be $\mathcal{J}_1$.  Then for any restricted monotone mapping $\mu_r: \mathcal{J}_1\rightarrow \cup_{w_i\in \mathcal{J}_1}\mathcal{\tilde{W}}_2^{|w_i|}\cup\{\perp\}\times \dots \times \cup_{w_i\in \mathcal{J}_1}\mathcal{\tilde{W}}_m^{|w_i|}\cup\{\perp\}$ satisfying $\mathcal{A}(s_1,\dots,s_m)\le \mathcal{A}(s_1,\mu_r(s_1))\le (1+12\epsilon)\mathcal{A}(s_1,\dots,s_m)$ for every window $w_i\in \mathcal{J}_1$ Algorithm \ref{alg:windowestimate} outputs a certified tuple $(w_i,\mu_r(w_i),c_i)$ such that the followings are satisfied.

\begin{enumerate}
    \item If $w_i$ is uniquely certifiable then $\mathcal{A}(w_i,\mu_r(w_i))\le c_i\le (1+\epsilon)\mathcal{A}(w_i,\mu_r(w_i)).$
    \item If $w_i$ is trivially approximable then $\mathcal{A}(w_i,\mu_r(w_i))=c_i=|w_i|.$
    \item If $w_i$ is large-cost certifiable then $\mathcal{A}(w_i,\mu_r(w_i))\le c_i\le (2-\frac{3p}{512}+\epsilon)\mathcal{A}(w_i,\mu_r(w_i)).$
    \item Otherwise $\mathcal{A}(w_i,\mu_r(w_i))\le c_i\le (|w_i|+\sum_{w_j\in \mu_r(w_i)}|w_j|)/m.$
\end{enumerate}
\end{lemma}

\begin{proof}

Consider a verified partition $\mathcal{J}_1$ of $s_1$ and a restricted monotone mapping $\mu_r: \mathcal{J}_1\rightarrow \cup_{w_i\in \mathcal{J}_1}\mathcal{\tilde{W}}_2^{|w_i|}\cup\{\perp\}\times \dots \times \cup_{w_i\in \mathcal{J}_1}\mathcal{\tilde{W}}_m^{|w_i|}\cup\{\perp\}$.
Let $w_i$ be uniquely certifiable. Then $|\sigma(w_i)|\ge |w_i|-\frac{p|w_i|}{8}$. Hence by Lemma \ref{lem:window1} $\mathcal{A}(w_i,\mu_r(w_i))\le (1+12\epsilon)\frac{p|w_i|}{8}\le \frac{p|w_i|}{4}$. Let $\theta\in \{\frac{p}{4},\frac{p}{(1+\epsilon)4},\dots,\frac{1}{\beta}\}$ be such that $\frac{\theta |w_i|}{(1+\epsilon)}<\mathcal{A}(w_i,\mu_r(w_i))\le \theta |w_i|$. Then $\mu_r(w_i)\in \mathcal{W}^s_{i,2,\theta}\times \dots \times \mathcal{W}^s_{i,m,\theta}$ and Algorithm \ref{alg:uniquematch} outputs a certified $m$-tuple $(w_i,\mu_r(w_i),\theta|w_i|)$. As $\frac{\theta |w_i|}{(1+\epsilon)}<\mathcal{A}(w_i,\mu_r(w_i))$ we get the required approximation guarantee. By the choice of $\theta$ $\mathcal{A}(w_i,\mu_r(w_i))\le \theta|w_i|=c_i$.

Next let $w_i$ is large-cost and $w_i=w_u\circ w_{u+1}\circ\dots\circ w_v$ where $\forall j\in[u,v]$, $|w_j|=\beta$. Here $|w_i|=k\beta$. As $prev(w_i)$ and $next(w_i)$ is uniquely certifiable their matches in each string $s_j$, put a restriction on the substring of $s_j$ to which $w_i$ has a match. Formally, for each string $s_j$ we enumerate all potential matches for $prev(w_i)$ and $next(w_i)$ by trying all possible pairs $(\bar{w}_j,\bar{\bar{w}}_j)\in \mathcal{\tilde{W}}_{p,j}^s\times \mathcal{\tilde{W}}_{q,j}^s$. For some choice of $(\bar{w}_j,\bar{\bar{w}}_j)$ if $\sum_{j\in [2,\dots,m]} s(\bar{\bar{w}}_j)-e(\bar{w}_j)\ge 5k\beta m$ then it follows that for any $(\bar{w'}_j,\bar{\bar{w'}}_j)\in \mathcal{{W}}_{p,j}^s\times \mathcal{{W}}_{q,j}^s$ where $\bar{w'}_j\cap \bar{w}_j \neq \phi$ and $\bar{\bar{w'}}_j\cap \bar{\bar{w}}_j \neq \phi$, $\sum_{j\in [2,\dots,m]} s(\bar{\bar{w'}}_j)-e(\bar{w'}_j)\ge 5k\beta m-2\beta m\ge 3k\beta m.$ But in this case cost of $w_i$ under $\sigma$ is at least $\sum_{j\in [2,\dots,m]} s(\bar{\bar{w'}}_j)-e(\bar{w'}_j)-k\beta m\ge 2k\beta m$ and hence $w_i$ is trivially approximable. 
For this Algorithm \ref{alg:uniquematch} outputs a certified $m$-tuple $(w_i,\mu_r(w_i),|w_i|)$. Moreover Algorithm \ref{alg:uniquematch} ensures that each window of $\mu_r(w_i)$ has size at most $|w_i|$. Hence the approximation ration is $1$. Trivially $\mathcal{A}(w_i,\mu_r(w_i))= |w_i|=c_i$.

Otherwise assume $\sum_{j\in [2,\dots,m]} s(\bar{\bar{w}}_j)-e(\bar{w}_j)< 5k\beta m$. First consider the case when $w_i$ is large-cost certifiable. In this case by definition $|\mathcal{D}_i^2|\ge |\mathcal{D}_i^1|$ and cost of every window of $\mathcal{D}_i^2$ is at least $\frac{p\beta}{8}$. Hence, $|\sigma(w_i)|\le |w_i|-\frac{p\beta k}{16}=|w_i|-\frac{p|w_i|}{16}$. Let $\mu_r(w_i)=(w'_2,\dots,w'_m)$. Notice by construction of $\mu_r$ for each $j\in[2,m]$, $|i_s^{w_i}(j)-s(w'_j)|\le \frac{\epsilon p|w_i|}{16}$. Here, $i_s^{w_i}$ is the index of the first character of $w_i$ that is aligned under $\sigma$ and $i_s^{w_i}(j)$ is the index of the character of $s_j$ that is aligned with $w_i[i_s^{w_i}]$. Hence, $\mathcal{A}(w_i,\mu(w_i))\ge \frac{p|w_i|}{16}-\frac{2\epsilon p|w_i|}{16}>\frac{p|w_i|}{32}$. In this case algorithm \ref{alg:largecostmatch} returns a tuple $(w_i,\mu_r(w_i),LargeAlign(w_i,\mu_r(w_i)))$ where $LargeAlign(w_i,\mu_r(w_i))$ returns $(2-\frac{3p}{512}+\epsilon)$ approximation of $\mathcal{A}(w_i,\mu_r(w_i))$.

 Otherwise if $w_i$ is not large-cost certifiable then as $\mathcal{A}(w_i,\mu(w_i))$ can be small, algorithm LargeAlign() can't ensure a good approximation ($<2$) for $\mathcal{A}(w_i,\mu(w_i))$. But it returns an alignment of cost $c$ where $c\le (|w_i|+\sum_{w_j\in \mu_r(w_i)}|w_j|)/m$ as this is the maximum possible cost. Hence Algorithm \ref{alg:largecostmatch} returns a tuple $(w_i,\mu_r(w_i),c)$ where $c\le (|w_i|+\sum_{w_j\in \mu_r(w_i)}|w_j|)/m$. Trivially $\mathcal{A}(w_i,\mu_r(w_i))\le (|w_i|+\sum_{w_j\in \mu_r(w_i)}|w_j|)/m=c_i$.

\end{proof}

\begin{claim}
\label{claim:disjoint}
For $w_i,w_j\in \mathcal{J}_1$, such that $w_i\neq w_j$, for each $k\in[2,m]$, $\mathcal{\tilde{W}}^s_{i,k}\cap\mathcal{\tilde{W}}^s_{j,k}=\phi$.
\end{claim}
\begin{proof}
As otherwise assume $\exists j\in[2,m]$ such that $w\in \mathcal{\tilde{W}}^s_{i,k}\cap\mathcal{\tilde{W}}^s_{j,k}$. But then $\mathcal{A}(w,w_i)\le \frac{p\beta}{4}$ and $\mathcal{A}(w,w_j)\le \frac{p\beta}{4}$. Here, $\beta=|w_i|=|w_j|$. By triangular inequality $\mathcal{A}(w_i,w_j)\le \frac{p\beta}{2}$ and we get a contradiction.
\end{proof}

\begin{claim}
\label{claim:triapprox}
If $\sum_{\begin{subarray} ww_i\in \mathcal{J}_1\\ w_i \textit{ is large-cost uncertifiable } \end{subarray} }|w_i|=\ell$, then $\mathcal{A}(s_1,\dots,s_m)\ge \frac{\ell}{2\epsilon}$. 
\end{claim}
\begin{proof}
Define $\mathcal{D}^2=\cup_{w_i\mathcal{J}_1}\mathcal{D}_i^2$ and $\mathcal{D}^1=\cup_{w_i\mathcal{J}_1}\mathcal{D}_i^1$. Here for each $w\in \mathcal{D}^1,\mathcal{D}^2$, $|w|=\beta$. As for each $w_i\in \mathcal{J}_1$, if $w_i$ is large-cost uncertifiable then  $|\mathcal{D}_i^1|\ge |\mathcal{D}_i^2|$. Hence, $|\mathcal{D}^1|\ge  \frac{\ell}{2\beta}$. By definition for each $w_j\in \mathcal{D}^1$, $\sum_{k\in[2,m]}|\mathcal{\tilde{W}}_{j,k}^s|\ge \frac{16m}{p\epsilon}$. As except one tuple $(w_j,w_2,\dots,w_m)$ where $(w_2,\dots,w_m)\in \mathcal{\tilde{W}}_{j,2}^s\times \dots \times \mathcal{\tilde{W}}_{j,m}^s$ all others have cost at least $\frac{p\beta}{8}$, total number of unaligned characters in $\cup_{k\in[2,m]}\mathcal{\tilde{W}}_{j,k}^s$ and $w_j$ is at least $(\frac{16m}{p\epsilon}-m)\frac{p\beta}{8}\ge \frac{m\beta}{\epsilon}$. By Lemma \ref{claim:disjoint} for two different $w_j,w_{j'}\in \mathcal{D}^1$ if $w_j\neq w_{j'}$, $\forall k\in[2,m]$, $\mathcal{\tilde{W}}_{j,k}^s\cap\mathcal{\tilde{W}}_{j',k}^s=\phi$. Hence the total number of unaligned characters in all windows in $\cup_{\substack {ww_j\in \mathcal{D}^1\\k\in[2,m]} }\mathcal{\tilde{W}}_{j,k}^s$ is at least $\frac{m\beta}{\epsilon}|\mathcal{D}^1|\ge \frac{m\beta}{\ell}\times \frac{\ell}{2\beta}=\frac{m\ell}{2\epsilon}$. Hence $\mathcal{A}(s_1,\dots,s_m)\ge \frac{\ell}{2\epsilon}$.
\end{proof}

\begin{definition}
\label{def:costestimation}
Given a $(p,B)$-pseudorandom string $s_1$ and $m-1$ adversarial strings $s_2,\dots,s_m$ Algorithm \ref{alg:windowestimate} outputs a set $S=\{(w_1^1,\dots,w_m^1,c^1),\dots,(w_1^k,\dots,w_m^k,c^k)\}$ of $(m+1)-$tuples. We define the cost estimation function  $\mathcal{E}:\{(w_1,\dots,w_m)| \textit{ where } (w_1,\dots,w_m,c^j)\in S  \textit{ for some } c^j\in\{c^1,\dots, c^k\}\}\rightarrow \{0,1,\dots,max_{j\in[k]}c^j\}$ as follows: 
$$\mathcal{E}(w_1,\dots,w_m)=min_{(w_1,\dots,w_m,c^j)\in S} c^j.$$
\end{definition}

\begin{theorem}
\label{thm:mainrandom}
Given a $(p,B)$-pseudorandom string $s_1$ and $m-1$ adversarial strings $s_2,\dots,s_m$ Algorithm \ref{alg:windowestimate} outputs a set $S=\{(w_1^1,\dots,w_m^1,c^1),\dots,(w_1^k,\dots,w_m^k,c^k)\}$ of $(m+1)-$tuples where each $(w_1^j,\dots,w_m^j,c^j)$ is a certified $m$-tuple and 

\begin{itemize}
    \item For every verified partition $\mathcal{J}_1\subseteq \{w_1^1,\dots,w_1^k\}$ and every monotone map $\mu_r:\mathcal{J}_1\rightarrow \{{w}_2^1,\dots,{w}_2^{k}\} \times \dots \times \{{w}_m^1,\dots,{w}_m^{k}\}$, $\mathcal{A}(s_1,\dots,s_m)\le \mathcal{A}_{\mathcal{E}}(s_1,\mu_r(s_1))$.

\item There exists a a subset $\bar{S}\subseteq S$ where, $\bar{S}=\{(\bar{w}_1^1,\dots,\bar{w}_m^1,c^1),\dots,(\bar{w}_1^{k'},\dots,\bar{w}_m^{k'},c^{k'})\}$ such that the following holds.
\begin{enumerate}
    \item $\{\bar{w}_1^1,\dots,\bar{w}_1^{k'}\}$ is a verified partition of $s_1$.
    \item $\mu_r:\{\bar{w}_1^1,\dots,\bar{w}_1^{k'}\}\rightarrow \{\bar{w}_2^1,\dots,\bar{w}_2^{k'}\} \times \dots \times \{\bar{w}_m^1,\dots,\bar{w}_m^{k'}\}$ is a restricted monotone map satisfying $ \mathcal{A}_{\mathcal{E}}(s_1,\mu_r(s_1))\le (2-\frac{3p}{512}+89\epsilon)\mathcal{A}(s_1,\dots,s_m)$.
    \end{enumerate}
\end{itemize}
Moreover the running time of Algorithm \ref{alg:windowestimate} is $\tilde{O}_m(n\beta^{m-1}+ n^{ \lfloor m/2\rfloor+3}+mn^3)$ and 

\noindent $|S|=\tilde{O}(\frac{n96^m\log^m n \beta^{m-1}}{p^m\epsilon^{2m}}+\frac{716^{2m}\log^m nn^{m/2+1}}{\epsilon^{4m}p^{3m}}).$    
\end{theorem}

\begin{proof}
For the analysis purpose fix an optimal alignment $\sigma$ of $s_1,\dots,s_m$ and the corresponding unique verified partition $\mathcal{J}_1$.
\begin{itemize}[leftmargin=.5cm]
\item For a given monotone map $\mu_r$ we can define 
\begin{align*}
   \mathcal{A}_{\mathcal{E}}(s_1,\mu_r(s_1)) & \ge \sum_{w\in \mathcal{J}_1}\mathcal{E}(w,\mu_r(w))\\
   &\ge \sum_{w\in \mathcal{J}_1}\mathcal{A}(w,\mu_r(w))\\
   &= \mathcal{A}(s_1,\mu_r(s_1))\\
   &\ge \mathcal{A}(s_1,\dots,s_m)
\end{align*}

The first inequality follows trivially from the definition. The second inequality follows from Lemma \ref{lem:randomcostapprox}. The third equality follows from the fact that $\mu_r$ is monotone and the last inequality follows from Lemma \ref{lem:window2}.

\item Given a $(p,B)$-pseudorandom string $s_1$ and $m-1$ adversarial strings $s_2,\dots,s_m$ and an optimal alignment $\sigma$, if $\mathcal{J}_1$ is the verified partition then by Claim \ref{claim:verpartition}, for every window $w\in \mathcal{J}_1$ if $w$ is uniquely certifiable then $w\in \mathcal{W}_1^s$ otherwise if $w$ is large-cost then $w\in \mathcal{W}_1^{large}$.

By Lemma \ref{lem:searchspace}, for every window $w_i\in \mathcal{J}_1$, $\mathcal{\tilde{J}}_j^{|w_i|}\subseteq \mathcal{\tilde{W}}_j^{|w_i|}$ and for every tuple $(w_1,\dots,w_m)\in \mathcal{J}_1\times \mathcal{\tilde{W}}_2^{|w_1|}\times \dots \times \mathcal{\tilde{W}}_m^{|w_1|}$ Algorithm \ref{alg:windowestimate} outputs a certified tuple. 
Moreover by Lemma \ref{claim:cover}
there exists a restricted monotone mapping $\mu_r: \mathcal{J}_1\rightarrow \cup_{w_i\in \mathcal{J}_1}\mathcal{\tilde{W}}_2^{|w_i|}\cup\{\perp\}\times \dots \times \cup_{w_i\in \mathcal{J}_1}\mathcal{\tilde{W}}_m^{|w_i|}\cup\{\perp\}$ satisfying $\mathcal{A}(s_1,\dots,s_m)\le \mathcal{A}(s_1,\mu_r(s_1))\le (1+12\epsilon)\mathcal{A}(s_1,\dots,s_m)$. 
Let $\bar{S}\subseteq S$ where, $\bar{S}=\{(\bar{w}_1^1,\dots,\bar{w}_m^1,c^1),\dots,(\bar{w}_1^{k'},\dots,\bar{w}_m^{k'},c^{k'})\}$ be such that $\mathcal{J}_1=\{\bar{w}_1^1,\dots,\bar{w}_1^{k'}\}$ and for each $j\in[k']$, $\mu_r(\bar{w}_1^{j})=(\bar{w}_2^j,\dots,\bar{w}_m^j)$.
Next we upper bound the cost of the tuples in $\bar{S}$.

Let $\mathcal{M}_1\subseteq \mathcal{J}_1$ be the set of uniquely certifiable windows. Each window $w$ in $\mathcal{J}_1\setminus \mathcal{M}_1$ is large cost. Let $\mathcal{M}_2 \subseteq \mathcal{J}_1\setminus \mathcal{M}_1$ be the set of trivially approximable windows, $\mathcal{M}_3 \subseteq \mathcal{J}_1\setminus \mathcal{M}_1$ be the set of large-cost certifiable windows and $\mathcal{M}_4 \subseteq \mathcal{J}_1\setminus \mathcal{M}_1$ be the set of large-cost uncertifiable windows. Note $\cup_{j\in[4]} \mathcal{M}_j=\mathcal{J}_1$ and $\mathcal{M}_j$s are pairwise disjoint. 
Note as $\mu_r$ is a monotone map, $\mathcal{A}_{\mathcal{E}}(s_1,\dots,s_m)= \sum_{j\in[4]}\sum_{w \in \mathcal{M}_j}\mathcal{E}(w,\mu_r(w)).$

Assume for each $(\bar{w}_1^j,\dots,\bar{w}_m^j,c^j)$, $c_j=\mathcal{E}(\bar{w}_1^j,\dots,\bar{w}_m^j)$.
By Lemma \ref{lem:randomcostapprox},
\begin{enumerate}
\item $\sum_{w \in \mathcal{M}_1}\mathcal{E}(w,\mu_r(w))\le (1+\epsilon)\sum_{w \in \mathcal{M}_1}\mathcal{A}(w,\mu_r(w))$.
\item  $\sum_{w \in \mathcal{M}_2}\mathcal{E}(w,\mu_r(w))\le \sum_{w \in \mathcal{M}_2}\mathcal{A}(w,\mu_r(w))$.
\item  $\sum_{w \in \mathcal{M}_3}\mathcal{E}(w,\mu_r(w))\le (2-\frac{3p}{512}+\epsilon)\sum_{w \in \mathcal{M}_3}\mathcal{A}(w,\mu_r(w))$. 
\end{enumerate}

Let $\ell=\sum_{w \in \mathcal{M}_4}|w|$. Then, $\sum_{w \in \mathcal{M}_4}\mathcal{E}(w,\mu_r(w))\le 2\ell$. By Lemma \ref{claim:triapprox}, as $\mathcal{A}(s_1,\dots,s_m)\ge \frac{\ell}{2\epsilon}$, $\sum_{w \in \mathcal{M}_4}\mathcal{E}(w,\mu_r(w))\le 2\ell\le 4\epsilon \mathcal{A}(s_1,\dots,s_m)$.Therefore, 
$$\sum_{w\in\mathcal{J}_1}\mathcal{E}(w,\mu_r(w))\le (2-\frac{3p}{512}+\epsilon)\sum_{w \in \mathcal{J}_1}\mathcal{A}(w,\mu_r(w))+4\epsilon \mathcal{A}(s_1,\dots,s_m)$$

As $\mathcal{A}(s_1,\dots,s_m)\le \sum_{w \in \mathcal{J}_1}\mathcal{A}(w,\mu_r(w))$,

$$\sum_{w\in\mathcal{J}_1}\mathcal{E}(w,\mu_r(w))\le (2-\frac{3p}{512}+5\epsilon)\sum_{w \in \mathcal{J}_1}\mathcal{A}(w,\mu_r(w))$$
 As $\mu_r$ is a monotone map we can define,
\begin{align*}
   \mathcal{A}_{\mathcal{E}}(s_1,\mu_r(s_1)) & = \sum_{w\in \mathcal{J}_1}\mathcal{E}(w,\mu_r(w))\\
   &\le (2-\frac{3p}{512}+5\epsilon)\sum_{w \in \mathcal{J}_1}\mathcal{A}(w,\mu_r(w))\\
   &= (2-\frac{3p}{512}+5\epsilon)\mathcal{A}(s_1,\mu_r(s_1))\\
   &\le (1+12\epsilon)(2-\frac{3p}{512}+5\epsilon)\mathcal{A}(s_1,\dots,s_m)\\
   &\le (2-\frac{3p}{512}+89\epsilon)\mathcal{A}(s_1,\dots,s_m)
\end{align*}

\item Next we analyze the running time of Algorithm \ref{alg:windowestimate}. We start with Algorithm \ref{alg:uniquematch}. The algorithm starts by creating $\mathcal{W}_1^s$, the set of disjoint windows of size $\beta$ of string $s_1$ in time $O(n)$. Next for each $w_i\in \mathcal{W}_1^s$ and $j\in [2,m]$, the algorithm initializes set $\mathcal{W}_{i,j}^s$ in time $O(nm)$. After this the algorithm creates the set of window $\mathcal{\bar{W}}_{j,\theta}^s$ and $\mathcal{\bar{W}}^s_j$ for each $\theta\in \{\frac{p}{4},\dots,\frac{1}{\beta}\}$ and $j\in[2,m]$ in time $\tilde{O}(\frac{nm\log n}{\theta \beta})=\tilde{O}(nm)$. Note $|\mathcal{\bar{W}}^s_j|=\tilde{O}(\frac{n}{\theta \beta})$ and $|\mathcal{W}_1^s|=\frac{n}{\beta}$.

Next for each $w_i\in \mathcal{W}_1^s$, the algorithm creates set $\mathcal{W}_{i,j}^s$ by detecting all window of $\mathcal{\bar{W}}^s_j$ which are at distance at most $\frac{p\beta}{4}$. Each computation of $\mathcal{A}(w_i,w_j)$ can be done using dynamic program in time $O(|w_i|^2)=O(\beta^2)$. Hence the total time taken to construct all $\mathcal{W}_{i,j}^s$ is $\tilde{O}(\frac{n}{\beta}\times \frac{nm}{\theta \beta}\times \beta^2)=\tilde{O}(\frac{n^2m}{\theta})=\tilde{O}(n^3m)$ as $\theta \ge \frac{1}{\beta}\ge \frac{1}{n}$. 

After this the algorithm computes the set of disjoint windows $\mathcal{\tilde{W}}_{i,j}^s$ for each $\mathcal{W}_{i,j}^s$. This can be done using a greedy algorithm in time $O(|\mathcal{W}_{i,j}^s|^2)$. Note here given any two windows $w,w'$ each of size $\beta$, we can compute $w\cap w'$, in time $O(1)$ just by checking $s(w), e(w), s(w'), e(w')$. As there are $\frac{nm}{\beta}$ many different $\mathcal{W}_{i,j}^s$ and $|\mathcal{W}_{i,j}^s|\le |\mathcal{\bar{W}}_j^s|=\tilde{O}(\frac{n}{\theta \beta})$, total time taken to compute all $\mathcal{\tilde{W}}_{i,j}^s$ is $O(\frac{nm}{\beta}|\mathcal{W}_{i,j}^s|^2)=\tilde{O}(\frac{n^3m}{\theta^2\beta^3})=\tilde{O}(\frac{n^3m}{\beta})=\tilde{O}(n^{2.5}m)$ as $\theta \ge \frac{1}{\beta}$ and $\beta\ge \sqrt{n}$.

Next for each $w_i\in \mathcal{W}_1^s$ and each $j\in[2,m]$ and each $\theta\in\{\frac{p}{4},\frac{p}{4(1+\epsilon)},\dots,\frac{1}{\beta}\}$, the algorithm sets $\mathcal{W}_{i,j,\theta}^s=\phi$ if $\sum_{j\in[2,m]}|\mathcal{\tilde{W}}_{i,j}^s|\ge \frac{16m}{p\epsilon}$. Hence otherwise we can assume $\sum_{j\in[2,m]}|\mathcal{\tilde{W}}_{i,j}^s|< \frac{16m}{p\epsilon}$ and therefore $\sum_{j\in[2,m]}|\mathcal{W}_{i,j,\theta}^s|\le \frac{48m\log n}{p\epsilon^2\theta}$. 
But this implies $\Pi_{j\in[2,m]}|\mathcal{W}_{i,j,\theta}^s|\le (\frac{48m\log n}{p\epsilon^2\theta(m-1)})^{m-1}<(\frac{96\log n}{p\epsilon^2\theta})^m$. Next for each tuple 
$(w_i,w_2,\dots,w_m)\in w_i\times\mathcal{W}_{i,2,\theta}^s\times \dots \times \mathcal{W}_{i,m,\theta}^s$ the algorithm checks if
$\mathcal{A}(w_i,w_2,\dots,w_m)\le \theta|w_i|$ in time 
$\tilde{O}(m^2\theta^m|w_i|^m)=\tilde{O}(m^2\theta^m\beta^m)$ (by Section \ref{sec:linear}). As there are $O(\log n)$ different choices for $\theta$, time required for each $w_i$ is $\tilde{O}((\frac{96\log n}{p\epsilon^2\theta})^m\times(m^2\theta^m\beta^m))=\tilde{O}(\frac{96^mm^2\beta^m\log^m n}{p^m\epsilon^{2m}})$. Hence time required for all $w_i\in \mathcal{W}_1^s$ is $\tilde{O}(\frac{nm^296^m\beta^{m-1}\log^m n }{p^m\epsilon^{2m}})$. 

Hence the total time taken by Algorithm \ref{alg:uniquematch} is $\tilde{O}(\frac{nm^296^m\beta^{m-1}\log^m n }{p^m\epsilon^{2m}}+n^3m)$.

Next we analyze the running time of Algorithm \ref{alg:largecostmatch}. The algorithm starts by creating a set of windows $\mathcal{W}_1^\ell$ for each $\ell\in[\frac{n}{\beta}].$ Total time required for this is $O(\frac{n^2}{\beta^2})=O(n)$ as $\beta\ge \sqrt{n}$. Also the total number of windows in each $\mathcal{W}_1^\ell$ is at most $\frac{n}{\beta}$ and the number of windows in $\mathcal{W}_1^{large}$ is $O(\frac{n^2}{\beta^2})$. 

For each window $w_i\in \mathcal{W}_1^{large}$ (where, $|w_i|=\ell \beta$), if the corresponding $\mathcal{W}_{p,j}^s=\phi$ or $\mathcal{W}_{q,j}^s=\phi$ or $\sum_{j\in[2,m]}|\mathcal{\tilde{W}}_{p,j}^s|\ge \frac{16m}{p\epsilon}$ or $\sum_{j\in[2,m]}|\mathcal{\tilde{W}}_{q,j}^s|\ge \frac{16m}{p\epsilon}$ discard $w_i$. This checking can be done in time $O(\frac{n^2m}{\beta^2})=O(nm)$. Note the sets $\mathcal{W}_{p,j}^s,\mathcal{W}_{q,j}^s, \mathcal{\tilde{W}}_{p,j}^s, \mathcal{\tilde{W}}_{q,j}^s$ are already created in Algorithm \ref{alg:uniquematch}. 

Otherwise consider the set $T=\{(\bar{w}_2,\bar{\bar{w}}_2,\dots,\bar{w}_m,\bar{\bar{w}}_m);(\bar{w}_2,\bar{\bar{w}}_2,\dots,\bar{w}_m,\bar{\bar{w}}_m)\in \mathcal{\tilde{W}}_{p,2}^s\times\mathcal{\tilde{W}}_{q,2}^s\times \dots\times\mathcal{\tilde{W}}_{p,m}^s\times\mathcal{\tilde{W}}_{q,m}^s \}$. Note as $\sum_{j\in[2,m]}|\mathcal{\tilde{W}}_{p,j}^s|< \frac{16m}{p\epsilon}$, $\Pi_{j\in[2,m]}|\mathcal{\tilde{W}}_{p,j}^s|\le  (\frac{16m}{p\epsilon(m-1)})^{m-1}<(\frac{32}{p\epsilon})^m$. Similarly as $\sum_{j\in[2,m]}|\mathcal{\tilde{W}}_{q,j}^s|< \frac{16m}{p\epsilon}$, $\Pi_{j\in[2,m]}|\mathcal{\tilde{W}}_{q,j}^s|< (\frac{32}{p\epsilon})^m$. Hence size of set $T$ is at most $(\frac{32}{p\epsilon})^{2m}$. 

Next for every tuple of set $T$ we do the following. For every string $s_j$, define substring $s_j^{\bar{w},\bar{\bar{w}}}$ in $O(1)$ time. If the sum of the length of these substrings over all $s_j$s is $\ge 5\ell\beta m$, create set of windows $\mathcal{W}_{i,j}^{\bar{w},\bar{\bar{w}}}$ of substring $s_j^{\bar{w},\bar{\bar{w}}}$ by setting $\theta=1$. As $|s_j^{\bar{w},\bar{\bar{w}}}|\le n$, $|\mathcal{W}_{i,j}^{\bar{w},\bar{\bar{w}}}|\le \frac{2n\log n}{\epsilon \ell \beta}\le \frac{2n\log n}{\epsilon\beta}$. Next for each tuple $(w_2,\dots,w_m)\in \mathcal{W}_{i,2}^{\bar{w},\bar{\bar{w}}}\times \dots \times \mathcal{W}_{i,m}^{\bar{w},\bar{\bar{w}}}$, the algorithm outputs a certified tuple $(w_i,w_2,\dots,w_m,|w_i|)$. As $\Pi_{j\in[2,m]}|\mathcal{W}_{i,j}^{\bar{w},\bar{\bar{w}}}|\le \tilde{O}(\frac{n\log n}{\epsilon \beta})^m=\tilde{O}(\frac{n^{m/2}\log^m n}{\epsilon^m})$ (as $\beta \ge \sqrt{n}$), total time required is $\tilde{O}(\frac{n^{m/2}\log^m n}{\epsilon^m})$.

Otherwise if $\sum_{j\in[2,m]}s(\bar{\bar{w}}_j)-e(\bar{w}_j)<5\ell\beta m$, the algorithm creates set of windows $\mathcal{W}_{i,j}^{\bar{w},\bar{\bar{w}}}=\cup_{\theta\in\{1,\dots,\frac{p}{16}\}}\mathcal{P}(s_j^{\bar{w},\bar{\bar{w}}},\ell\beta,\theta)$. Here total number of windows in all $\mathcal{W}_{i,j}^{\bar{w},\bar{\bar{w}}}$ is $\sum_{j\in[2,m]}|\mathcal{W}_{i,j}^{\bar{w},\bar{\bar{w}}}|\le \frac{250m}{p\epsilon}$. Hence, $\Pi_{j\in[2,m]}|\mathcal{W}_{i,j}^{\bar{w},\bar{\bar{w}}}|\le (\frac{500}{p\epsilon})^m$. Next for each tuple $(w_2,\dots,w_m)\in\Pi_{j\in[2,m]}\mathcal{W}_{i,j}^{\bar{w},\bar{\bar{w}}}$, call algorithm $LargeAlign(w_i,w_2,\dots,w_m,\frac{p}{32})$. As for any $\ell$, $\ell\beta\le n$, time required for each call is $\tilde{O}(2^{\lfloor m/2\rfloor +1}mn^{\lfloor m/2\rfloor+2})$. 
Hence total time required $\frac{1000^m}{p^{m}\epsilon^m}mn^{\lfloor m/2\rfloor+2}$. For all $w_i\in \mathcal{W}_1^{large}$ and all choices of $(\bar{w}_2,\bar{\bar{w}}_2,\dots,\bar{w}_m,\bar{\bar{w}}_m)\in \mathcal{\tilde{W}}_{p,2}^s\times \mathcal{\tilde{W}}_{q,2}^s\times \dots \times \mathcal{\tilde{W}}_{p,m}^s\times \mathcal{\tilde{W}}_{q,m}^s$ time taken $\tilde{O}(\frac{n^2}{\beta^2}(\frac{32}{p\epsilon})^{2m}\frac{1000^m}{p^{m}\epsilon^m}mn^{\lfloor m/2\rfloor+2})=\tilde{O}(\frac{1012^{2m}mn^{ \lfloor m/2\rfloor+3}}{p^{3m}\epsilon^{3m}})$. 
Hence total time taken by Algorithm \ref{alg:largecostmatch} is $\tilde{O}(\frac{1012^{2m}mn^{ \lfloor m/2\rfloor+3}}{p^{3m}\epsilon^{3m}}+\frac{n^{m/2\log^m n}}{\epsilon^m})=\tilde{O}(\frac{1012^{2m}\log^m nmn^{ \lfloor m/2\rfloor+3}}{p^{3m}\epsilon^{3m}})$. 

We can conclude that Algorithm \ref{alg:windowestimate} runs in time $$\tilde{O}(\frac{nm^296^m\beta^{m-1}\log^m n }{p^m\epsilon^{2m}}+n^3m+\frac{1012^{2m}\log^m nmn^{ \lfloor m/2\rfloor+3}}{p^{3m}\epsilon^{3m}})$$

Hiding the exponential term over the constants and $\log$ factors we claim that the running time of Algorithm \ref{alg:windowestimate} is 
$$\tilde{O}_m(nm^2\beta^{m-1}+n^3m+ mn^{ \lfloor m/2\rfloor+3})=\tilde{O}_m(n\beta^{m-1}+ n^{ \lfloor m/2\rfloor+3}+mn^3)=\tilde{O}_m(n\beta^{m-1}+ n^{ \lfloor m/2\rfloor+3}).$$

\item Next we bound $|S|$. Note every window has size at least $\beta$. Hence $|\mathcal{W}_1^s|=\frac{n}{\beta}$. Also for each $\theta$ and $j$, if $\mathcal{W}^s_{i,j,\theta}\neq \phi$, then 
$\sum_{j\in[2,m]}|\mathcal{\tilde{W}}_{i,j}^s|< \frac{16m}{p\beta}$ and therefore $\sum_{j\in[2,m]}|\mathcal{W}_{i,j,\theta}^s|\le \frac{48m\log n}{p\epsilon^2\theta}$. But this implies 
$\Pi_{j\in[2,m]}|\mathcal{W}_{i,j,\theta}^s|\le (\frac{48m\log n}{p\epsilon^2\theta(m-1)})^{m-1}<(\frac{96\log n}{p\epsilon^2\theta})^m=(\frac{96\beta}{p\epsilon^2})^m$ as $\theta \ge \frac{1}{\beta}$. 
As there are $O(\log n)$ different choices for $\theta$, 
for each $w_i\in \mathcal{W}_1^s$, number of certified tuples added to $S_1$ is $\tilde{O}((\frac{96\beta\log n }{p\epsilon^2})^m)$. Therefore, $|S_1|= \tilde{O}(\frac{n96^m\log^m n \beta^{m-1}}{p^m\epsilon^{2m}})$.

Next for each $\ell\in [\frac{n}{\beta}]$, $|\mathcal{W}_1^\ell|\le \frac{n}{\beta}$. In Algorithm \ref{alg:windowestimate} for each $w_i\in \mathcal{W}_1^\ell$, $\sum_{j\in[2,m]}|\mathcal{\tilde{W}}^s_{p,j}|\le \frac{16m}{p\epsilon}$. Hence $\Pi_{j\in[2,m]}|\mathcal{\tilde{W}}^s_{p,j}|\le (\frac{16m}{p\epsilon(m-1)})^{m-1}<(\frac{32}{p\epsilon})^m$. Similarly, $\Pi_{j\in[2,m]}|\mathcal{\tilde{W}}^s_{q,j}|<(\frac{32}{p\epsilon})^m$. Hence size of set 
$T=\{(\bar{w}_2,\bar{\bar{w}}_2,\dots,\bar{w}_m,\bar{\bar{w}}_m);(\bar{w}_2,\bar{\bar{w}}_2,\dots,\bar{w}_m,\bar{\bar{w}}_m)\in\mathcal{\tilde{W}}^s_{p,2}\times  \mathcal{\tilde{W}}^s_{q,2}\times \dots\times  \mathcal{\tilde{W}}^s_{p,m}\times  \mathcal{\tilde{W}}^s_{p,m}\}$ is at most $(\frac{32}{p\epsilon})^{2m}$. 
Next for each tuple in $T$, if $\sum_{j\in[2,m]}s(\bar{\bar{w}}_j)-e(\bar{w}_j)\ge 5\ell\beta m$, 
$\Pi_{j\in[2,m]}|\mathcal{W}_{i,j}^{\bar{w},\bar{\bar{w}}}|= \tilde{O}((\frac{n\log n}{\epsilon \beta})^m)=\tilde{O}(\frac{n^{m/2}\log^m n}{\epsilon^m})$. Otherwise, $\Pi_{j\in[2,m]}|\mathcal{W}_{i,j}^{\bar{w},\bar{\bar{w}}}|= \tilde{O}((\frac{500}{p\epsilon})^m)$. 
 Hence total number of certified tuple in $S_2$ is $\tilde{O}(\frac{n^2}{\beta^2}(\frac{32}{p\epsilon})^{2m}(\frac{n^{m/2}\log^m n}{\epsilon^m}+\frac{500^m}{p^m\epsilon^m}))=\tilde{O}(\frac{716^{2m}\log^m nn^{m/2+1}}{\epsilon^{4m}p^{3m}})$.

Therefore, $|S|=\tilde{O}(\frac{n96^m\log^m n \beta^{m-1}}{p^m\epsilon^{2m}}+\frac{716^{2m}\log^m nn^{m/2+1}}{\epsilon^{4m}p^{3m}}).$ 

\end{itemize}

\end{proof}

\begin{proof}{Proof of Theorem \ref{thm:align2}}
Given a $(p,B)$ pseudorandom string $s_1$ and $m-1$ adversarial strings $s_2,\dots,s_m$, to approximate $\mathcal{A}(s_1,\dots,s_m)$ first run Algorithm \ref{alg:windowestimate} that outputs a set $S=\{(w_1^1,\dots,w_m^1,c^1),\dots,(w_1^k,\dots,w_m^k,c^k)\}$ of $(m+1)-$tuples where each $(w_1^j,\dots,w_m^j,c^j)$ is a certified $m$-tuple such that There exists a a subset $\bar{S}\subseteq S$ where, $\bar{S}=\{(\bar{w}_1^1,\dots,\bar{w}_m^1,c^1),\dots,(\bar{w}_1^{k'},\dots,\bar{w}_m^{k'},c^{k'})\}$ such that 
$\{\bar{w}_1^1,\dots,\bar{w}_1^{k'}\}$ is a verified partition of $s_1$ and there exists a restricted monotone map  $\mu_r:\{\bar{w}_1^1,\dots,\bar{w}_1^{k'}\}\rightarrow \{\bar{w}_2^1,\dots,\bar{w}_2^{k'}\} \times \dots \times \{\bar{w}_m^1,\dots,\bar{w}_m^{k'}\}$ satisfying $ \mathcal{A}_{\mathcal{E}}(s_1,\mu_r(s_1))\le (2-\frac{3p}{512}+89\epsilon)\mathcal{A}(s_1,\dots,s_m)$.
Given set $S$ we use dynamic program to compute $\mathcal{A}_{\mathcal{E}}(s_1,\mu_r(s_1))$.
Let $S_j$ denotes the set of windows from string $s_j$ for which $S$ contains a certified $m$-tuple. For a given window $w_j\in S_j$, $e(w_j)$ represents the index of the last character of $w_j$. Let $e(S_j)$ be the set of indices of the last characters of all windows in $S_j$. For $i\in e(S_j)$ define $pre(i)=max\{i'\in e(S_j)\cup\{0\}; i'<i\}$. 
 For each $(i_1,\dots,i_m)\in e(S_1)\times \dots \times e(S_m)$ we compute $\mathcal{A}_\mathcal{E}(i_1,\dots,i_m)$ that represents the minimum cost of an alignment ending at $(i_1,\dots,i_m)$ under cost estimation $\mathcal{E}$. We can define $\mathcal{A}_\mathcal{E}(i_1,\dots,i_m)$ to be the minimum among the followings.
\begin{enumerate}
    \item For each subset $Q\subseteq [m]\setminus \{\phi\}$, sum of the cost of deleting some characters of each string $s_j$, for $j\in Q$ and deleting the last window of $s_1$ if $1\in Q$: $\sum_{j\in Q} i_j- pre(i_j)+\mathcal{A}_\mathcal{E}(i'_1,\dots,i'_m)$ where $i'_j=pre(i_j)$ if $j\in Q$ and $i'_j=i_j$ otherwise.
    \item Cost of matching the last $m$ windows: $min_{(w_1,\dots,w_m,c)\in S; \forall j\in[m], e(w_j)=i_j} \mathcal{E}(w_1,\dots,w_m)+\mathcal{A}_\mathcal{E}(s(w_1)-1,\dots,s(w_m)-1)$
\end{enumerate}

Output $\mathcal{A}_\mathcal{E}(n,\dots,n)$. As number of windows ending at an index is $\tilde{O}(1)$, the running time of the dynamic program is $$\tilde{O}(2^m|S|)=\tilde{O}(2^m(\frac{n96^m\log^m n \beta^{m-1}}{p^m\epsilon^{2m}}+\frac{716^{2m}\log^m nn^{m/2+1}}{\epsilon^{4m}p^{3m}}))=\tilde{O}_m(n\beta^{m-1}+n^{m/2+1})$$

Moreover by Theorem \ref{thm:mainrandom} the running time of Algorithm \ref{alg:windowestimate} is $\tilde{O}_m(n\beta^{m-1}+ n^{ \lfloor m/2\rfloor+3})$. Hence we design an algorithm that computes $(2-\frac{3p}{512}+89\epsilon)$ approximation of $\mathcal{A}(s_1,\dots,s_m)$ in time $\tilde{O}_m(n\beta^{m-1}+ mn^{ \lfloor m/2\rfloor+3})$.
\end{proof}

\section{$\frac{\lambda}{2+\epsilon}$-approximation for Multi-sequence LCS}
\label{sec:multilcs}

In this section we provide an algorithm that given $m$ strings $s_1,\dots,s_m$ each of length $n$, such that $\mathcal{L}(s_1,\dots,s_m)=\lambda n$, where $\lambda\in(0,1)$ computes an $\frac{\lambda}{2+\epsilon}$ approximation of $\mathcal{L}(s_1,\dots,s_m)$. The algorithm is nearly identical to the algorithm described in Section~\ref{thm:align1gap}, and in fact slightly simpler which helps us to improve the running time further. In particular, we get the following theorem.

\begin{theorem*}[\ref{thm:lcs}]
For any constant $\epsilon>0$, given $m$ strings $s_1,\dots,s_m$ of length $n$ over some alphabet set $\Sigma$ such that $\mathcal{L}(s_1,\dots,s_m)=\lambda n$, where $\lambda\in(0,1)$, there exists an algorithm that computes an $\frac{\lambda}{2+\epsilon}$ approximation of $\mathcal{L}(s_1,\dots,s_m)$ in time $\tilde{O}_m(n^{\lfloor m/2\rfloor +1}+mn^2)$.
\end{theorem*}

Since we do not have any prior knowledge of $\lambda$, we solve a gap version: given $m$ strings $s_1,\dots,s_m$ of length $n$ over some alphabet set $\Sigma$, $\lambda\in (0,1)$ and a constant $c> 1$, the objective is to decide whether $\mathcal{L}(s_1,\dots,s_m)\ge \lambda n$ or $\mathcal{L}(s_1,\dots,s_m)< \frac{\lambda^2 n}{c}$. More specifically if $\mathcal{L}(s_1,\dots,s_m)\ge \lambda n$ we output $1$, else if $\mathcal{L}(s_1,\dots,s_m)< \frac{\lambda^2 n}{c}$ we output $0$ otherwise output any arbitrary answer. We design an algorithm that decides this gap version for $c=2$. This immediately implies an $(\frac{\lambda}{2+\epsilon})$ approximation of $LCS(s_1,s_2,..,s_m)$ following a similar logic of going from Theorem~\ref{thm:align1gap} to Theorem~\ref{thm:align1}. We now prove Theorem~\ref{thm:lcs}.

\subsection{Algorithm for $\frac{\lambda}{2+\epsilon}$-approximation of $LCS(s_1,\dots,s_m)$}

We partition the input strings into two groups $G_1$ and $G_2$ where $G_1$ contains the strings $s_1,\dots,s_{\lceil m/2\rceil}$ and $G_2$ contains the strings $s_{\lceil m/2\rceil+1},\dots,s_m$. 
Next compute a longest common subsequence $L_1$ of $G_1$ with $|L_1|\ge \lambda n$. Remove all aligned characters of $s_1$ in $L_1$. We represent the modified $s_1$ by $s_1^{[n]\setminus L_1(s_1)}$: string $s_1$ restricted to the characters with indices in $[n]\setminus L_1(s_1)$. Compute an LCS $L_2$ of $s_1^{[n]\setminus L_1(s_1)}, s_2,\dots,s_{\lceil m/2 \rceil}$. At $i$th step given $i-1$ common subsequences $L_1,\dots,L_{i-1}$, we compute an LCS $L_i$ of $s_1^{[n]\setminus \cup_{j\in[i-1]}L_j(s_1)}, s_2,\dots,s_{\lceil m/2 \rceil}$. 
We continue this process until it reaches a round $k$ such that 
$|L_k|<\lambda n-\frac{\lambda^2n(k-1)}{2}$. 
Let $L_1,\dots,L_k$ be the sequences generated. 

Next for each $L_i$ returned, we compute a longest common subsequence $L'_i$ of $L_i, s_{\lceil m/2\rceil+1},\dots,s_m$. If there exists an $i\in[k]$ such that $|L'_i|\ge \frac{\lambda^2n}{2}$, output $L'_i$.

\begin{lemma}
\label{lem:enumlcs}
Given strings $s_1,\dots,s_{|G_1|}$ of length $n$ and a parameter $\lambda n$ as input, where $\lambda\in (0,1]$, there exists a set of $k \leq 2/\lambda$ different common subsequences $L_1,\dots,L_k$ of $s_1,\dots,s_{|G_1|}$ each of length at least $\lambda n-\lambda^2n(k-1)/2$ such that for any common subsequence $\sigma$ of $s_1,\dots,s_m$ of length at least $\lambda n$ there exists a $L_i$, where $|\sigma(s_1)\cap L_i(s_1)|\ge \frac{\lambda^2 n}{2}$. The running time of the algorithm is $\tilde{O}_m(n^{|G_1|}+mn^2)$.
\end{lemma}

\begin{proof}

Given $k$ subsequences $L_1, L_2, \dots, L_k$ of $s_1$ such that for each $i \in [k]$, $|L_i|\ge \lambda n-\frac{n\lambda^2(k-1)}{4}$, and  $L_1(s_1),L_2(s_1),\dots$ are disjoint,
we have 
\begin{align*}
  |\cup_{i\in[k]}L_i(s_1)|
  &\ge \lambda n+(\lambda n-\frac{\lambda^2n}{2})+\dots+(\lambda n-\frac{(k-1)\lambda^2 n}{2})
  \\
  &=k\lambda n-\frac{\lambda^2n}{2}(1+2+\dots+(k-1))
 \\ 
  &> k\lambda n-\frac{k^2\lambda^2 n}{4}
\end{align*}

Substituting $k=2/\lambda$ we get $|\cup_{i\in[k]}L_i(s_1)|>n$. 
Now if we compute $L_1,\dots, L_j$ and $j <2/\lambda$, then we know for each common subsequence $\sigma$ of $s_1,\dots,s_m$ with length at least $\lambda n$  if $\sigma\notin \{L_1,\dots,L_{j-1}\}$, then
$|\cup_{i\in [j-1]}(\sigma(s_1)\cap L_j(s_1))|\ge \frac{(j-1)\lambda^2n}{2}$. Hence there exists at least one $i\in [j-1]$ such that $|L_i(s_1)\cap \sigma(s_1)|\ge \frac{\lambda^2n}{2}$. Otherwise if the algorithm runs for $2/\lambda$ rounds then $\cup_{j\in[2/\lambda]}L_j(s_1)=[n]$. Then for any common subsequence $\sigma$ of $s_1,\dots,s_m$ with length at least $\lambda n$, $\sigma$ will have an intersection at least $\frac{\lambda^2n}{2}$ with at least one $L_i$. 

As $|k|\le 2/\lambda$ the algorithm runs for at most $2/\lambda$ rounds where at each round it computes the LCS of $G_1$ strings such that $\sum_{j} |L_j| \leq n$ where $L_j$s are pairwise disjoint. This can be performed using Theorem \ref{thm:hsmulti} in $\tilde{O}_m(n^{|G_1|}+mn\lambda)=\tilde{O}_m(n^{|G_1|}+mn^2)$ time. 
\end{proof}
By setting $|G_1|=\lceil m/2 \rceil\le \lfloor \frac{m}{2}\rfloor +1$, we get a running time of $\tilde{O}_m(n^{\lfloor m/2 \rfloor +1}+mn^2)$.
Now by Lemma~\ref{lem:enumlcs} if $\sigma$ is an LCS of $s_1, s_2, .., s_m$, then there exists a $L_i$ such that $|L_i(s_1) \cap \sigma(s_1)| \ge \frac{\lambda^2 n}{2}$. Thus, when we compute the LCS of $L_i(s_1), s_{\lceil m/2 \rceil +1},..,s_m$, we are guaranteed to return a common subsequence of $s_1, s_2,...,s_m$ of length at least $\frac{\lambda^2 n}{2}$. Hence, taking the right choice of $\lambda$ following the gap version we get the claimed approximation bound. Using Theorem \ref{thm:hsmulti}, the running time to compute a common subsequence of $L_j(s_1), s_{\lceil m/2 \rceil +1},..,s_m$ for all $j$ is $\tilde{O}_m(n^{\lfloor m/2 \rfloor +1}+mn^2)$.
This completes the proof of Theorem~\ref{thm:lcs}.\qed

\bibliographystyle{alpha}
\bibliography{ref}

\newpage
\appendix

\section{An $\tilde{O}(m^2k^m)$ algorithm for Alignment Distance of Multiple Sequences with $\tilde{O}(mn)$ Preprocessing}
\label{sec:linear}

We first recall an algorithm developed in~\cite{Ukkonen85,LMS98,LV88,Myers86} that computes edit distance between two strings in $O(n+k^2)$ time.

\subsection*{Warm-up: An $O(n+k^2)$ algorithm for Edit Distance.} The well-known dynamic programming algorithm computes an $(n+1) \times (n+1)$ edit-distance matrix $D[0...n][0...n]$ where entry $D[i,j]$ is the edit distance, $\ED(A^i,B^j)$ between the prefixes $A[1,i]$ and $B[1,j]$ of $A$ and $B$, where $A[1,i]=a_1a_2...a_i$ and $B[1,j]=b_1b_2...b_j$. The following is well-known and easy to verify coupled with the boundary condition $D[i,0] = D[0,i] = i$ for all $i \in [0,n]$. 

For all $i, j \in [0,n]$
\[ D[i,j] = \min \left\{ \begin{array}{ll}
         D[i-1,j]+1 & \mbox{if $i > 0$};\\
        D[i,j-1]+1 & \mbox{if $j > 0$};\\
        D[i-1,j-1]+1 (a_i\neq b_j) & \mbox{if $i,j > 0$}.
        \end{array} \right. \]

The computation cost for this dynamic programming is $O(n^2)$. To obtain a significant cost saving when $\ED(A,B) \leq k <<n$, the $O(n+k^2)$ algorithm works as follows. It computes the entries of $D$ in a greedy order, computing first the entries with value 0, $1,2,...k$ respectively. Let diagonal $d$ of matrix $D$, denotes all $D[i,j]$ such that $j=i+d$. Therefore, the entries with values in $[0,k]$ are located within diagonals $[-k,k]$.  Now since the entries in each diagonal of $D$ are non-decreasing, it is enough to identify for every $d \in [-k,k]$, and for all $h \in [0,k]$, the last entry of diagonal $d$ with value $h$. The rest of the entries can be inferred automatically. Hence, we are overall interested in identifying at most $(2k+1)k$ such points. 
The $O(n+k^2)$ algorithm shows how building a suffix tree over a combined string $A\$B$ (where $\$$ is a special symbol not in $\Sigma$) helps identify each of these points in $O(1)$ time, thus achieving the desired time complexity.

Let $L^h(d)=\max\{i : D[i,i+d]=h\}$. The $h$-wave is defined by $L^h=\langle L^h(-k),...,L^h(k) \rangle$. Therefore, the algorithm computes $L^h$ for $h=0,..k$ in the increasing order of $h$ until a wave $e$ is computed such that $L^e(0)=n$ (in that case $\ED(A,B) = e$), or the wave $L^k$ is computed in the case the algorithm is thresholded by $k$. Given $L^{h-1}$, we can compute $L^h$ as follows. 

Define $$Equal(i,d)=\max_{q \geq i}{(q \mid A[i,q]=B[i+d,q])}$$
Then, $L^0(0)=Equal(0,0)$ and 
\[ L^h(d) = \max \left\{ \begin{array}{ll}
         Equal(L^{h-1}(d)+1,d) & \mbox{if $h-1 \geq 0$};\\
        Equal(L^{h-1}(d-1),d) & \mbox{if $d-1 \geq -k , h-1 \geq 0$};\\
        Equal(L^{h-1}(d+1)+1,d) & \mbox{if $d+1, h+1 \leq k$}.
        \end{array} \right. \] 

Using a suffix tree of the combined string $A\$B$, any $Equal(i,d)$ query can be answered in $O(1)$ time, and we get a running time of $O(n+k^2)$. 

\subsection*{An $\tilde{O}(m^2k^m)$ algorithm for Alignment Distance of Multiple Sequences with $\tilde{O}(nm)$ Preprocessing}

We now extend the above $\tilde{O}(n+k^2)$ algorithm to computing alignment distance of $m$ strings. Recall that we are given $m$ strings $s_1,s_2,...,s_m$ each of length $n$. The following is an $O(n^m)$ time-complexity dynamic programming to obtain the edit distance of $m$ strings. We fill up an $m$-dimensional dynamic programming matrix $D[[0,n]....[0,n]]$ where the entry $D[i_1,i_2,..,i_m]$ computes the edit distance among the prefixes $s_1[1,i_1], s_2[1,i_2],...,s_m[1,i_m]$. As a starting condition, we have $D[0,...,0]=0$. Let  $\vec{e_i}=[0,0,..,\underbrace{1}_{i\text{th index}},0,0..]$, and $\mathbb{V}_{j}$ represents an $m$-dimensional vector $\langle j, j, j, ..., j \rangle$.  The dynamic programming is given by the following recursion.
For all $i_1,i_2,...,i_m \in [0,n]$
\[ D[i_1,i_2,...,i_m] = \min \left\{ \begin{array}{ll}
         D[\langle i_1, i_2,.., i_m \rangle - \vec{e_j}]+ 1&~~\mbox{for all $j=1,2,..,m$}\\
         & \mbox{ if $\langle i_1, i_2,.., i_m \rangle - \vec{e_j} \geq \langle 0,0,...,0\rangle$};\\
         D[i_1-1,i_2-1,..,i_m-1]
         &\mbox{if $s_1[i_1]=s_2[i_2]=...=s_m[i_m]$ and} \\ &  \mbox{ $i_1, i_2, ..., i_m >0$}.
        \end{array} \right. \] 
        
In order to compute $D[i_1,i_2,...,i_m]$, we can either delete an element $s_j[i_j]$, or align $s_1[i_1], s_2[i_2],..,s_m[i_m]$ if they all match. 
The time to compute each entry $D[i_1,i_2,...,i_m]$ is $m$. The overall running time is $O(mn^m)$. 
 
\noindent{\bf Observation.}  In order to design an $\tilde{O}(m^2k^m)$ algorithm with preprocessing $\tilde{O}(mn)$, first observe that if $\mathcal{A}(s_1,s_2,...,s_m) \leq k$ then it is not possible that in the final alignment, we have $m$ indices $i_1, i_2,...,i_m$ aligned to each other such that $\max{|i_j-i_k|, j,k \in [1,2,..,m]} > k$. Since all strings have equal length, this would imply a total number of deletions $> mk$, or $\mathcal{A}(s_1,s_2,...,s_m) > k$.
 
\noindent{\bf Algorithm.} Let diagonal $\vec{d}$ of matrix $D$ denotes an $(m-1)$ dimensional vector, and contains all $D[i_1,i_2,..i_m]$ such that $\langle i_2, i_3,..,i_m \rangle=\mathbb{V}_{i_1}+\vec{d}$. Let $D_k=\{\vec{d} \mid \max_j |d[j]| \leq k\}$.  Then $|D_k|= (2k+1)^{m-1}$ since each entry $i_j \in i_i+\{ -k,-k+1,..,0, 1,...,k\}$, for $j=2,3,..,k$. We want to identify the entries with values in $[0,km]$ located within diagonals $D_k$.
 
 We similarly define $L^h(d)=\max\{i : D[i, \mathbb{V}_i+\vec{d}]=h\}$. The $h$-wave is defined by $L^h=\{ L^h(\vec{d}) \mid \vec{d} \in D_k\}$. Therefore, the algorithm computes $L^h$ for $h=0,..,km$ in the increasing order of $h$ until a wave $e$ is computed such that $L^e(\vec{0})=n$ (in that case $\mathcal{A}(A,B) = e$), or the wave $L^{km}$ is computed in the case the algorithm is thresholded by $k$. Given $L^{h-1}$, we can compute $L^h$ as follows. 

Define $$Equal(i,\vec{d})=\max_{q \geq i}{(q \mid s_1[i,q]=s_2[i+d[1],q])=s_3[i+d[2],q]=....=s_m[i+d[m-1],q]}).$$ That is $Equal(i,\vec{d})$ computes the longest prefix of the first string starting at index $i$ that can be matched to all the other strings following diagonal $\vec{d}$.

Next, we define the neighboring diagonals $N_1(\vec{d})$ and $N_2(\vec{d})$ of $\vec{d}$.

$$N_1(\vec{d})=\{\vec{d'} \mid ||d-d'||_1=1 ~\&~ \vec{d'} < \vec{d}  \}.$$ 
$$N_2(\vec{d})=\{ \vec{d'}=\vec{d}+\mathbb{V}_{+1}\}.$$
 

Then, $L^0(0)=Equal(0,\vec{0})$ and 
\[ L^h(\vec{d}) = \max \left\{ \begin{array}{ll}
        Equal(L^{h-1}(\vec{d'} \in N_1(\vec{d})), \vec{d}) & \mbox{if $\vec{d'} \in D_k$ , $h-1 \geq 0$};\\
        Equal(L^{h-1}(\vec{d'} \in N_2(\vec{d}))+1,\vec{d}) & \mbox{if $\vec{d'} \in D_k$, $h-1 \geq 0$}.
        \end{array} \right. \]

Next, we show that it is possible to preprocess $s_i$, $i=1,2,..m$ separately so that even then each $Equal(i,\vec{d})$ query can be implemented in $O(m\log{n})$ time. 

\subsubsection*{Preprocessing Algorithm}
\label{sec:small-preprocess}

The preprocessing algorithm constructs $\log(n)+1$ hash tables for each string $s$.
The $\ell$-th hash table corresponds to window size $2^{\ell}$;
we use a rolling hash function (e.g. Rabin fingerprint) to construct a hash table 
of all contiguous substrings of $s$ of length $2^{\ell}$ in time $O(n)$. Since there are $\log{n}+1$ levels, the overall preprocessing time for $s$ is $O(n\log{n})$. Let $H_{s_i}[\ell]$ store all the hashes for windows of length $2^\ell$ of $s_i$ for $i=1,2,..,m$. Hence the total preprocessing time is $\tilde{O}(nm)$.

\subsubsection*{Answering $Equal(i,\vec{d})$ in $O(m\log{n})$ time}
\label{sec:small-query}
$Equal(i,\vec{d})$ queries can be implemented by doing a simple binary search over the presorted hashes in $O(m\log{n})$ time. Suppose $Equal(i,\vec{d}[j])=q_j$. We identify the smallest $\ell \geq 0$ such that $q_j < 2^{\ell}$, and then do another binary search for $q_j$ between $i+2^{\ell-1}$ to $i+2^{\ell}$. Finally, we set 
$Equal(i,\vec{d})=\min{(q_2,...,q_m)}$.

\section{An $\tilde{O}_m(\lambda n^m)$ Algorithm for Multi-sequence LCS} 

\begin{theorem}
\label{thm:hsmulti}
Given $m$ strings $s_1,\dots,s_m$ each of length $n$ such that $\mathcal{L}(s_1,\dots,s_m)=\lambda n$ where $\lambda\in (0,1)$, there exists an algorithm that computes $\mathcal{L}(s_1,\dots,s_m)$ in time $\tilde{O}_m(\lambda n^m+nm)$.
\end{theorem}

The algorithm is build over the algorithm of \cite{Hirschberg77}, that given two strings $x,y$ of length $n$ such that $\mathcal{L}(x,y)=\lambda n$, computes $\mathcal{L}(x,y)$ in time $\tilde{O}(\lambda n^2)$. 
Though the algorithm of \cite{Hirschberg77} shares a similar flavor with the classical quadratic time dynamic program algorithm, the main contribution of this work is that it introduces the concept of minimal $\ell$-candidates that ensure that to compute the LCS, instead of enumerating the whole DP, it is enough to compute some selective entries that are important. Moreover they show if the LCS is small then the total number of minimal $\ell$-candidates can be bounded. Also they can be constructed efficiently. 

\paragraph{An $\tilde{O}(\lambda n^2)$ Algorithm for LCS.} We first provide a sketch of the algorithm of \cite{Hirschberg77}. We start with a few notations. Given two indices $i,j\in[n]$, let $\mathcal{L}(i,j)$ denotes the length of the LCS of $x[1,i]$ and $y[1,j]$ and $x_i$ denotes the $i$th character of string $x$. 

Given indices $i,j$ we call $<i,j>$ an $\ell$-candidate if $x_i=y_j$ and $\exists i',j'\in[n]$ such that $i'<i$, $j'<j$ and $<i',j'>$ is an $(\ell-1)$-candidate. We say that $<i,j>$ is generated over $<i',j'>$. Also define $<0,0>$ to be the $0$-candidate. (for this purpose add a new symbol $\alpha$ at the beginning of both $x,y$. Hence $x[0]=y[0]=\alpha$)
With this definition using induction we can claim that $<i,j>$ is an $\ell$-candidate \emph{iff} $\mathcal{L}(i,j)\ge \ell$ and $x_i=y_j$. Moreover as $\mathcal{L}(x,y)=\lambda n$, the maximum value of $\ell$ for which there exists an $\ell$-candidate is $\lambda n$. Hence to compute the LCS what we need to do is to construct a sequence of  $0$-candidate, $1$-candidate, $\dots$, $(\lambda n-1)$-candidate and a $\lambda n$-candidate such that the $i$th candidate can be generated from the $(i-1)$th candidate. Note as for each $i$, there can be many $\ell$-candidates enumerating all of them will be time consuming. 

Therefor the authors bring the notion of minimal $\ell$-candidate that are generated as follows. Consider two $\ell$-candidates $<i_1,j_1>$ and $<i_2,j_2>$. If $i_1\ge i_2$ and $j_1\ge j_2$, then it is enough to keep only $<i_2,j_2>$ as any $(\ell+1)$-candidate that is generated from $<i_1,j_1>$, can be generated from $<i_2,j_2>$ as well. Call $<i_1,j_1>$ a spurious candidate. 

\begin{lemma}
\label{lem:a1}
Let the set $\{<i_\ell,j_\ell>, \ell\in\{1,2,\dots\}\}$ denotes the set of $\ell$-candidates. After discarding all spurious $\ell$-candidates it can be claimed that $i_1<i_2<\dots$ and $j_1>j_2>\dots$.
\end{lemma}
\begin{proof}
For any two $\ell$-candidates $<i_1,j_1>$ and $<i_2,j_2>$, either 1) $i_1<i_2$ and $j_1\le j_2$ or 2) $i_1<i_2$ and $j_1> j_2$ or 3) $i_1=i_2$ and $j_1\le j_2$ or 4) $i_1=i_2$ and $j_1> j_2$. In the first and third case $<i_2,i_2>$ is spurious and in the forth case $<i_1,j_1>$ is spurious. Hence after removing all spurious candidates it can be ensured that $i_1<i_2<\dots$ and $j_1>j_2>\dots$.
\end{proof}

The $\ell$-candidates which are left after the removal of all spurious candidates are called minimal $\ell$-candidates. Notice as for each $i$ there is at most one minimal $\ell$ candidate, total number of minimal $\ell$-candidates for all choices of $i$ and $\ell$ is at most $\lambda n^2$. Using this bound and Lemma 3 in \cite{Hirschberg77}, an algorithm can be designed to compute all the minimal $\ell$-candidates and thus $\mathcal{L}(x,y)$ in time $\tilde{O}(\lambda n^2)$.

\paragraph{Generalization for $m$ strings.}
Now we provide an upper bound on the number of minimal $\ell$-candidates for $m$ strings given $\mathcal{L}(s_1,\dots,s_m)=\lambda n$. 
Given indices $i_1,\dots,i_m$ we call $<i_1,\dots,i_m>$ an $\ell$-candidate if $s_1[i_1]=\dots=s_m[i_m]$ and $\exists i'_1,\dots,i'_m\in[n]$ such that $i'_j<i_j$, and $<i'_1,\dots,i'_m>$ is an $(\ell-1)$-candidate. We say that $<i_1,\dots,i_m>$ is generated over $<i'_1,\dots, i'_m>$. Similar to the two string case using induction we can prove $<i_1,\dots,i_m>$ is an an $\ell$-candidate \emph{iff} $\mathcal{L}(s_1,\dots,s_m)\ge \ell$ and $s_1[i_1]=\dots=s_m[i_m]$. Therefore to compute $\mathcal{L}(s_1,\dots,s_m)$ it will be enough to generate a sequence of  $0$-candidate, $1$-candidate, $\dots$, $(\lambda n-1)$-candidate and a $\lambda n$-candidate such that the $i$th candidate can be generated from the $(i-1)$th candidate. Next we describe the notion of spurious candidates and bound the total number of minimal $\ell$-candidates. 

For two $\ell$-candidates $<i_1,\dots,i_{m-2},i_{m-1},i_m>$ and $<i_1,\dots,i_{m-2},i'_{m-1},i'_m>$, if $i_{m-1}\ge i'_{m-1}$ and $i_m\ge i'_m$ then we call the tuple $<i_1,\dots,i_{m-2},i_{m-1},i_m>$ spurious and discard it as any $(\ell+1)$ tuple that is generated from $<i_1,\dots,i_{m-2},i_{m-1},i_m>$ can be generated from $<i_1,\dots,i_{m-2},i'_{m-1},i'_m>$ as well. The tuples that survives are called minimal $\ell$-candidates.
Hence following a similar argument as given for Lemma \ref{lem:a1}, we can claim the following.

\begin{lemma}
\label{lem:a2}
Let the set $\{<i_1,\dots,i_{m-2},i_{m-1}^\ell,i_{m}^\ell>, \ell\in\{1,2,\dots\}\}$ denotes the set of $\ell$-candidates for fixed values of $i_1,\dots,i_{m-2}$. After discarding all spurious $\ell$-candidates it can be claimed that $i_{m-1}^1<i_{m-1}^2<\dots$ and $i_{m}^1>i_{m}^2>\dots$.
\end{lemma}
Note this implies that for a fixed choice of $i_1,\dots,i_{m-1}$, there exists at most one minimal $\ell$-candidate. As $\ell=\lambda n$, total number of minimal $\ell$-candidates over all choices of $i_1,\dots,i_{m-1}$ and $\ell$ is at most $\lambda n^m$.

Next we state a lemma that is a generalisation of Lemma 3 of \cite{Hirschberg77} for $m$ strings.

\begin{lemma}
For $\ell\ge 1$ $<i_1,\dots,i_{m-2},i_{m-1},i_m>$ is a minimal $\ell$-candidate \emph{iff} $<i_1,\dots,i_{m-2},i_{m-1},i_m>$ is a $\ell$-candidate with the minimum $m$th coordinate value such that $low<i_m<high$ where $high$ is the minimum $m$th coordinate value of all $\ell$-candidates having first $m-2$ coordinate values $i_1,\dots,i_{m-2}$ and the $(m-1)$th coordinate value less than $i_{m-1}$ and low is the minimum $m$th coordinate value of all $(\ell-1)$-candidates having first $m-2$ coordinate values $i_1,\dots,i_{m-2}$ and the $(m-1)$th coordinate value less than $i_{m-1}$.
\end{lemma}

Together with the above lemma and the bound on the number of minimal $\ell$-candidates following the algorithm of \cite{Hirschberg77}, we can design an algorithm that computes $\mathcal{L}(s_1,\dots,s_m)$ in time $\tilde{O}_m(\lambda n^m+nm)$.

\section{Finding Alignment with Minimum Deletion Similarity}
\label{sec:mindelappendix}

\begin{theorem}
\label{thm:thm1}
Given $m$ strings $s_1,\dots,s_m$, each of length $n$, $q$ sets $S_1,\dots, S_q\subseteq[n]$ and a parameter $0\le d\le n$, there exists an algorithm that computes a common alignment $\sigma_n$ such that $\sum_{k\in [m]}|\bar{\sigma}_n(s_k)|\le dm$ and $|\cup_{j\in [q]}(\bar{\sigma}_n(s_1)\cap S_j)|$ is minimized in time $\tilde{O}(2^mmn^{m+1})$.
\end{theorem}

Our algorithm MinDelSimilarAlignment() uses dynamic program to compute $\sigma_n$. The dynamic program matrix $D$ can be defined as follows: $D_{i_1,\dots,i_m,x}$ where, $i_1,\dots,i_m\in[n]$ and $x\in[mn]$ represents the minimum overlap of the unaligned characters of $s_1[1,i_1]$ with $\cup_{h=1}^{q} S_h$ in a common subsequence of $s_1[1,i_1],\dots,s_m[1,i_m]$, of cumulative alignment cost at most $x$ 
We can compute $D_{i_1,\dots,i_m,x}$ recursively as follows. We consider four different cases and output the one that minimizes the overlap. As the technique is similar to the one used in Section \ref{sec:maxdel} instead of giving a detailed description we just provide the formal definition of the dynamic program.

\paragraph{\textbf{Case 1.}} 
First consider the case where, $\exists j\in [2,m]$ such that $s_j[i_j]\neq s_1[i_1]$ and $i_1\in \cup_{j\in[q]}S_j$. We can define $D_{i_1,\dots,i_m,x}$ as follows.

 \begin{equation}
 D_{i_1,\dots,i_m,x}= 
  min
  \begin{cases}
 \begin{cases}
 {min}_{\begin{subarray} CC\subseteq[m]\setminus{\phi}\\ x'\le x-|C|\\ 1\in C \end{subarray}} D_{i'_1,\dots,i'_m,x'}+1  & i'_j=i_j-1 \text{ if } j\in C \\
& \text{ else } i'_j=i_j 
 \end{cases}\\
 \begin{cases}
 {min}_{\begin{subarray} CC\subseteq[m]\setminus{\phi}\\ x'\le x-|C|\\ 1\notin C \end{subarray}} D_{i'_1,\dots,i'_m,x'}  & i'_j=i_j-1 \text{ if } j\in C \\
& \text{ else } i'_j=i_j 
 \end{cases}
\end{cases}\\
\end{equation}

\paragraph{\textbf{Case 2.}} Next consider the scenario where 
$\exists j\in [2,m]$ such that $s_j[i_j]\neq s_1[i_1]$ and $i_1\notin \cup_{j\in[q]}S_j$. We can define $D_{i_1,\dots,i_m,x}$ as follows.

 \begin{equation}
 D_{i_1,\dots,i_m,x}= 
\begin{cases}
 {min}_{\begin{subarray} CC\subseteq[m]\setminus{\phi}\\ x'\le jx-|C| \end{subarray}} D_{i'_1,\dots,i'_m,x'}  & i'_j=i_j-1 \text{ if } x\in C \\
& \text{ else } i'_j=i_j 
 \end{cases}
\end{equation}

\paragraph{\textbf{Case 3.}} Consider the case where $s_1[i_1]=\dots=s_m[i_m]$ and $i_1\in \cup_{j\in[q]}S_j$. 
We can define $D_{i_1,\dots,i_m,x}$ as follows.

 \begin{equation}
 D_{i_1,\dots,i_m,x}= 
 min
  \begin{cases}
  {min}_{x'\le x}D_{i_1-1,\dots,i_m-1,x'} \\
  
 \begin{cases}
 {min}_{\begin{subarray} CC\subseteq[m]\setminus{\phi}\\ x'\le x-|C|\\ 1\in C \end{subarray}} D_{i'_1,\dots,i'_m,x'}+1  & i'_j=i_j-1 \text{ if } j\in C \\
& \text{ else } i'_j=i_j 
 \end{cases}\\

 \begin{cases}
 {min}_{\begin{subarray} CC\subseteq[m]\setminus{\phi}\\ x'\le x-|C|\\ 1\notin C \end{subarray}} D_{i'_1,\dots,i'_m,x'}  & i'_j=i_j-1 \text{ if } j\in C \\
& \text{ else } i'_j=i_j 
 \end{cases}
  \end{cases}\\
\end{equation}

\paragraph{\textbf{Case 4.}} Lastly consider the case where $s_1[i_1]=\dots=s_m[i_m]$ and $i_1\notin \cup_{j\in[q]}S_j$. 
Formally $D_{i_1,\dots,i_m,x}$ can be defined as follows.

 \begin{equation}
 D_{i_1,\dots,i_m,x}= 
 min
\begin{cases}

 {min}_{x'\le x}D_{i_1-1,\dots,i_m-1,x'} \\
    
 \begin{cases}
 {min}_{\begin{subarray} CC\subseteq[m]\setminus{\phi}\\ x'\le x-|C| \end{subarray}} D_{i'_1,\dots,i'_m,x'}  & i'_j=i_j-1 \text{ if } j\in C \\
& \text{ else } i'_j=i_j 
 \end{cases}
 \end{cases}\\
\end{equation}
Notice, by the definition of the DP the required subsequence $\sigma_n$ is the sequence corresponding to $min_{x\le d}D_{n+1,\dots,n+1,x}$ where again we append each string with a new symbol $\#$ at the end which must all be aligned. Therefore we can use a similar backtracking process as used in algorithm MaxDelSimilarAlignment() to compute $L, \bar{L},\sigma_n$.

\paragraph{Running Time Analysis.} 
The running time of algorithm MinDelSimilarAlignment() is $\tilde{O}(2^mmn^{m+1})$. We can use a similar analysis as used in algorithm MaxDelSimilarAlignment().

\end{document}